\documentclass[a4paper,11pt]{article}
\usepackage{jcappub} 
\makeatletter
\newcommand{\preprint}[1]{\gdef\@preprint{#1}}
\newcommand{\printpreprint}{%
\begin{flushright}\normalfont\small\@preprint\end{flushright}\vspace{-2.4em}}
\makeatother
\usepackage{bm}
\usepackage{slashed}
\usepackage{textcomp}
\usepackage[T1]{fontenc} 

\newcommand{\be}{\begin{equation}}
\newcommand{\ee}{\end{equation}}
\newcommand{\bes}{\begin{equation*}}
\newcommand{\ees}{\end{equation*}}
\newcommand{\Eq}[1]{Eq.~\eqref{#1}}
\newcommand{\Eqs}[2]{Eqs.~\eqref{#1} and \eqref{#2}}

\newcommand{\squared}[1]{\left| #1 \right|^2}

\preprint{CALT-TH-2025-022}

\title{\boldmath Probing Non-Minimal Dark Sectors via the 21 cm Line at Cosmic Dawn}

\author[a]{Federico Cima}
\author[b,c]{and Francesco D'Eramo}

\affiliation[a]{Walter Burke Institute for Theoretical Physics, California Institute of Technology,\\ Pasadena, CA 91125, USA}
\affiliation[b]{Dipartimento di Fisica e Astronomia,  Università degli Studi di Padova,\\ Via Marzolo 8, 35131 Padova, Italy}
\affiliation[c]{Istituto Nazionale di Fisica Nucleare (INFN), Sezione di Padova,\\ Via Marzolo 8, 35131 Padova, Italy}

\emailAdd{fcima@caltech.edu}
\emailAdd{francesco.deramo@pd.infn.it}

\abstract{Observations of the hydrogen hyperfine transition through the 21 cm line near the end of the cosmic dark ages provide unique opportunities to probe new physics. In this work, we investigate the potential of the sky-averaged 21 cm signal to constrain metastable particles produced in the early universe that decay at later times, thereby modifying the thermal and ionization history of the intergalactic medium. The study begins by extending previous analyses of decaying dark matter (DM), incorporating back-reaction effects and tightening photon decay constraints down to DM masses as low as 20.4 eV. The focus then shifts to non-minimal dark sectors with multiple interacting components. The analysis covers two key scenarios: a hybrid setup comprising a stable cold DM component alongside a metastable sub-component, and a two-component dark sector of nearly degenerate states with a metastable heavier partner. A general parameterization based on effective mass spectra and fractional densities allows for a model-independent study. The final part presents two explicit realizations: an axion-like particle coupled to photons, and pseudo-Dirac DM interacting via vector portals or electromagnetic dipoles. These scenarios illustrate how 21 cm cosmology can set leading bounds and probe otherwise inaccessible regions of parameter space.}

\begin{document}
\printpreprint
\maketitle
\flushbottom

\section{Introduction}

The standard cosmological model ($\Lambda \rm CDM$) relies on the existence of a pressureless fluid with an average mass density about five times larger than that of baryonic matter. Evidence for this form of dark matter (DM) has been accumulated across a wide range of length scales~\cite{ber:par}, from galaxies to clusters and up to the scale of the observable universe. The observational support for DM remains one of the most compelling motivations for physics beyond the standard model (SM). Despite its ubiquity in the cosmos, its extremely weak interactions with the visible sector make uncovering its particle nature a formidable challenge.

The landscape of particle candidates remains vast and diverse~\cite{Feng:2010gw,Cirelli:2024ssz,Bozorgnia:2024pwk}. A rather compelling and plausible scenario, grounded in well-motivated theories and rich with phenomenological implications, posits that DM is not absolutely stable but only sufficiently long-lived to align with the universe’s thermal history. In this framework, its lifetime may be extremely long yet finite, opening the possibility of detecting signals from its decays. Cosmological probes offer powerful constraints on such scenarios and present unique opportunities to uncover their particle nature. Key observables include anisotropies in the cosmic microwave background (CMB)~\cite{pad:det,sla:cmb,lop:con,dia:con,sla:gen,xu:cmb}, CMB spectral distortions~\cite{chl:the,bol:spe}, and the Ly$\alpha$ forest~\cite{liu:lym,cap:cmb}. More broadly, many well-motivated extensions of the SM predict long-lived metastable states not necessarily related to DM, which can be abundantly produced in the early universe and later inject energy through decays, leaving observable imprints in cosmological data.

This work focuses on a highly promising avenue in observational cosmology: probing the high-redshift universe via the 21 cm line of neutral hydrogen, arising from the hyperfine transition of its ground state. As neutral hydrogen dominates the intergalactic medium (IGM), this signal provides valuable insight into the thermal and ionization state of the gas. Consequently, 21 cm cosmology~\cite{pri:21c,fur:cos} opens a new observational window onto the early universe, with potential to reveal previously inaccessible aspects of structure formation and reionization, as well as to constrain fundamental physics. As a late-time probe, akin to Ly$\alpha$ forest measurements, the 21 cm signal is a powerful tool to constrain low-redshift energy injection, often setting stronger bounds on DM lifetime than earlier probes~\cite{sla:les}. Numerous studies have explored using both the monopole and power spectrum of the 21 cm signal to constrain DM properties~\cite{dam:bou,cla:21c,mit:bou,evo:unv,fac:21c,sit:the,liu:imp,men:con,jon:fuz,sch:con,val:the,Xu:2024uas, Mittal:2021egv,Cheung:2018vww}. Several experiments are currently underway aiming to detect either the sky-averaged signal~\cite{phi:pro,sin:sar,spi:ant,mon:cal,deLeraAcedo:2022kiu} or its fluctuations~\cite{pob:ope,tin:rea,rot:lof,deb:hyd}. While limits on the power spectrum are available, monopole measurements remain very challenging due to overwhelming foregrounds. The only claimed detection so far, by the EDGES collaboration in 2018~\cite{bow:ana}, has been disputed by SARAS~\cite{sin:ont}.

Given these premises, this paper updates and extends previous analyses by introducing several novel contributions to the use of 21 cm cosmology as a probe of physics beyond the SM. We begin with a concise review in Sec.~\ref{sec:setup}, outlining the fundamentals of the 21 cm line and our methodology for setting bounds in the event of a hypothetical monopole detection. In Sec.~\ref{sec:DMbounds}, we revisit the potential of probing DM decays with 21 cm cosmology. Earlier analyses~\cite{cla:21c,mit:bou} did not account for back-reaction effects in the efficiency factors describing energy deposition in the IGM. Moreover, theoretical limitations in modeling low-energy photon and electron deposition had restricted photon-channel constraints to DM masses above $10$ keV. Recent advances~\cite{liu:exo1} now allow these bounds to extend to lighter masses. Accordingly, we present updated constraints on DM decays using a state-of-the-art code that models both energy injection and deposition into the IGM, while incorporating back-reaction from the thermal bath on exotic inputs. This improved treatment yields more precise and stringent bounds across all SM channels. For photon decay channels, we extend the constraints down to masses as low as $2 \mathcal{R} \simeq 20.4~\mathrm{eV}$, with $\mathcal{R}$ the energy of the Ly$\alpha$ transition.

A distinctive feature of this work is the exploration of non-minimal dark sectors. In the second part of the paper, we relax the assumption of minimality and move beyond Occam's razor, allowing for a dark sector with multiple degrees of freedom. Sec.~\ref{sec:non-minimal} contains model-independent results, with Sec.~\ref{subsec:sub-component} offering a thorough analysis of hybrid scenarios featuring both a stable cold dark matter (CDM) particle and a decaying component. Building on this, Sec.~\ref{subsec:nearlydegenerate} explores scenarios with two nearly degenerate dark particles, where the lighter is stable and accounts for the observed DM abundance, while the heavier decays into the lighter and visible particles, leaving imprints on the 21~cm signal. In Sec.~\ref{sec:Lag}, we construct explicit microscopic models that realize such non-minimal scenarios within well-motivated extensions of the SM. The first case, discussed in Sec.~\ref{subsec:ALP}, involves an axion-like particle (ALP); this is a setting where our refinements are particularly relevant, as Refs.~\cite{liu:exo2, xu:cmb} have shown that CMB anisotropy bounds dominate in the low-mass regime, and we expect forthcoming 21~cm constraints to be even more stringent. The second framework, presented in Sec.~\ref{sec:pseudoDirac}, considers two nearly degenerate Majorana fermions coupled to the visible sector through distinct portals. Secs.~\ref{sec:DMbounds} to \ref{sec:Lag}, which form the core of the article, can be read independently. Readers interested in a specific aspect (e.g., the model-independent study or the analysis of microscopic theories) can focus on the section most relevant to them. Our main findings are summarized in Sec.~\ref{sec:summary}, with additional technical details provided in the appendices.

\section{General Setup and Strategy}
\label{sec:setup}

We focus on the sky-averaged 21 cm signal and adopt a formalism that tracks the ionization and thermal history of the IGM in the presence of exotic energy injections. This section opens with a concise overview of 21 cm cosmology and then outlines the methodology used to derive constraints from a hypothetical measurement of the monopole signal. The modeling of the gas temperature and ionization fraction under energy injection is summarized in App.~\ref{app:ion&therhistory}.

The sky-averaged 21 cm signal is sensitive to additional injections of electromagnetically interacting particles occurring between recombination and the onset of star formation. During the cosmic dark ages, it effectively functions as a precise calorimeter, enabling the detection of deviations from $\Lambda$CDM predictions. The key observable is the \emph{differential brightness temperature}, $\delta T_b$, which quantifies the intensity contrast between the observed radio signal and a reference background, taken here to be the CMB. The monopole signal as a function of redshift $z$ is given by~\cite{pri:21c,fur:cos}
\begin{equation}
\overline{\delta T_b} = 23 \,\text{mK} \left(1-\frac{T_\gamma(z)}{T_S(z)}\right)\left(\frac{\Omega_b(t_0)h^2}{0.02} \right) \left(\frac{0.15}{\Omega_m(t_0)h^2} \right)^{1/2} \sqrt{\frac{1+z}{10}}\,x_{\rm HI}(z) \ , 
\label{signal}
\end{equation}
where $T_\gamma$ is the CMB temperature, $\Omega_b(t_0)$ and $\Omega_m(t_0)$ are the present-day baryon and matter density parameters, $h$ is the reduced Hubble constant, and $x_{\rm HI} = n_H / (n_p + n_H)$ denotes the neutral hydrogen fraction. As evident from this expression, the so-called \emph{spin temperature}, $T_S$, plays a central role in determining the signal. Its value is set by the ratio of hydrogen atoms in the triplet and singlet states
\begin{equation}
\frac{n_1(z)}{n_0(z)}\equiv 3\exp\left( -\frac{E_{10}}{T_S(z)}\right),
\end{equation}
where we assume a Maxwell-Boltzmann distribution, and $E_{10} \simeq 5.9~\mu\text{eV}$ is the hyperfine splitting of the hydrogen ground state.

The evolution of the spin temperature is governed by three primary mechanisms that can induce spin-flip transitions: (1) absorption or emission of CMB photons; (2) collisions between hydrogen atoms and other gas particles; (3) resonant scattering of Ly$\alpha$ photons, which can lead to spin flips during de-excitation. This last process is known as the Wouthuysen-Field (WF) effect~\cite{wou:ont,fie:exc}. Each mechanism is associated with a characteristic temperature: the CMB temperature $T_\gamma$, the gas temperature $T_g$, and the Ly$\alpha$ color temperature $T_\alpha$. Assuming equilibrium and detailed balance, these temperatures combine according to  
\begin{equation}
T_S^{-1} = \frac{T_\gamma^{-1} + y_c T_g^{-1} + y_\alpha T_\alpha^{-1}}{1 + y_c + y_\alpha},
\end{equation}
where $y_c$ and $y_\alpha$ are the effective collisional and Ly$\alpha$ coupling coefficients respectively. These coefficients depend on both the rates of the underlying microscopic processes, which are well characterized in the literature~\cite{fur:cos, wou:ont, fie:exc, fur:spi, kuh:the, zyg:hyp}, and the temperature of the radio background.

The evolution of the signal can be understood by examining Eq.~\eqref{signal}. Before recombination, the signal is zero due to the absence of neutral hydrogen; that is, $x_{\rm HI}(z > z_{\rm rec}) \sim 0$, with $z_{\rm rec} \sim 1100$. After recombination, the signal can appear either in absorption (negative) or emission (positive), depending on the sign of $(1 - T_\gamma / T_S)$. The main phases of the signal’s evolution as predicted by the $\Lambda$CDM model are summarized below.
\begin{itemize}
    \item $150 \lesssim z \lesssim z_{\rm rec}$. Compton scattering between CMB photons and the residual free electrons remaining after recombination remains efficient enough to keep the gas and photon bath in thermal equilibrium, ensuring $T_g \simeq T_\gamma$ during this period. The high gas density leads to strong collisional coupling, with $y_c \gg \{1, y_\alpha\}$, which forces the spin temperature to closely follow the gas temperature, $T_S \simeq T_g$. As a result, the spin temperature effectively equals the CMB temperature, and no 21 cm signal is expected.
    \item $30 \lesssim z \lesssim 150$. In this phase, Compton scattering becomes inefficient, causing the gas to thermally decouple from the CMB. Being non-relativistic, the gas cools adiabatically with redshift as $T_g \propto (1+z)^2$, cooling faster than the CMB. Meanwhile, the gas density remains sufficiently high for the spin temperature to stay closely coupled to the gas temperature. Therefore, $T_S < T_\gamma$, and the 21 cm signal is expected in absorption.
    \item $z_\star \lesssim z \lesssim 30$. As the gas becomes increasingly diluted, collisional coupling weakens below the level of radiative coupling to the CMB (i.e., $y_c < 1$). Consequently, the spin temperature aligns with the CMB temperature, $T_S \simeq T_\gamma$, causing the 21 cm signal to vanish once more until the formation of the first stars at redshift $z_\star$.
    \item $z_h \lesssim z \lesssim z_\star$. The formation of the first stars marks a pivotal moment, as they emit substantial fluxes of Ly$\alpha$ photons. Although this radiation is not particularly efficient at heating the gas, atomic recoils induced by the WF effect drive the Ly$\alpha$ color temperature to align with the gas temperature, $T_\alpha \simeq T_g < T_\gamma$. In addition, the intense Ly$\alpha$ flux makes the Ly$\alpha$ coupling dominant, $y_\alpha \gg \{1, y_c\}$. As a result, the 21 cm signal during this epoch is expected to appear in absorption. The star formation redshift remains uncertain, currently constrained only to the range $10 \lesssim z_\star \lesssim 25$.
    \item $z_r\lesssim z\lesssim z_h$. At a redshift $z_h$ following the onset of star formation, astrophysical heating becomes significant, raising the gas temperature above that of the CMB, $T_g > T_\gamma$. During this phase, the 21 cm signal is expected to appear in emission, as the spin temperature remains strongly coupled to the Ly$\alpha$ color temperature $T_\alpha$, and thus to the gas temperature $T_g$, due to the dominant Ly$\alpha$ coupling $y_\alpha$.
    \item $z\lesssim z_r$. The 21 cm signal disappears once the universe completes reionization at redshift $z_r$, as the neutral hydrogen fraction approaches zero, $x_{\rm HI} \sim 0$. Any remaining signal is expected to originate primarily from neutral hydrogen within collapsed structures.
\end{itemize}

\begin{figure}
\centering
\includegraphics[width=0.48\textwidth]{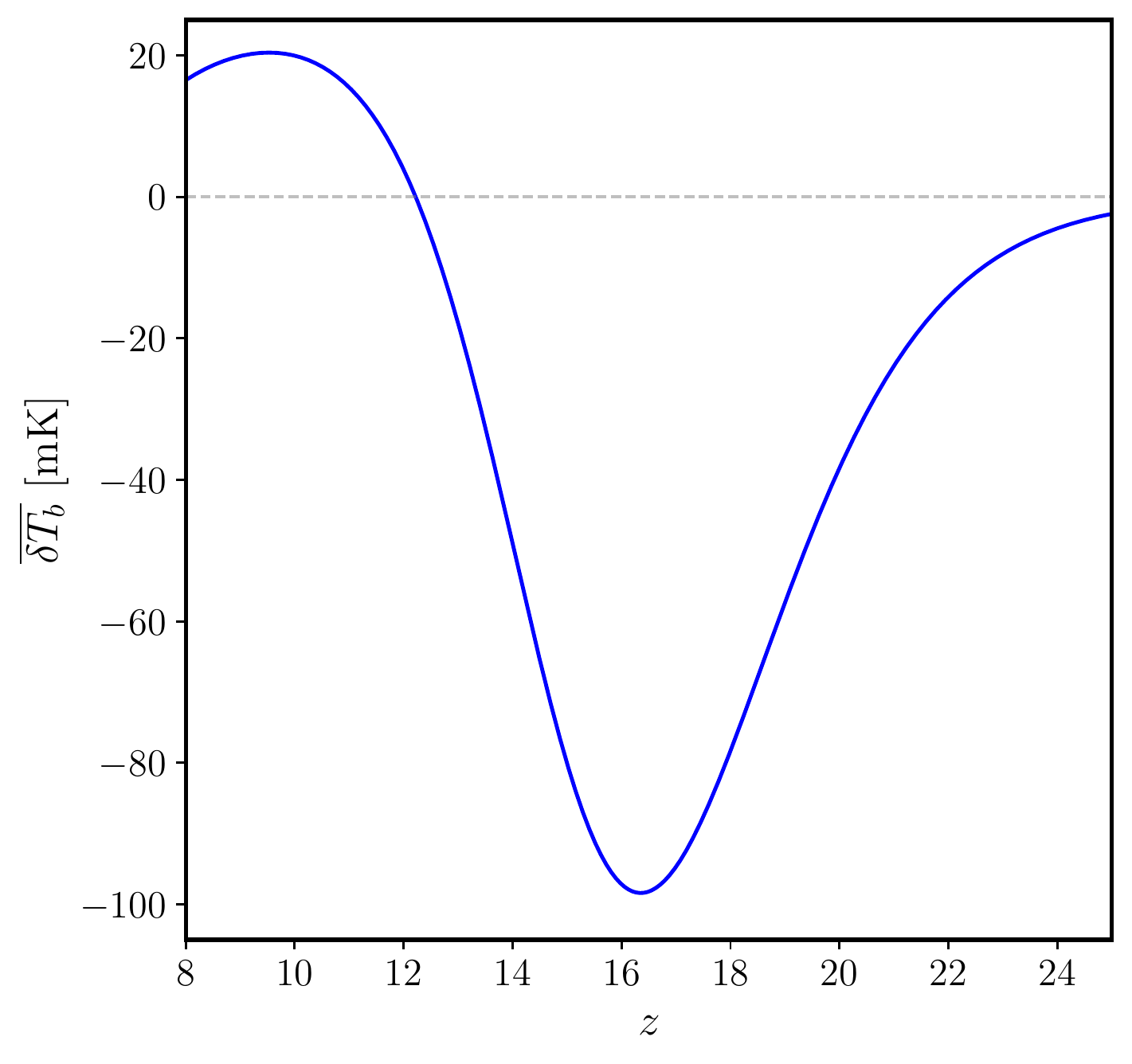} \quad 
\includegraphics[width=0.48\textwidth]{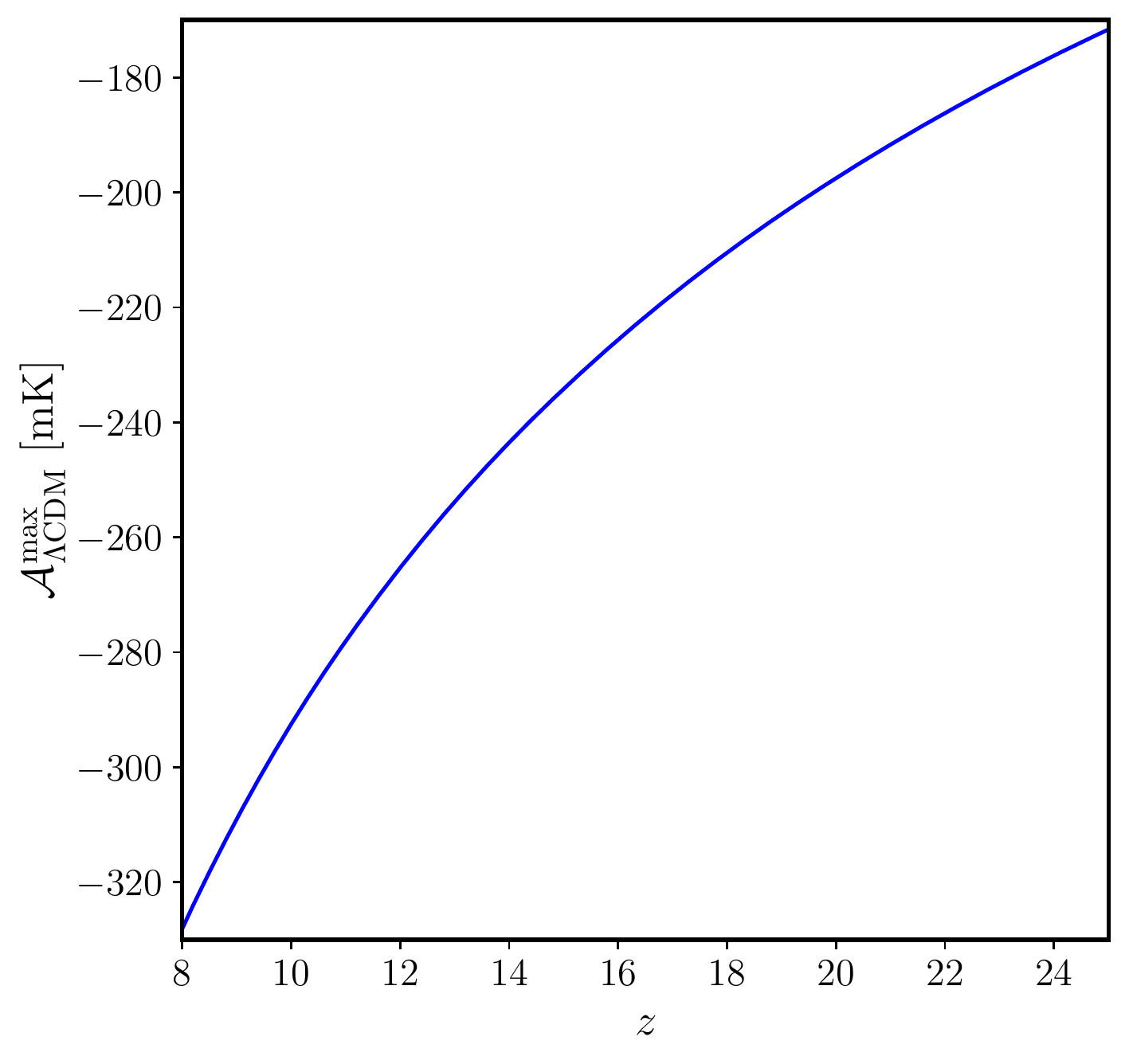}
\caption{Simulated monopole signal $\overline{\delta T_b}$ in the redshift range $8 < z < 25$, computed using the publicly available code \textsf{Zeus21}~\cite{mun:ane}, shown for two astrophysical scenarios. \textbf{Left panel:} Default fiducial parameters. \textbf{Right panel:} Maximally efficient Ly$\alpha$ coupling with no astrophysical heating.}
\label{fig:monopolesignal}
\end{figure}

The focus of our analysis is the absorption feature triggered by star formation, expected at lower redshifts and currently targeted by several experimental campaigns. The left panel of Fig.~\ref{fig:monopolesignal} presents a numerical simulation of the signal across the relevant redshift range. Using the fiducial astrophysical parameters of the \textsf{Zeus21} code~\cite{mun:ane}, which was employed to generate the signal, the spectral feature is clearly visible. 

In this work, we define the \emph{amplitude} of the signal as the minimum value of $\overline{\delta T_b}$. This minimum occurs at the \emph{trough redshift}, denoted $z_t$. All amplitudes are indicated by $\mathcal{A}$ and are negative quantities. Within the $\Lambda$CDM model, the brightness temperature reaches its lowest point around $z_t \sim 16$, with an amplitude $\mathcal{A}_{\Lambda \rm CDM} = \min_z \overline{\delta T_b}(z) \sim \overline{\delta T_b}(16) \sim -100\, \mathrm{mK}$.

The details of star formation remain highly uncertain, and the results presented here depend on the choice of fiducial astrophysical parameters. Nevertheless, it is possible to place a bound on the amplitude of the signal within the $\Lambda$CDM model. Since the amplitude is negative, this corresponds to an upper bound on its absolute value; for this reason, we refer to this quantity as the \textit{maximal amplitude}. This bound is obtained by considering the extreme case in which astrophysical heating is completely neglected, and $\overline{\delta T_b}$ is evaluated by imposing $T_S = T_g$. This approximation is commonly known as maximally efficient Ly$\alpha$ pumping. Physically, it corresponds to assuming that the timescale on which the WF effect couples the Ly$\alpha$ color temperature $T_\alpha$ to the gas temperature $T_g$, via atomic recoils, is much shorter than the timescale associated with astrophysical heating. Indeed, once $y_\alpha \gtrsim 1$ and the spin temperature begins to align with $T_\alpha$, then $T_\alpha \gtrsim T_g$, as can be verified both analytically and numerically~\cite{hir:wou, mun:ane}.

The plot in the right panel of Fig.~\ref{fig:monopolesignal} shows the evolution of the brightness temperature in the limit of maximally efficient Ly$\alpha$ pumping. Since astrophysical heating is neglected, no absorption dip appears in this case. Therefore, this function provides the maximal amplitude (in absolute value) of the signal within the $\Lambda$CDM model as a function of redshift
\begin{equation}
    \mathcal{A}_{\Lambda\rm CDM}^{\rm max}(z) \equiv \overline{\delta T_b}(T_S = T_g, z) \,.
    \label{eq:ALCDMmax}
\end{equation}
Any additional exotic energy injection would increase the gas temperature $T_g$, thereby reducing the amplitude (in absolute value) of the 21 cm signal. This observation underlies our strategy to set conservative bounds once experimental measurements of the 21 cm monopole signal become available.

Our strategy to set bounds shares similarities with previous approaches developed in the context of DM decays~\cite{cla:21c,mit:bou}. The underlying idea is straightforward. We start from the fact that the maximal amplitude compatible with the $\Lambda$CDM model is achieved through the approximation of maximally efficient Ly$\alpha$ pumping. Any exotic source of heating in the IGM would raise the brightness temperature, making the 21 cm signal less negative. Consequently, the following chain of inequalities holds
\begin{equation}
    \mathcal{A}_{\Lambda{\rm CDM}}^{\rm max}(z) < \mathcal{A}^{\rm max}_\chi(z) < \mathcal{T}(z) \,.
\end{equation}
Here, $\mathcal{A}_{\Lambda\rm CDM}^{\rm max}(z)$ is defined in Eq.~\eqref{eq:ALCDMmax} and shown in the right panel of Fig.~\ref{fig:monopolesignal}. The function $\mathcal{A}^{\rm max}_{\chi}(z)$ denotes the amplitude of the signal including energy deposition from the decays of a generic metastable state $\chi$, also assuming maximum Ly$\alpha$ efficiency. Finally, we denote $\mathcal{T}(z)$ as a threshold value: if a detection occurs, it should be set to the experimentally measured amplitude; otherwise, as is currently the case, its choice is arbitrary and various options are explored here.

The approach described above represents the most conservative strategy, as it attributes all the heating of the IGM to $\chi$ decays into SM particles while neglecting the astrophysical heating expected within the $\Lambda \rm CDM$ model. It is important to note that this reasoning assumes no exotic processes are cooling the gas or increasing the background CMB brightness temperature, which would produce an observed amplitude smaller than $\mathcal{A}_{\Lambda\rm CDM}^{\rm max}$.\footnote{Following the EDGES collaboration's reported detection~\cite{bow:ana}, several proposals along these lines were put forward to resolve or ease the apparent tension between the data and theoretical predictions~\cite{ber:sev,liu:imp,Barkana:2018qrx,pos:roo,mor:axi,mun:21c,liu:rev,Katz:2024ayw,Katz:2025sie}.}
While CMB spectral distortions caused by exotic decays could in principle alter the CMB brightness temperature at the relevant frequency of $1.42$ GHz, these effects are always negligible and can safely be ignored.

\begin{table}
    \centering
    \begin{tabular}{c|c c c c c}
         & $z_t=12$ & $z_t=15$ & $z_t=18$ & $z_t=21$ & $z_t=24$ \\ \hline
         $\mathcal{T}/\mathcal{A}_{\Lambda\rm CDM}^{\rm max}=0.25$ & $-66$ mK & $-59$ mK & $-52$ mK& $-48$ mK & $-44$ mK\\
         $\mathcal{T}/\mathcal{A}_{\Lambda\rm CDM}^{\rm max}=0.50$ & $-133$ mK & $-117$ mK & $-105$ mK & $-96$ mK & $-88$ mK\\
         $\mathcal{T}/\mathcal{A}_{\Lambda\rm CDM}^{\rm max}=0.75$ & $-199$ mK & $-176$ mK & $-158$ mK & $-144$ mK & $-132$ mK
    \end{tabular}
    \caption{Explicit values of the threshold amplitude $\mathcal{T}(z)$, rounded to the nearest integer, for each choice of the ratio $\mathcal{T}/\mathcal{A}_{\Lambda\rm CDM}^{\rm max}$ and the redshift of the trough $z_t$. For comparison, commonly used fixed threshold values in the literature are $-50$ mK and $-100$ mK.}
    \label{tab:threshold values in mK}
\end{table}

Our approach differs from previous studies in several key aspects. First, we use a more accurate treatment of energy deposition with the \textsf{DarkHistory} package \cite{liu:cod, liu:exo1}. Second, we adopt a more flexible strategy by varying not only the threshold amplitude $\mathcal{T}(z)$ but also the redshift of the trough $z_t$. As explained above, $z_t$ roughly corresponds to the redshift at which the first stars formed, a value that remains uncertain. To capture this, we scan a plausible range $z_t = \{12, 15, 18, 21, 24\}$ when studying the decay of a single-component DM particle.
Concretely, in order to perform such scan, we evaluate Eq. \eqref{eq:ALCDMmax} at each of these redshifts. The only input parameter that needs to be computed numerically is the gas temperature at the chosen redshift, $T_g(z_t)$, which we calculate running \textsf{DarkHistory}.
This allows us to explore how bounds depend on this parameter and gain insight into the qualitative dynamics of energy deposition in the IGM. Regarding the threshold amplitude, instead of fixing a single value as often done in the literature (e.g. $-50$ or $-100$ mK), we set $\mathcal{T}(z)$ to a constant fraction of the maximal amplitude predicted by the $\Lambda\mathrm{CDM}$ model at each chosen redshift. Specifically, we consider the three values $\mathcal{T}/\mathcal{A}_{\Lambda\rm CDM}^{\rm max} = \{0.25, 0.5, 0.75\}$. This choice reflects the significant variation of the IGM gas temperature across the relevant redshift range, where $T_g$ changes from about $3.6\,\mathrm{K}$ at $z=12$ to roughly $13.0\,\mathrm{K}$ at $z=24$. It also conveniently expresses the threshold as a percentage deviation from the baseline $\Lambda\mathrm{CDM}$ scenario with maximally efficient Ly$\alpha$ pumping. For ease of comparison with other works, the explicit threshold amplitudes $\mathcal{T}(z)$ are given in Tab.~\ref{tab:threshold values in mK} for all these choices.

We stress that measuring the 21-cm monopole remains extremely challenging, due to astrophysical uncertainties and the overwhelming foregrounds. These foregrounds exceed the expected signal by several orders of magnitude, making their subtraction a dominant source of systematic uncertainty in sky-averaged measurements. Because our bounds rely on the amplitude of the measured signal, they are unavoidably sensitive to the accuracy of this subtraction, which we assume can be achieved with sufficient precision. For these reasons, our bounds should be viewed as complementary to those obtained from 21-cm power spectrum measurements, where foreground subtraction is expected to be less severe; see, e.g., Ref.~\cite{pob:ope}.

\section{Decaying Dark Matter: Updated Analysis}
\label{sec:DMbounds}

We consider the \emph{single component} DM scenario, where a metastable particle $\chi$ accounts for all of the DM. We update and refine the constraints for this case. In contrast to earlier studies~\cite{mit:bou, cla:21c}, we incorporate back-reaction effects using the \textsf{DarkHistory} package~\cite{liu:cod}, which are expected to strengthen the bounds by at least $10$--$50\%$. Notably, exotic energy injection raises the free electron fraction, thus increasing the ionization of hydrogen and helium.\footnote{This, in turn, leads to a smaller $f^i(z)$ and a larger $f^h(z)$, both defined in Eq.~\eqref{eq:fczdef}.}

We examine two-body decay channels of the metastable particle $\chi$ into SM particles. The injection spectra are taken from Ref.~\cite{cir:ppp}, except for decays into photons, electrons, and muons, where we use spectra that omit electroweak corrections. This choice is motivated by the fact that Ref.~\cite{cir:ppp} provides spectra only down to $m_\chi = 10$ GeV, while our analysis aims to cover the full kinematically allowed mass range. In particular, this allows us to extend the prospective bounds for DM decaying into photons to lower masses. The new release of the \textsf{DarkHistory} code, \textsf{DHv2.0}, incorporates additional physical effects such as excitations to higher hydrogen levels and CMB spectral distortions~\cite{liu:exo1, liu:exo2}. However, as noted in the original papers, for masses above $10$ keV, these effects have negligible impact on the ionization and thermal history. To accelerate computations, we therefore use the earlier version, \textsf{DHv1.0}, for most of our analysis. While these additional effects may be relevant for lighter masses, \textsf{DHv2.0} is considerably slower than \textsf{DHv1.0}, and Ref.~\cite{liu:exo2} shows that \textsf{DHv1.0} remains accurate down to $m_\chi = 2\mathcal{R} \simeq 20.4$ eV. Accordingly, we scan the parameter space with \textsf{DHv1.0} and validate results below $10$ keV using \textsf{DHv2.0}.\footnote{This validation is necessary since Ref.~\cite{liu:exo2} focuses on constraints from CMB anisotropies at redshifts near $z \sim 300$, while our analysis targets much later epochs around $z \sim 15$.} As discussed in \cite{liu:exo2, xu:cmb}, even for masses below $2\mathcal{R}$, which produce photons with energies below the Ly$\alpha$ threshold, there remains an effect on the ionization and thermal history because such photons can excite higher hydrogen levels, facilitating ionization by CMB photons. Nevertheless, we limit our scan to $m_\chi \geq 2\mathcal{R}$, leaving detailed studies at lower masses for future work. Exploratory computations with \textsf{DHv2.0} suggest that decaying DM with masses of $3$, $5$, $10$, and $15$ eV and lifetime $\tau = 10^{19}$ s reduce the maximal amplitude of the 21 cm monopole signal by roughly $30\%$, while for $m_\chi = 1$ eV and the same lifetime, the effect drops below $5\%$.

\begin{figure}
    \centering
    \includegraphics[width=.48\linewidth]{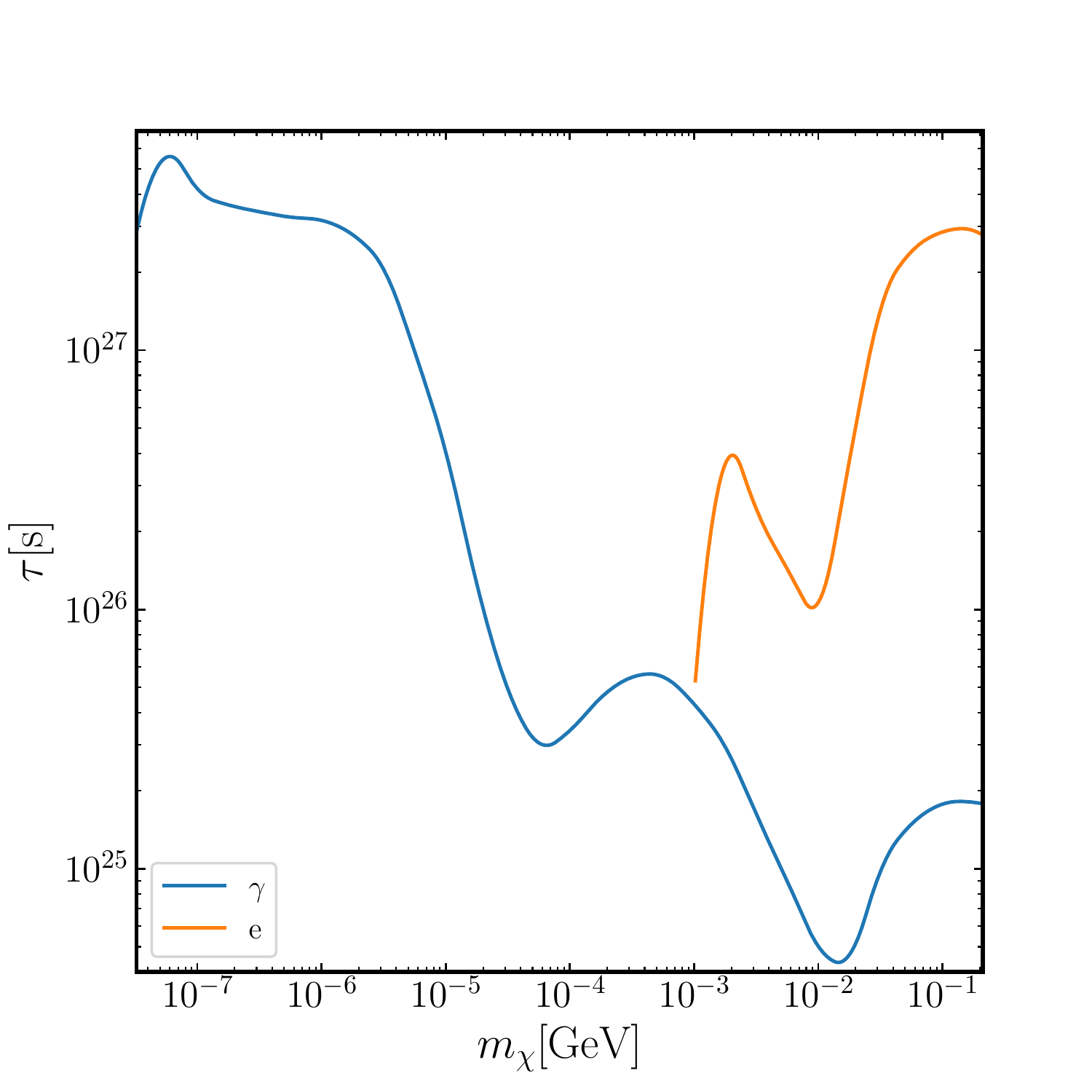}
    \includegraphics[width=.48\linewidth]{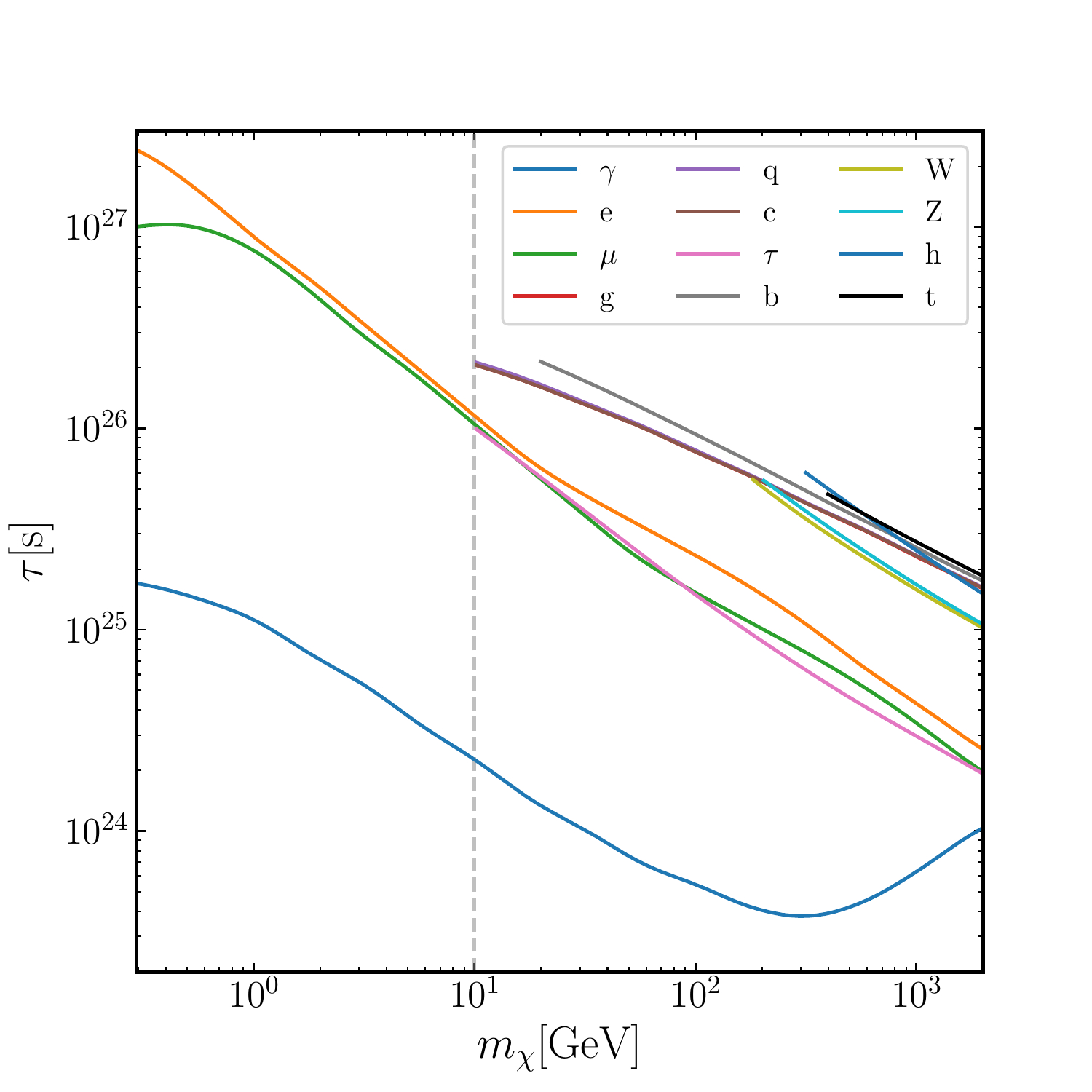}
    \caption{Bounds on the DM lifetime $\tau$ as a function of the DM mass $m_\chi$ in the single-component scenario. These limits assume a signal at redshift $z_t = 15$ with a threshold $\mathcal{T}/\mathcal{A}_{\rm std} = 0.25$. \textbf{Left panel:} Mass range $2\mathcal{R} < m_\chi < 2m_\mu$, where the only allowed final states are photons and $e^\pm$ pairs. \textbf{Right panel:} Mass range $2m_\mu < m_\chi < 2$~TeV. The vertical gray dashed line at $10$~GeV marks the lower bound for hadronically decaying particles to avoid perturbativity issues.}
    \label{fig:DMdecaysFULL}
\end{figure}

DM decays inject energy at a rate per unit volume given by
\begin{equation}
    \left(\frac{dE}{dV dt}\right)_{inj} = \frac{\rho_{\rm DM}(t)}{\tau} \simeq \frac{\rho_{\rm DM}(t_0)}{\tau} (1+z)^3,
    \label{decay injection single component}
\end{equation}
where $\tau$ is the DM lifetime and we have assumed $\tau \gg t$. The results for this scenario are shown in Fig.~\ref{fig:DMdecaysFULL}, covering all SM final states listed in the legend. We report bounds on the DM lifetime for the specific choice $z_t = 15$, consistent with the redshift favored by numerical simulations of the monopole signal, and for $\mathcal{T}/\mathcal{A}_{\Lambda\rm CDM}^{\rm max} = 0.25$, which corresponds to the most conservative parameter choice considered in this work. The left panel displays results for the mass range $2\mathcal{R} < m_\chi < 2m_\mu$, while the right panel shows the range $2m_\mu < m_\chi < 2$ TeV. The overall magnitude of the bounds and their relative ordering among channels are in agreement with those reported in Ref.~\cite{mit:bou}. These limits can be directly compared to the CMB anisotropy bounds from Ref.~\cite{sla:gen}, which provide the only other available cosmological constraints covering all the decay channels considered here. As expected from basic scaling arguments~\cite{sla:les}, the 21 cm bounds are stronger by roughly two to three orders of magnitude.

\begin{figure}
    \centering
    \includegraphics[width=.48\linewidth]{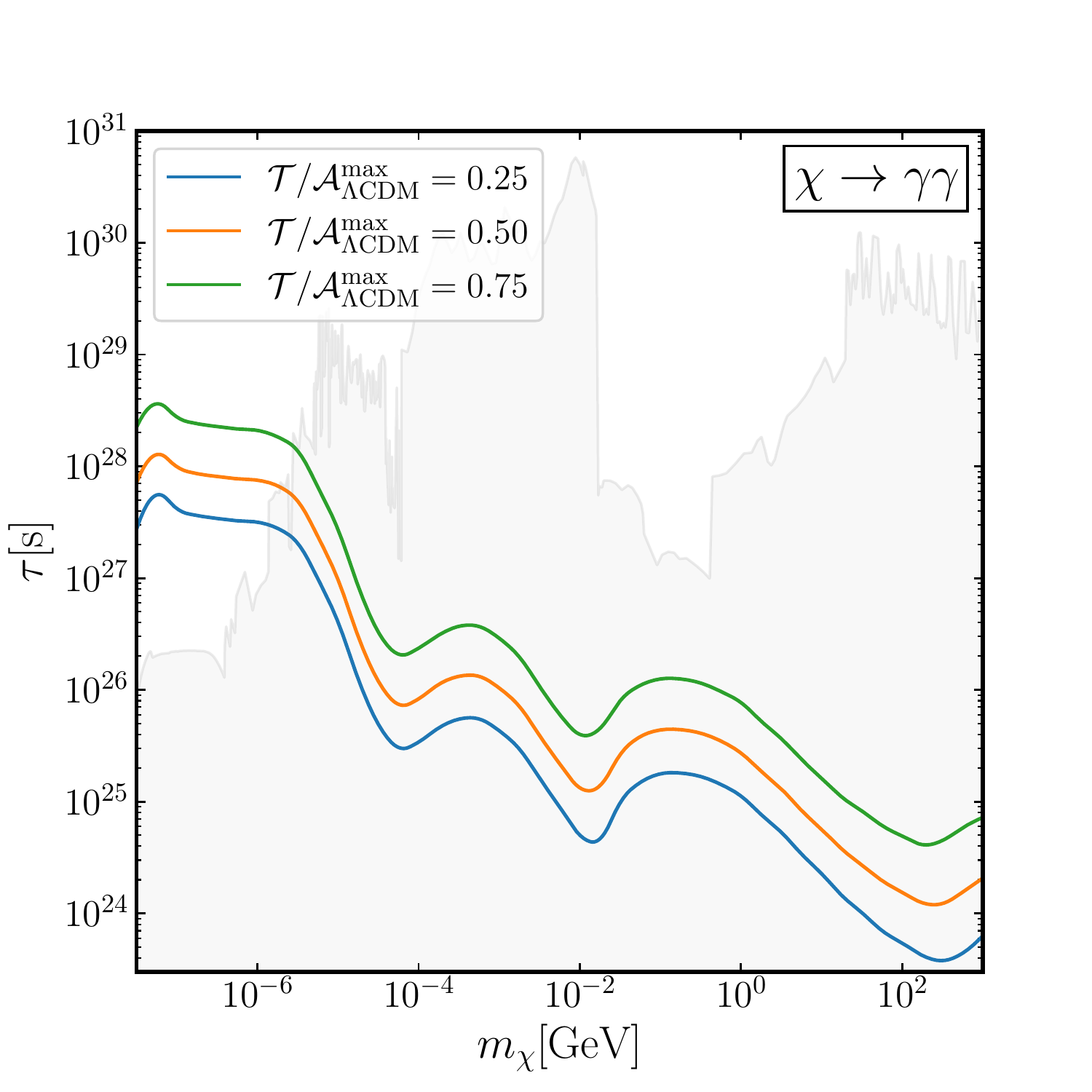}
    \includegraphics[width=.48\linewidth]{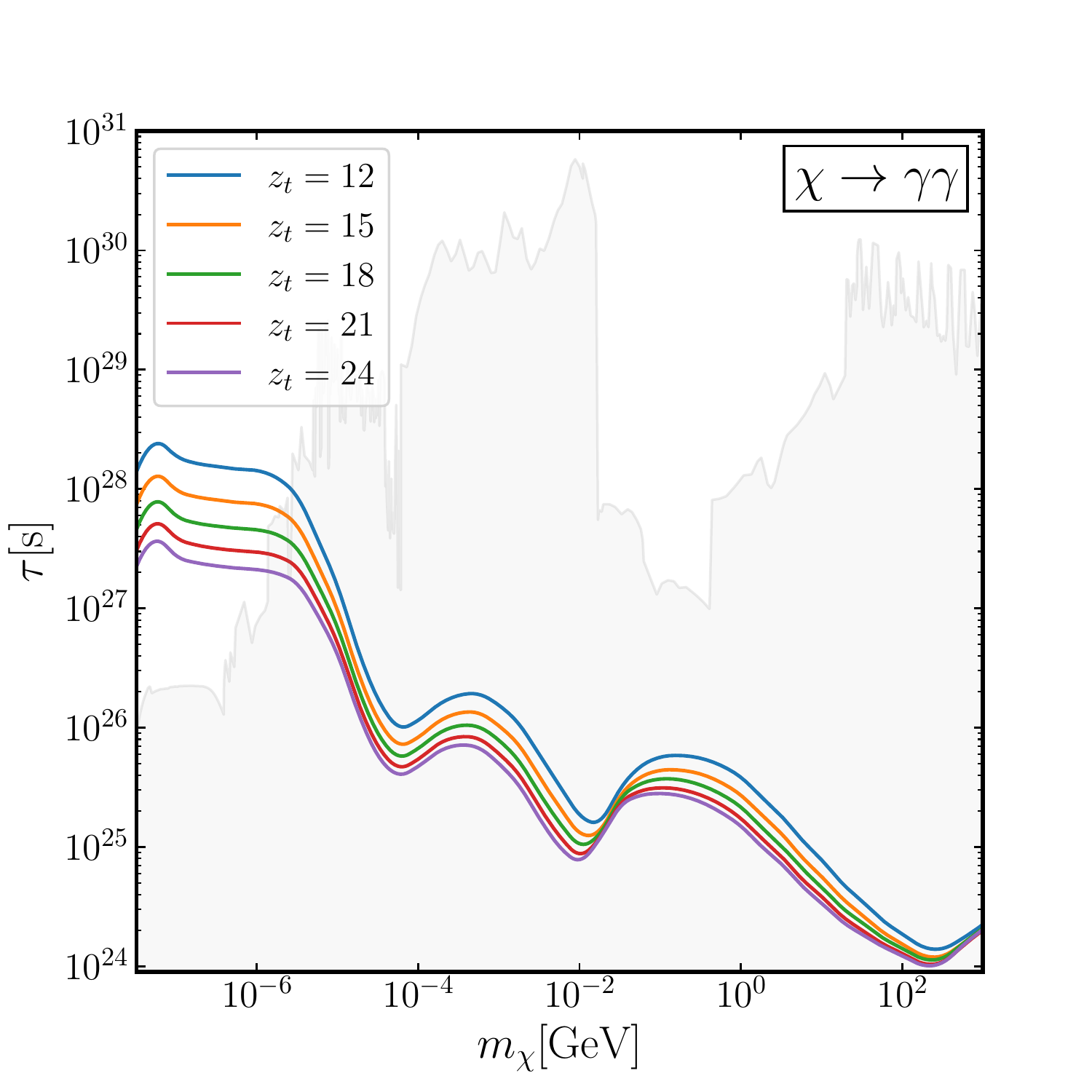}
    \caption{Bounds on the DM lifetime $\tau$ as a function of the DM mass $m_\chi$ for the  channel $\chi \to \gamma \gamma$. We investigate the dependence of the bounds on the redshift of the trough $z_t$ and the threshold $\mathcal{T}/\mathcal{A}_{\Lambda\rm CDM}^{\rm max}$. Existing constraints are shown shaded in light gray ~\cite{wad:str, Arias:2012az, Essig:2013goa, Horiuchi:2013noa, Ng:2019gch, Sicilian:2020glg, Foster:2021ngm, Foster:2022nva, Roach:2022lgo, Cirelli:2023tnx, Massari:2015xea,Giesen:2015ufa, Boudaud:2016mos}. \textbf{Left panel:} The redshift is fixed at $z_t = 15$, and three different values of the threshold are shown. \textbf{Right panel:} The threshold is fixed at $\mathcal{T}/\mathcal{A}_{\Lambda\rm CDM}^{\rm max} = 0.50$, while the redshift $z_t$ is varied.}
    \label{fig:singlecomponentphoton}
\end{figure}

We now turn to a detailed study of decay into photons. In Fig.~\ref{fig:singlecomponentphoton}, we show the corresponding bounds obtained by varying $\mathcal{T}/\mathcal{A}_{\Lambda\rm CDM}^{\rm max}$ at fixed $z_t = 15$ (left panel), and by varying $z_t$ at fixed $\mathcal{T}/\mathcal{A}_{\Lambda\rm CDM}^{\rm max} = 0.50$ (right panel).\footnote{As mentioned above, for DM masses below $10$~keV, it may be important to treat carefully the photon spectrum below the hydrogen ionization threshold. In this regime, using \textsf{DHv2.0} instead of \textsf{DHv1.0} can be relevant. To validate our computation with \textsf{DHv1.0} for $m_\chi < 10$~keV, we computed point-wise the expected signal $\mathcal{A}_\chi(z)$ using \textsf{DHv2.0}, setting $\tau$ to the limiting value obtained with \textsf{DHv1.0} for a fixed threshold $\mathcal{T}(z)$. We verified that the resulting amplitude remained sufficiently close to the specified threshold. More precisely, we found that $\mathcal{A}_{\Lambda\rm CDM}^{\rm max}(z)/\mathcal{T}(z)$ matched the originally specified value with discrepancies well below $1\%$.}
 The qualitative trends are as expected. Lower values of $z_t$ yield stronger bounds, since DM decays have more time to heat the IGM, thereby increasing the relative deviation from the $\Lambda$CDM prediction. Similarly, increasing $\mathcal{T}/\mathcal{A}_{\Lambda\rm CDM}^{\rm max}$ also tightens the bounds, as it corresponds to a more restrictive condition on the allowed heating. In practice, varying $\mathcal{T}/\mathcal{A}_{\Lambda\rm CDM}^{\rm max}$ approximately results in a rescaling of the bounds. 

All minima and maxima in the curves of Fig.~\ref{fig:singlecomponentphoton} are shifted toward higher masses compared to bounds derived from CMB anisotropies. This shift arises from the greater dilution of gas at lower redshifts due to cosmic expansion, which increases the characteristic length scale over which photons and electrons deposit energy into the IGM. Because the shape of the bounds depends on the energy dependence of the cross sections that govern the cooling cascades of electrons and photons, achieving a similar relative enhancement or suppression of energy deposition requires higher masses at lower redshifts. This behavior is clearly illustrated in the right panel of Fig.~\ref{fig:singlecomponentphoton}. At very low masses, a small peak appears due to injected photons lacking sufficient energy to ionize hydrogen or excite Ly$\alpha$ transitions. A similar feature, shifted to slightly lower masses, was also observed in Ref.~\cite{xu:cmb}, where CMB bounds below the ionization threshold were recently obtained using the full capabilities of \textsf{DHv2.0}.

\begin{figure}
    \centering
    \includegraphics[width=.48\linewidth]{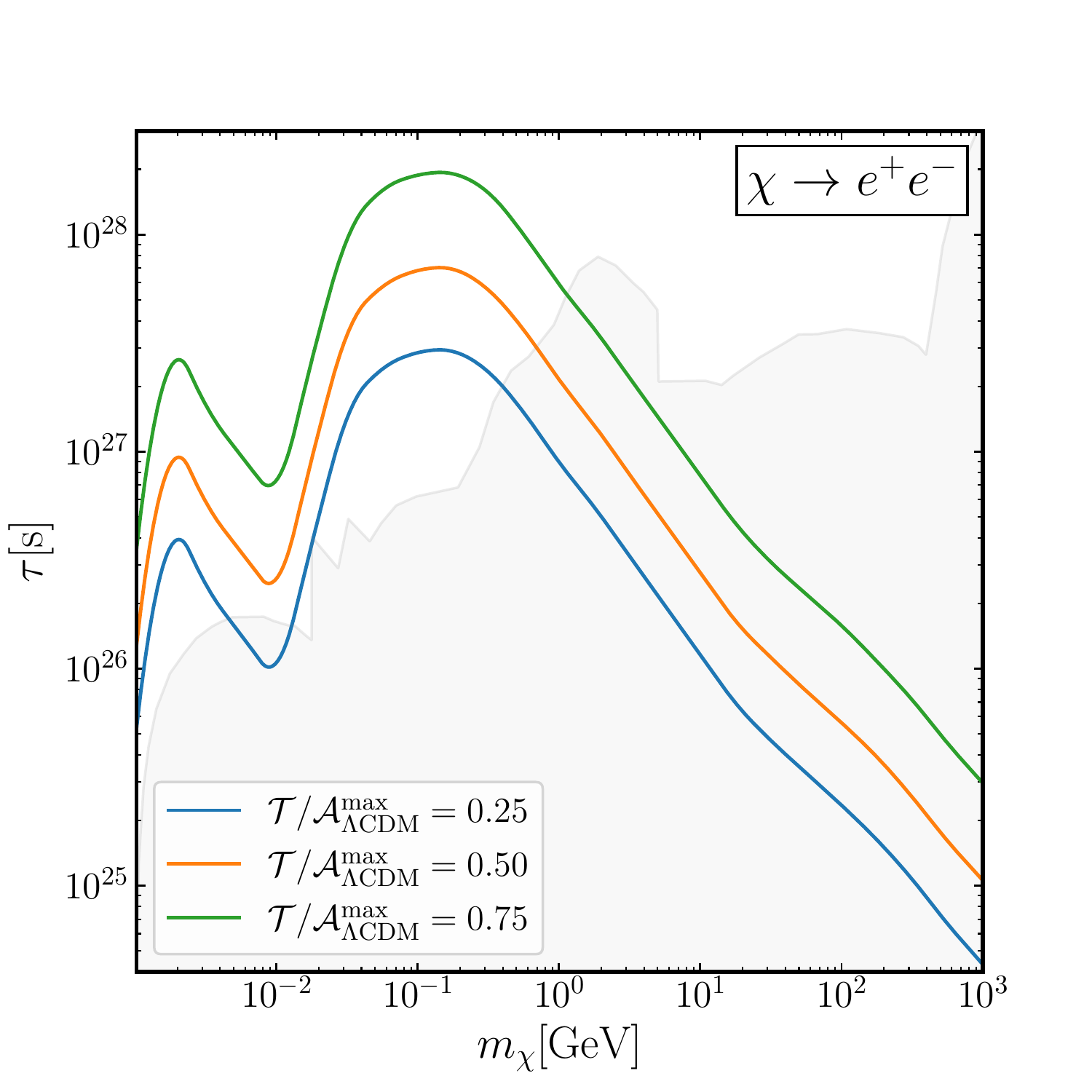}
    \includegraphics[width=.48\linewidth]{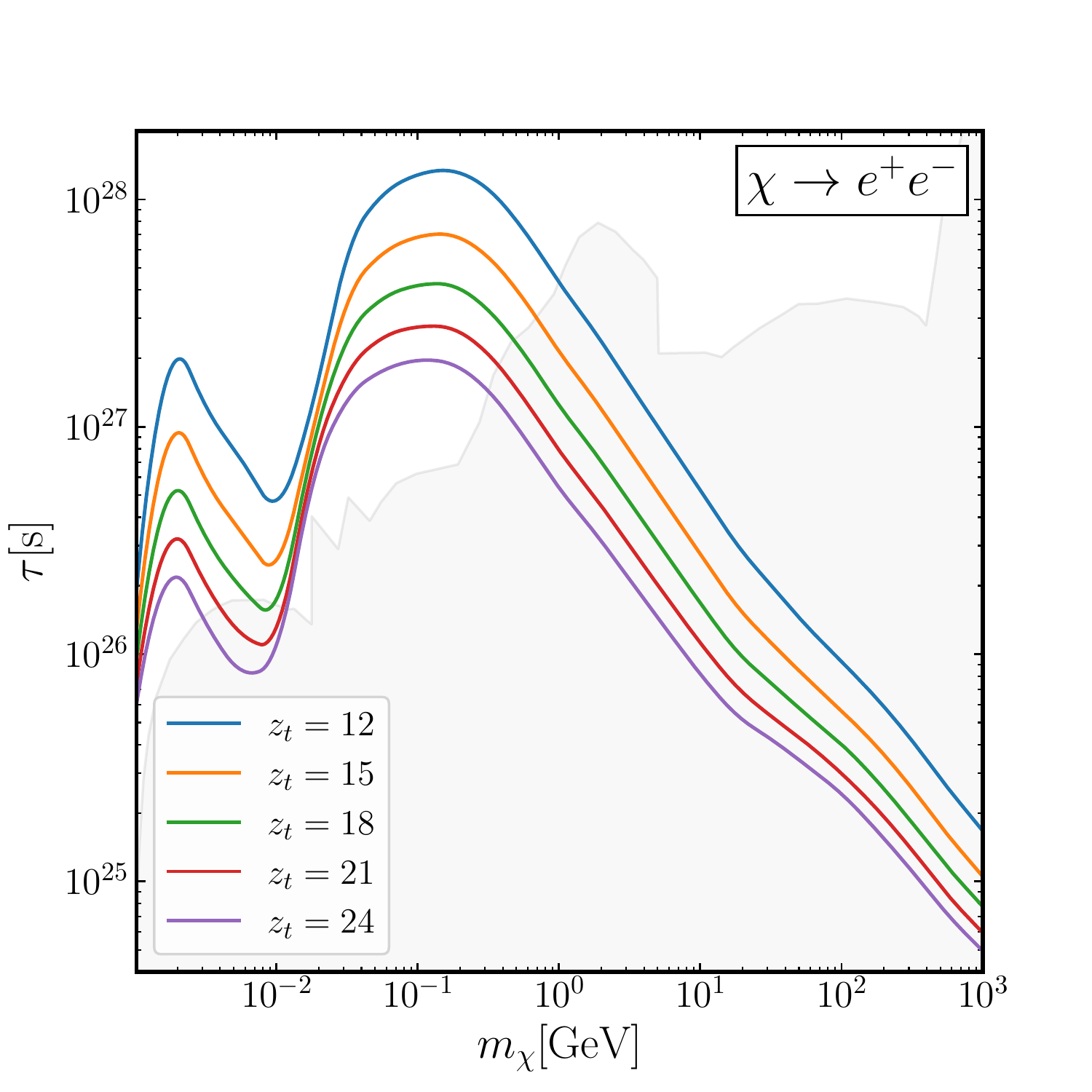}
    \caption{Same as Fig.~\ref{fig:singlecomponentphoton} but for the decay channel $\chi\to e^+ e^-$.}
    \label{fig:singlecomponentsboundelectrons}
\end{figure}

Beyond photons, the decay channel into electron/positron has notable phenomenological implications, with corresponding bounds shown in Fig.~\ref{fig:singlecomponentsboundelectrons} that follow a similar qualitative trend to those of the photon case. A comprehensive review of how the bounds depend on the choice of 21 cm signal for other decay channels is presented in App.~\ref{app: additional results}.

We conclude this section with an obvious yet particularly useful remark for the remainder of this article and for future analyses. While we treat $\chi$ as DM and thus adopt $\rho_{\mathrm{DM}}(t_0) \simeq 1.26~\mathrm{GeV/cm}^3$, it is straightforward to reinterpret our bounds for any long-lived relic $\phi$ with a present-day abundance $\rho_{\phi}(t_0)$. Specifically, the lower bound on the lifetime in this case can be obtained by rescaling $\tau$ to $\left[\rho_{\phi}(t_0)/\rho_{\mathrm{DM}}(t_0)\right] \tau$.

\section{Probing Non-Minimal Dark Sectors}
\label{sec:non-minimal}

We now turn to dark sectors comprising multiple degrees of freedom and consider two distinct scenarios. In Sec.~\ref{subsec:sub-component}, we analyze a sub-dominant, unstable DM component that decays into visible states, thereby altering the 21 cm signal. In Sec.~\ref{subsec:nearlydegenerate}, we explore a dark sector consisting of two nearly degenerate states, where decays of the heavier, unstable state into the lighter, stable one inject energy that affects 21 cm cosmology. Our treatment remains model-independent, focusing on effective parameters such as the density fraction of the unstable sub-component and the mass spectrum of the dark sector states. Explicit Lagrangian realizations of these scenarios are discussed in the subsequent section.

\subsection{Metastable DM sub-component}
\label{subsec:sub-component}

The first scenario we consider features a dark sector composed of both a stable CDM component and a metastable DM sub-component that eventually decays into visible states. The analysis in Sec.~\ref{subsec:ALP} provides an explicit Lagrangian realization of this scenario, in which the metastable state is an ALP coupled to photons. Contrary to naive expectations, the resulting bounds are not a simple rescaling of those obtained for a single-component DM model. In particular, the lifetime of the sub-dominant unstable component can be shorter than the age of the universe.

Let us denote the decaying component by $\chi$. Assuming the presence of a stable CDM component, the total DM energy density evolves as  
\begin{equation}
    \rho_{\rm DM}(t) = \rho_{\rm CDM}(t) + \rho_\chi(t) = \left[\rho_{\rm CDM}(\bar{t}) + \rho_\chi(\bar{t})e^{-(t-\bar{t})/\tau}\right]\left(\frac{a(\bar{t})}{a(t)}\right)^3 \ .
    \label{exact DM mass density evolution}
\end{equation}
The first equality follows directly from the hybrid DM framework under consideration, while the second requires further clarification. Here, $\bar{t}$ denotes a reference time in the early universe, $a(t)$ is the scale factor, and the expression is valid for $t \geq \bar{t}$. This reference time must satisfy several conditions. First, the production mechanisms of both CDM and $\chi$ (e.g., freeze-out, freeze-in, misalignment) must have concluded by $\bar{t}$. Since the two components may originate from different processes at different epochs, we take $\bar{t}$ to be later than the time of the last production mechanism. No number-changing processes occur for $t \geq \bar{t}$, and both CDM and $\chi$ evolve by free-streaming along geodesics in the expanding universe. However, this condition alone does not justify the second equality in Eq.~\eqref{exact DM mass density evolution}, which assumes that both components are non-relativistic at $\bar{t}$, as reflected by the $a(t)^{-3}$ scaling. While dark sector particles may initially be relativistic, the equation remains valid provided they have redshifted and cooled sufficiently to behave as non-relativistic matter by $\bar{t}$. We therefore choose $\bar{t}$ late enough to ensure that both CDM and $\chi$ can be treated as non-relativistic fluids. This construction remains consistent as long as $\chi$ becomes non-relativistic well before it decays, i.e., if $\tau \gg \bar{t}$. 

We characterize the sub-component by its relative abundance $F_\chi$ prior to decay
\begin{equation}
    F_\chi \equiv \frac{\rho_\chi(\bar{t})}{\rho_{\rm CDM}(\bar{t}) + \rho_\chi(\bar{t})} \ .
    \label{relative abundance parameterization}
\end{equation}
The single-component scenario is recovered in the limit $F_\chi \to 1$, with the additional requirement that the DM lifetime exceeds the age of the universe, i.e., $\tau \gg t_0$. 

Observations of the 21 cm signal are primarily sensitive to energy injection occurring after recombination. Earlier energy injections would induce CMB spectral distortions that modify the background photon distribution relevant for the cooling cascade rates. However, such early energy depositions do not significantly affect the thermal and ionization history of the IGM. This effect is negligible for our purposes and is not studied further here. Therefore, we safely set \(\bar{t} = 0\) in all subsequent expressions.

The power injected per unit physical volume due to $\chi$ decays is obtained from the expression in Eq.~\eqref{decay injection single component}, replacing $\rho_{\rm DM}(t)$ with $\rho_\chi(t)$. Expressing the latter in terms of the relative abundance defined in Eq.~\eqref{relative abundance parameterization}, we find
\begin{equation}
    \left(\frac{dE}{dV\, dt}\right)_{\rm inj}^{\chi_{\rm subDM}} = \frac{F_\chi\, e^{(t_0 - t)/\tau}}{e^{t_0/\tau}(1 - F_\chi) + F_\chi} \, \frac{\rho_{\rm DM}(t_0)\, (1+z)^3}{\tau} 
    \simeq F_\chi\, e^{-t/\tau} \, \frac{\rho_{\rm DM}(t_0)\, (1+z)^3}{\tau} \ .
    \label{generalpowerinjectedfromsubcomponent}
\end{equation}
Here, the superscript indicates that we are considering the case of a DM sub-component, and the redshift of energy injection is related to the scale factor by $1 + z = a(t_0)/a(t)$. This expression assumes that $\chi$ decays entirely into SM particles; a scenario where this assumption does not hold is discussed in the next subsection. The first equality in Eq.~\eqref{generalpowerinjectedfromsubcomponent} is completely general and does not rely on any particular value of the lifetime $\tau$ or the relative abundance $F_\chi$. The final equality, on the other hand, follows the standard approximation commonly adopted in the literature~\cite{sla:gen}, which is valid when the exponential in the denominator dominates. This simplification holds to better than $\mathcal{O}(1\%)$ accuracy for lifetimes $\tau \gtrsim 10^{19}$~s or for abundances $F_\chi \lesssim 10^{-6}$.

We adopt a procedure similar to that employed in the previous section. The main difference is the presence of an additional parameter: the relative abundance $F_\chi$. To incorporate this dependence, we minimally modified the \textsf{DH} code by replacing the standard decay-induced energy injection with the expression in Eq.~\eqref{generalpowerinjectedfromsubcomponent}, and introducing the parameter $F_\chi$ in the function \texttt{main.evolve}. To reduce computational cost, we restrict the calculation of bounds to the lifetime range $10^{13.2}~\text{s} < \tau < 10^{18}~\text{s}$. For $\tau \gtrsim 10^{18}~\text{s}$, the exponential in Eq.~\eqref{generalpowerinjectedfromsubcomponent} approaches unity, and the bounds become a trivial rescaling of those obtained in the single-component scenario. As fiducial redshift for the dip, we adopt $z_t = 15$, a value consistent with numerical simulations (see e.g.~\cite{mun:ane}). For the threshold, we take $\mathcal{T}/\mathcal{A}_{\Lambda\rm CDM}^{\rm max} = 0.5$. According to Tab.~\ref{tab:threshold values in mK}, this corresponds to a threshold of approximately $-100$~mK, which is a reference value commonly used in the literature and close to the expected signal amplitude.

\begin{figure}
    \centering
    \includegraphics[width=.48\linewidth]{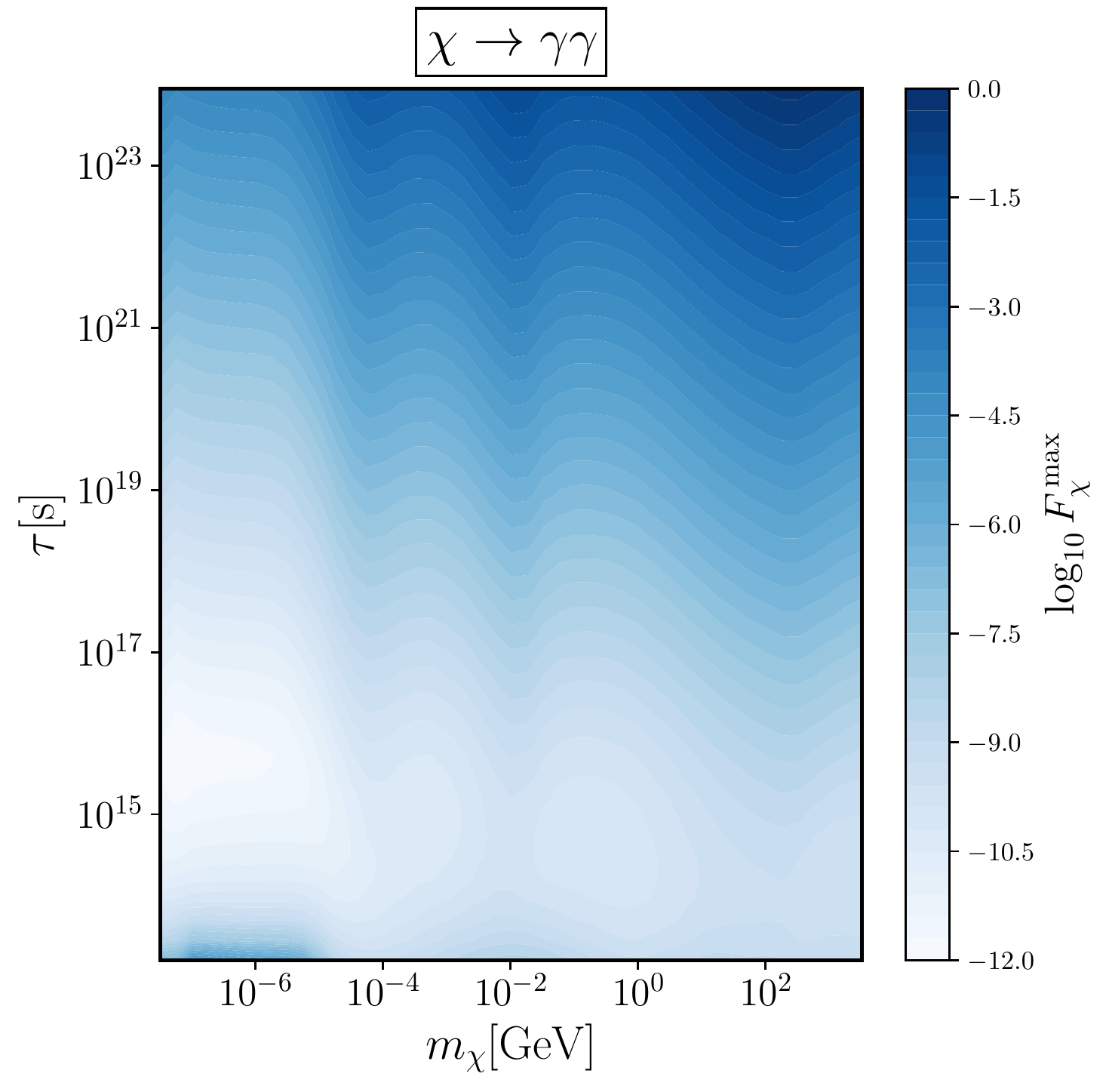}
    \includegraphics[width=.48\linewidth]{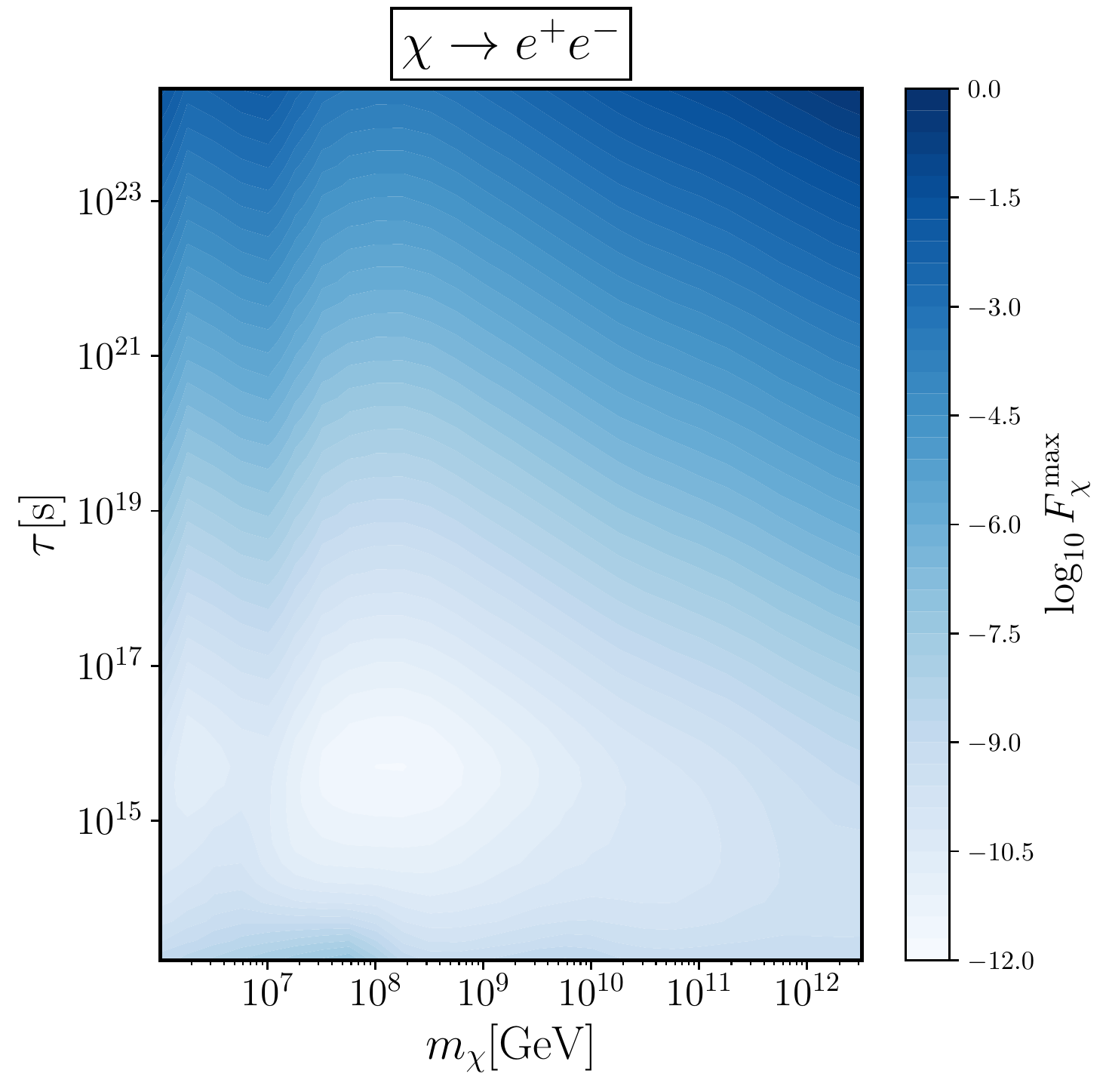}
    \caption{Maximal allowed fraction $F^{\rm max}_\chi$ of the unstable component $\chi$ in the $(m_\chi, \tau)$ plane, as constrained by a 21 cm monopole observation with $z_t = 15$ and $\mathcal{T}/\mathcal{A}_{\Lambda\rm CDM}^{\rm max} = 0.5$. \textbf{Left panel:} Decay mode $\chi \to \gamma\gamma$. \textbf{Right panel:} Decay mode $\chi \to e^+ e^-$.}
    \label{fig:sub-componentheatmaps}
\end{figure}

We explore the three-dimensional parameter space $(\tau, m_\chi, F_\chi)$, as illustrated in Fig.~\ref{fig:sub-componentheatmaps}. Specifically, we present the bounds as a heat map for the decay channels into photons (left panel) and electron--positron pairs (right panel). This approach differs from previous studies in the literature, particularly those focusing on CMB constraints~\cite{pou:cos, sla:gen}, as it preserves the full dependence of the bounds on all three parameters.

\begin{figure}
    \centering
    \includegraphics[width=.48\linewidth]{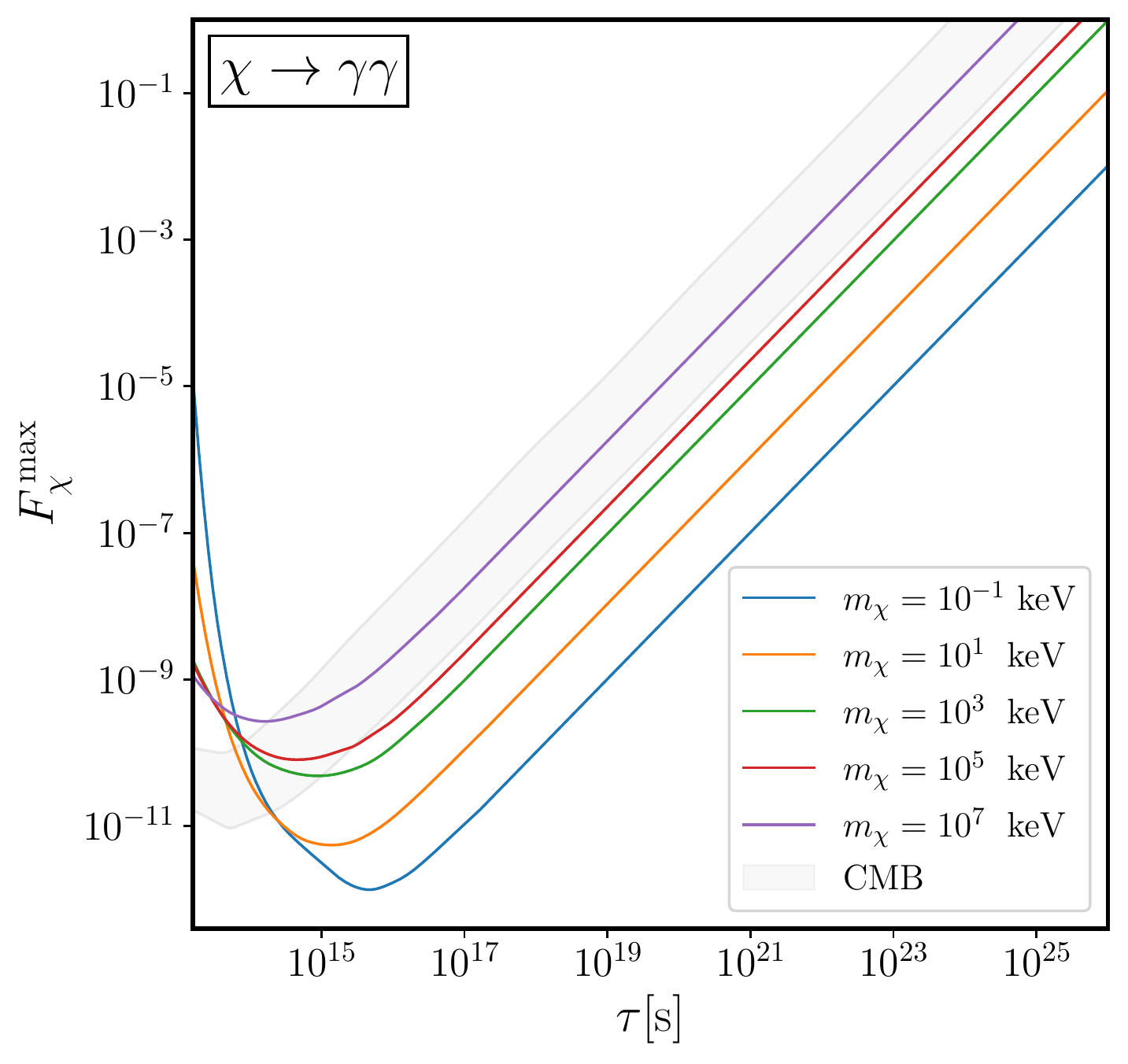}
    \quad
    \includegraphics[width=.48\linewidth]{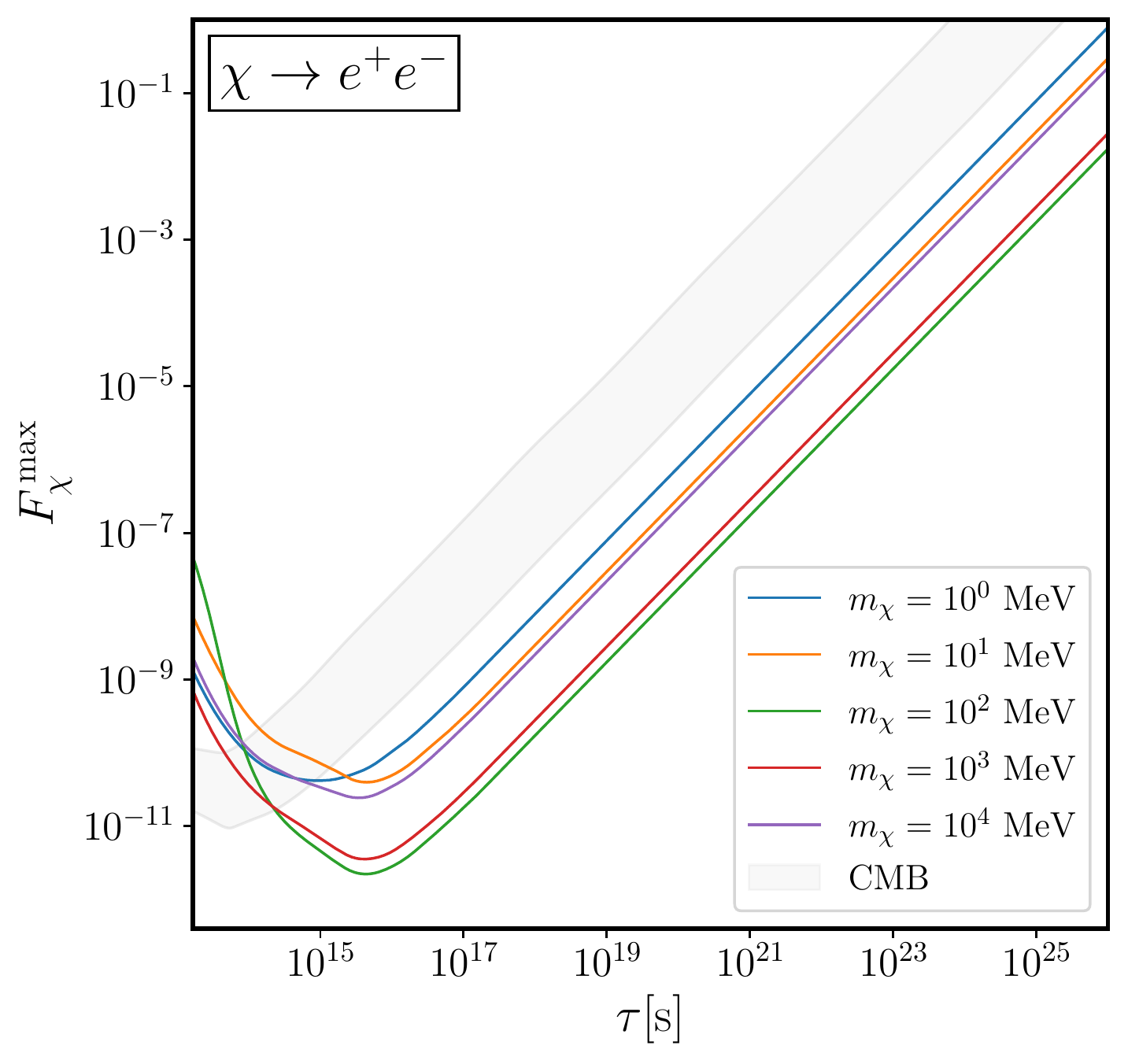}
    \caption{Fixed-mass sections of the results shown in Fig.~\ref{fig:sub-componentheatmaps}. Colored lines indicate the maximal allowed fraction $F_\chi^{\rm max}$ of the unstable DM sub-component $\chi$ as a function of the DM lifetime $\tau$. Different colors correspond to different DM masses $m_\chi$. For comparison, the light gray band shows the range of upper bounds on the DM mass fraction derived from CMB anisotropy constraints \cite{sla:gen}, with the band width reflecting variations across injection species and DM masses. \textbf{Left panel:} Decay mode $\chi \to \gamma\gamma$. \textbf{Right panel:} Decay mode $\chi \to e^+ e^-$.}
    \label{fig: sub-component mass section}
\end{figure}

Looking at fixed-mass sections of our limiting surface facilitates a direct comparison with existing CMB bounds. This is illustrated in the two panels of Fig.~\ref{fig: sub-component mass section}, where, for the same decay channels, we show the maximal mass fraction $F_\chi$ of DM allowed to decay as a function of its lifetime $\tau$. Two key features, which also serve as useful cross-checks, are readily apparent. First, the minimum of each curve occurs near $\tau \sim 10^{16}$~s, corresponding to the timescale probed by 21 cm observations at redshift $z_t = 15$. Second, for $\tau \gtrsim 10^{17}$~s, the relationship between $F_\chi$ and $\tau$ becomes linear, as expected when the exponential suppression in the energy injection effectively vanishes. While 21 cm bounds are weaker than those from CMB anisotropies at short lifetimes ($\tau \lesssim 10^{14}$~s), they quickly become significantly more stringent across the remaining parameter space. 

The results shown in Fig.~\ref{fig: sub-component mass section} can also be compared with existing constraints beyond those from the CMB anisotropy spectrum. For lifetimes longer than the age of the universe, $\tau \gtrsim 10^{17}~\text{s}$, the hierarchy of bounds is expected to match that of the single-component scenario shown in Figs.~\ref{fig:singlecomponentphoton} and~\ref{fig:singlecomponentsboundelectrons}, as the limits in this regime correspond to a simple linear rescaling. In particular, 21 cm observations are expected to provide the strongest constraints for $m_\chi \lesssim 10~\text{keV}$ in the $\chi \to \gamma\gamma$ channel and for $m_\chi \lesssim 1~\text{GeV}$ in the $\chi \to e^+ e^-$ channel. For lifetimes shorter than $10^{17}~\text{s}$, the situation is more subtle: the constraints do not scale linearly due to the exponential suppression of the energy injection rate. A dedicated analysis is therefore required. We carry out such an analysis for the case of ALPs in Sec.~\ref{subsec:ALP}.

\subsection{Nearly Degenerate Two-Component DM}
\label{subsec:nearlydegenerate}

Building on the results of the previous subsection, we now turn to a scenario that has been extensively studied in the literature. We consider a dark sector consisting of two particles, denoted $\chi_1$ and $\chi_2$. The lighter state, $\chi_1$, is stable on cosmological timescales and accounts for the observed DM abundance. The heavier state, $\chi_2$, is nearly degenerate with $\chi_1$, is produced in the early universe with a sizable abundance, and is metastable with a long lifetime. Its late-time decays into the lighter state and visible particles, $\chi_2 \rightarrow \chi_1 + \text{visible}$, can impact the 21 cm monopole signal. 

Popular realizations of this general framework include scenarios where only the inelastic channel is active, thereby suppressing direct detection rates~\cite{smi:ine}, as well as models with non-standard production mechanisms that yield detectable indirect detection signals for sub-GeV DM without conflicting with CMB constraints~\cite{DEramo:2018khz,Berlin:2023qco}. In this section, we derive general bounds on this class of scenarios using a minimal phenomenological approach. A concrete Lagrangian realization is presented in Sec.~\ref{sec:pseudoDirac}, where the two states are described as Weyl fermions with a large Dirac mass split by small Majorana mass terms. We emphasize that our results remain valid even in scenarios where elastic scattering between the dark sector and the SM is unsuppressed.

We parameterize the mass spectrum using an overall DM mass scale $m_\chi$ and a dimensionless relative mass splitting $\delta$. Specifically, the masses of the two dark sector particles are
\begin{subequations}
\begin{align}
m_{\chi_1} \equiv & \, m_\chi \ , \\
m_{\chi_2} \equiv & \, m_\chi (1 + \delta) \ .
\end{align}
\end{subequations}
Our analysis focuses on the regime where $\delta$ is small. The evolution of the total DM abundance in this scenario is more subtle than in the sub-component case, since the decays of $\chi_2$ increase the abundance of $\chi_1$. If we focus on what happens in a comoving volume, the total number of dark sector particles remains constant, while the total mass decreases due to the mass difference between the two states. The corresponding mass deficit is transferred to the SM sector through the decays of $\chi_2$. As a result, the total DM density evolves as
\begin{equation}
    \rho_{\rm DM}(t) = \left[\rho_{\chi_1}(\bar{t}) + \rho_{\chi_2}(\bar{t}) 
    \frac{1 + \delta \, e^{-(t - \bar{t})/\tau}}{1+\delta} \right] \left(\frac{a(\bar{t})}{a(t)}\right)^3,
    \label{exact DM mass density evolution iDM}
\end{equation}
where the reference time $\bar{t}$ satisfies the same properties discussed below Eq.~\eqref{exact DM mass density evolution}. The expression in Eq.~\eqref{exact DM mass density evolution iDM} can be interpreted as comprising three contributions to the total DM energy density. The first term in the square brackets corresponds to the energy density stored in $\chi_1$. The second term captures the portion of energy in $\chi_2$ that will eventually be transferred to $\chi_1$ upon decay. The prefactor $(1+\delta)^{-1}$ effectively converts $\rho_{\chi_2}$ from being expressed in terms of $m_{\chi_2}$ to $m_{\chi_1}$, reflecting the energy that remains in the dark sector after the decay. Finally, the third term accounts for the energy that will be injected into the SM sector, proportional to the mass splitting $\delta$ between $\chi_2$ and $\chi_1$. Furthermore, we parameterize the relative abundance similarly to what we have done in Eq.~\eqref{relative abundance parameterization} and define
\begin{equation}
    F_{\chi_2} \equiv \frac{\rho_{\chi_2}(\bar{t})}{\rho_{\chi_1}(\bar{t}) + \rho_{\chi_2}(\bar{t})} \ .
\end{equation}

The power injected per unit physical volume due to $\chi_2$ decays can be written as
\begin{equation}
\left(\frac{dE}{dV dt}\right)_{\rm inj}^{\chi_{2}\rightarrow \chi_1} = \frac{\mathcal{E}_{\rm SM}}{m_{\chi_2}} \frac{\rho_{\chi_2}(t)}{\tau},
\label{inelastic energy injection}
\end{equation}
where the superscript identifies the scenario under consideration. The quantity $\mathcal{E}_{\rm SM}$ denotes the average energy transferred to SM degrees of freedom per decay. We find it convenient to normalize this to $m_{\chi_2}$, as this would correspond to the total energy injection into the SM in the sub-component DM scenario discussed in the previous subsection. In general, the ratio $\mathcal{E}_{\rm SM}/m_{\chi_2}$ can vary depending on whether the decay proceeds via two-body or three-body channels. However, a rough estimate up to order one numerical factors can be obtained by neglecting the DM recoil and expanding at leading order in $\delta$. This yields $\mathcal{E}_{\rm SM}/m_{\chi_2} \sim \delta$.

We evaluate the power injected in Eq.~\eqref{inelastic energy injection} using Eq.~\eqref{exact DM mass density evolution iDM}, obtaining
\begin{align}
\left(\frac{dE}{dV dt}\right)_{\rm inj}^{\chi_{2}\rightarrow \chi_1} & =
\frac{\mathcal{E}_{\rm SM}}{m_{\chi_2}} \; \frac{(1+\delta) F_{\chi_2}  e^{(t_0-t)/\tau}}{e^{t_0/\tau} \left[1 + (1 - F_{\chi_2}) \delta \right] + F_{\chi_2}\delta} \; 
\frac{\rho_{\rm DM}(t_0)(1+z)^3}{\tau}\nonumber\\
&=\frac{\mathcal{E}_{\rm SM}}{m_{\chi_2}} \; F_{\chi_2} e^{-t/\tau} \; \frac{\rho_{\rm DM}(t_0)(1+z)^3 }{\tau}+\mathcal{O}\left(\delta^2\right).
\label{energy injection iDM}
\end{align}
The first line is fully general and makes no assumptions about the parameters, while the second is derived by expanding the expression to leading order in $\delta$. This approximation is justified because $\chi_1$, as the stable DM candidate, must retain a sufficiently low velocity after the decay to avoid disrupting structure formation~\cite{Wang:2013zba,Aoyama:2014tga,Cheng:2015bqa,Buch:2016qkv}, and the relative mass splitting is constrained to be $\delta \lesssim 10^{-3} \text{--} 10^{-4}$. Strictly speaking, this bound on $\delta$ applies if the relative abundance $F_{\chi_2}$ is of order one. If instead $F_{\chi_2} \ll 1$, larger mass splittings could be allowed. Nonetheless, our focus is on the small-$\delta$ regime, where meaningful constraints can be obtained for realistic models. Larger values of $\delta$ typically imply much faster decay rates that lie beyond the sensitivity of 21 cm observations. For this reason, we will retain just the zeroth order in $\delta$ of Eq. \eqref{energy injection iDM} in the following computations.

Comparing Eqs.~\eqref{energy injection iDM} and~\eqref{generalpowerinjectedfromsubcomponent}, we see that the two scenarios can be related up to quadratic corrections in the relative mass splitting
\be
\left(\frac{dE}{dV\,dt}\right)_{\rm inj}^{\chi_{2}\rightarrow \chi_1} =  \frac{\mathcal{E}_{SM}}{m_{\chi_2}}  \left(\frac{dE}{dV\,dt}\right)_{\rm inj}^{\chi_{\rm subDM}} + \mathcal{O}\left(\delta^2\right) \ .
\ee
It is thus manifest that the energy injection in the two-component DM scenario corresponds to a simple rescaling of the general sub-component case discussed above. The analysis can be translated by identifying the appropriate dictionary between the two frameworks. In particular, one constrains the combination $F_{\chi_2}\mathcal{E}_{SM}/m_{\chi_2} \sim F_{\chi_2} \delta$, rather than $F_{\chi}$ alone. Additionally, the relevant parameter space becomes $(\delta m_{\chi}, \tau)$ instead of $(m_{\chi}, \tau)$.

We must now specify the visible part of the final states produced in the decay in order to proceed with our analysis. If we focus on decays with at most three final state particles, the small mass splitting implies that the $\chi_2\to\chi_1$ transition can proceed just through three different channels in the mass range we consider: $\chi_2\to \chi_1 \gamma$, $\chi_2 \to \chi_1 \gamma\gamma$ and $\chi_2\to\chi_1 e^+ e^-$. Indeed, for $2\mathcal{R}<m_{\chi_2}<2\;\text{TeV}$ and relative mass splittings $\delta\lesssim 10^{-3}$, there is no sufficient energy to produce any other SM final state. In other words, $ 2\delta\; \text{TeV}< 2 m_\mu$, since the muon would be the lightest particle other than the electron.

The actual projected constraints that can be set on this class of models depends on $\mathcal{E}_{SM}$, that hinge on the kinematics of the decay. Some useful results about the kinematics of two- and three-body decays, which are all we need for our purposes, are collected in App.~\ref{app:rates}. In particular, the general expression for the differential decay rate is provided in Eq.~\eqref{eq:GammaGeneral} with the Lorentz invariant phase space defined in Eq.~\eqref{eq:LIPS}. 

\begin{figure}
    \centering
    \includegraphics[width=.48\linewidth]{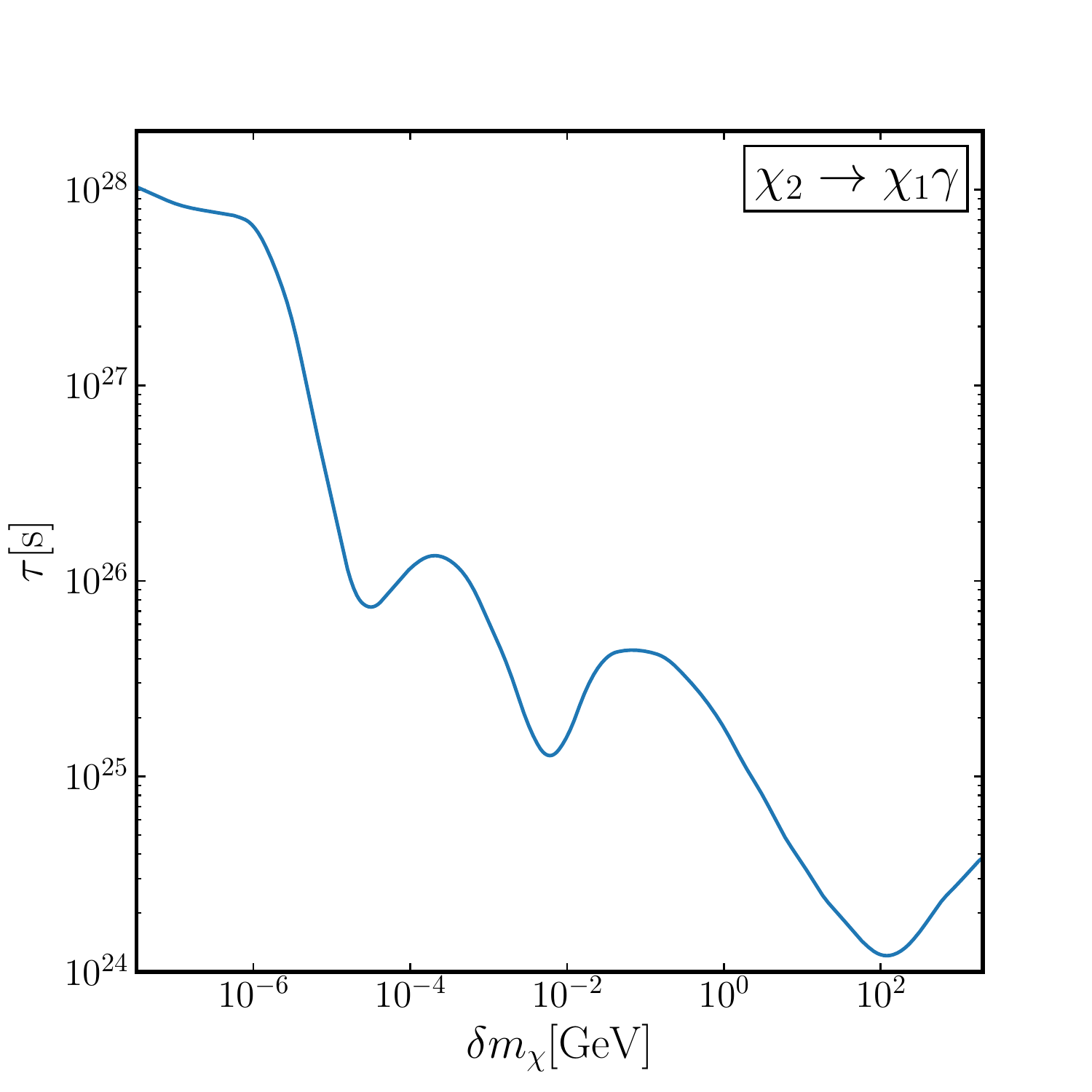}
    \quad
    \includegraphics[width=.48\linewidth]{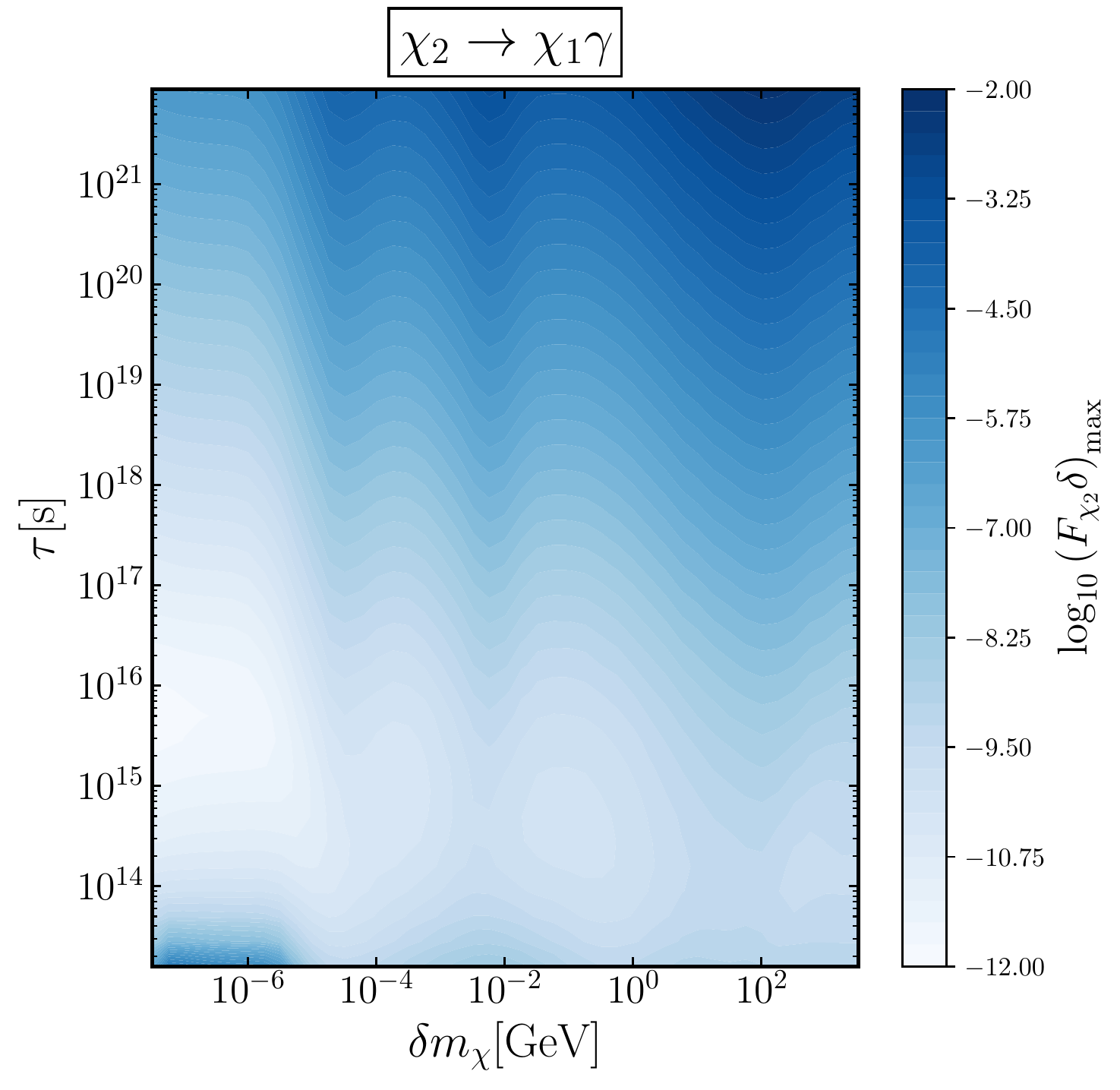}
    \caption{Model independent bounds on the nearly degenerate two-component DM for the decay channel $\chi_2 \to \chi_1 \gamma$. We assume the trough redshift to be at $z_t = 15$, and value of the threshold is set to $\mathcal{A}_{\Lambda\rm CDM}^{\rm max}/\mathcal{T}=0.50$.
    \textbf{Left panel:} Bounds on the $\chi_2$ lifetime $\tau$ as a function of the product  $\delta m_{\chi}$. 
    \textbf{Right panel:} Maximal allowed value for the quantity $F_{\chi_2}\delta$ in the $(\delta  m_{\chi}, \tau)$ plane.}
    \label{fig: sub-component single photon}
\end{figure}

As far as two-body decays are concerned, we have just the $\chi_2\to\chi_1\gamma$ channel. In the rest frame of the decaying $\chi_2$ particle, the energies of the decay products are
\begin{subequations}
    \begin{align}
         E_{\chi_1} &=\frac{m_{\chi_2}^2+m_{\chi_1}^2}{2m_{\chi_2}}= m_{\chi}+\mathcal{O}\left(\delta^2\right), \\ 
         E_\gamma &=\frac{m_{\chi_2}^2-m_{\chi_1}^2}{2m_{\chi_2}}= \delta m_{\chi} +\mathcal{O}\left(\delta^2\right) \ ,
    \end{align}
\end{subequations}
thus, $\mathcal{E}_{SM}=E_\gamma\simeq \delta m_{\chi}$. The differential two-body phase space has the general expression provided in Eq.~\eqref{eq:2bPS}. We plug this into the expression for the differential decay width and expand the final result in the small-$\delta$ limit
\begin{equation}
    \frac{d\Gamma_{\chi_2\to\chi_1\gamma}}{d\Omega}=\frac{|\mathcal{M}_{\chi_2\to\chi_1\gamma}|^2}{32 \pi^2 }\frac{\delta}{m_{\chi}}+\mathcal{O}\left(\delta^2\right).
\end{equation}
Strictly speaking, to get the bounds for this case a few additional steps would be required beyond the simple rescaling. As discussed in App.~\ref{app:ion&therhistory}, the energy deposition efficiencies $f^c(z)$ depend on the spectra of the injected particles. In the case of the $\chi_2 \to \chi_1 \gamma$ channel, the spectrum per decay is a single monochromatic photon. This final state was not considered in our analysis of the previous subsection, and for this specific channel we are effectively forced to re-run the code. We report the resulting bounds in Fig.~\ref{fig: sub-component single photon}. As we could have expected, the results are qualitatively very similar to the ones obtained for the channel $\chi\to \gamma\gamma$ in the single component DM scenario, with two key differences. First, the variable that is setting the energy deposition into the IGM is not the mass of the decaying particle, but the combination $\delta m_{\chi}$; this is nothing but the dark sector energy that is converted into SM degrees of freedom in a single decay process. Second, all the bounds are shifted to the left by roughly a factor of $2$, due to the fact that we are injecting a single photon per decay event instead of two.

In principle, the situation for three-body decays would be equally concerning, but in practice it poses no additional obstacles beyond the rescaling. For the three-body final states we consider, a proper treatment would also require rerunning the code with the correct energy spectra of the injected primaries as input. However, since $\delta$ is small, the decay kinematics is quasi-monochromatic, allowing us to safely apply the bounds derived in the previous section as we are about to show. More specifically, we have two possible channels: $\chi_2 \to \chi_1 \gamma \gamma$ and $\chi_2 \to \chi_1 e^+ e^-$. In both cases, the final states are not monochromatic. Let $E_{1,2}$ denote the energies of the two visible final-state particles; the total energy deposited into SM degrees of freedom in a single decay is then $E_{\rm SM} = E_1 + E_2$. The average deposited energy, which is crucial for the 21 cm signal, is given by
\begin{equation}
    \mathcal{E}_{\rm SM} \equiv \langle E_{\rm SM} \rangle = \langle E_1 \rangle + \langle E_2 \rangle = 2\langle E \rangle = \frac{2}{\Gamma_3} \int dE\, \frac{d\Gamma_3}{dE} \, E \ .
\end{equation}
Here, $\Gamma$ is the total decay rate, and we have used the fact that for both channels of interest, the two visible particles (photons or $e^+ e^-$) have an identical energy spectrum, implying $\langle E_1 \rangle = \langle E_2 \rangle \equiv \langle E \rangle$. Thus, $d\Gamma_3/dE$ denotes the differential three-body decay rate with respect to the energy of either SM particle in the final state.

In order to compute both the differential and total decay rates, we must integrate the differential rate over the Lorentz-invariant phase space (defined in Eq.~\eqref{eq:LIPS}, with $n=3$ in this case). Strictly speaking, this integration requires knowledge of the explicit form of the matrix element. A practical approach is to assume that the amplitude is constant. While this assumption is not exact, it yields the correct parametric scaling and provides a good estimate for small $\delta$, since in this regime the kinematics closely approximates that of a two-body decay. This further implies that the SM final-state particles are nearly monochromatic and emitted back-to-back in the rest frame of the decaying particle. If needed, a more refined treatment of the three-body kinematics can be implemented without significant difficulty, although it is unlikely to be necessary for most applications. Nevertheless, we will adopt such a treatment in Sec.~\ref{sec:pseudoDirac}, where we present concrete Lagrangian realizations.

Proceeding this way and calling $m_{\rm SM}$ the common mass of final-state SM particles ($m_{\rm SM} = 0$ for photons or $m_{\rm SM} = m_e$ for $e^+ e^-$), we can get the differential and the total decay rates 
\begin{subequations}
    \begin{align}
        \frac{d\Gamma_3}{dE}&= \frac{(m_{\chi_2}-m_{\chi_1}-2m_{\rm SM})(m_{\chi_2}+m_{\chi_1})}{128 \pi^3 m_{\chi_2}^2}|\mathcal{M}_3|^2 \\
        \Gamma_3 &=\frac{(m_{\chi_2}-m_{\chi_1}-2m_{\rm SM})^2(m_{\chi_2}+m_{\chi_1})^2}{256 \pi^3 m_{\chi_2}^3}|\mathcal{M}_3|^2 
    \end{align}
\end{subequations}
The average energy deposited into SM degrees of freedom turns out to be 
\begin{equation}
    \mathcal{E}_{\rm SM}=\frac{(m_{\chi_2}-m_{\chi_1})(m_{\chi_2} +m_{\chi_1}+2m_{\rm SM})}{2m_{\chi_2}} = \delta m_{\chi_1}-\left(\frac{\delta m_{\chi}}{2}-m_{\rm SM}\right)\delta+\mathcal{O}\left(\delta^3\right),
\end{equation}
where we have written the result this way to emphasize that $m_{\rm SM}\sim \delta m_{\chi}$. Thus, at leading order in $\delta$, in our phenomenological model we have $\mathcal{E}_{\rm SM}\simeq\delta m_{\chi}$ for both two and three-body decays. 

In light of this discussion, it should be clear that the bounds in Fig. \ref{fig:sub-componentheatmaps} hold also for our phenomenological two-component scenario with the following substitution of variables
\begin{equation}
    (m_{\chi}, \tau, F_{\chi}) \to (\delta m_{\chi}, \tau, F_{\chi_2}\delta) \ .
\end{equation}
This is the natural way to map to a three-dimensional representation the four-dimensional parameter space spanned by $(m_{\chi}, \tau, F_{\chi_2}, \delta)$. In particular, $\delta m_{\chi}$ is the energy proceeding into the SM sector, while $F_{\chi_2} \, \delta$ is the product of the mass fraction at production and the branching ratio of the energy transfer from the DM to SM particles. Once we specify a model, we can fix one or more parameters to get the bounds that apply to that specific case.

\section{Non-Minimal Dark Sectors at the Microscopic Level}
\label{sec:Lag}

Having completed a model-independent analysis of non-minimal dark sectors, we now focus on explicit realizations where our results can be directly applied. These scenarios typically involve long-lived states whose decays leave distinctive imprints on the 21\,cm signal. In each case considered here, we analyze both the scenario where the relic density of the metastable state is specified a priori and the one where it is computed self-consistently. For the latter, we track the detailed evolution of these metastable species throughout cosmic history, thoroughly accounting for all relevant production mechanisms. The framework for calculating decay widths and cross sections is summarized in App.~\ref{app:rates}, while the Boltzmann formalism used to model particle production in the early universe is reviewed in App.~\ref{app:BE}.

In Sec.~\ref{subsec:ALP}, we explore a scenario featuring an ALP coupled to photons. This framework includes, but is not limited to, cases where the dominant DM component is a distinct, stable cold relic, and the ALP forms a subdominant population that eventually decays into photons. In Sec.~\ref{sec:pseudoDirac}, we investigate a dark sector composed of two nearly degenerate Majorana fermions. The lighter state is cosmologically stable and accounts for the entire present-day DM abundance, while the heavier state is metastable and decays into the lighter one alongside visible particles. We consider scenarios where these interactions are mediated either by a spin-one vector boson or by dimension-5 magnetic and electric dipole operators.

\subsection{ALP coupled to photons}
\label{subsec:ALP}

A pseudoscalar coupled to photons offers a compelling target for 21 cm cosmology. The most natural origin for such a field is as a Nambu–Goldstone boson from a spontaneously broken Abelian symmetry. When this is identified with the Peccei–Quinn (PQ) symmetry~\cite{Peccei:1977ur,Peccei:1977hh}, the resulting particle is the QCD axion~\cite{Weinberg:1977ma,Wilczek:1977pj}, whose color anomaly induces a dimension-5 coupling to gluons, providing a dynamical solution to the strong CP problem. While we do not explicitly impose PQ symmetry here, we assume the Abelian symmetry is anomalous with respect to the electromagnetic gauge group. The heavy fermions responsible for this anomaly are integrated out at low energies, yielding the effective dimension-5 ALP interaction
\begin{equation}
    \mathcal{L}_{a\gamma\gamma} = \frac{g_{a\gamma\gamma}}{4} \, a F_{\mu\nu} \tilde{F}^{\mu\nu} \ ,
    \label{eq:aFFdual}
\end{equation}
where $g_{a\gamma\gamma}$ is a coupling of mass dimension $-1$, $F_{\mu\nu}$ is the electromagnetic field strength tensor, and $\tilde{F}^{\mu\nu} = \epsilon^{\mu\nu\rho\sigma} F_{\rho\sigma}/2$ denotes its dual. Here, we focus on processes occurring below the Fermi scale, where electroweak symmetry is spontaneously broken and electromagnetism remains the sole unbroken gauge group. Extending this effective theory to a fully electroweak-invariant framework is straightforward. To maintain phenomenological viability, we assume the Abelian symmetry responsible for the ALP is softly broken, generating a nonzero ALP mass $m_a$, which we treat as a free parameter.

The presence of electrically charged particles in the primordial plasma enables ALP production via processes mediated by the interaction in \Eq{eq:aFFdual}. The dominant production channels involve charged SM fermions $\psi^\pm$ through two main processes: resonant photon conversions into ALPs ($\psi^\pm \gamma \rightarrow \psi^\pm a$, known as the \emph{Primakoff effect}), and fermion-antifermion annihilations into ALP-photon pairs ($\psi^+ \psi^- \rightarrow \gamma a$). We neglect ALP production via inverse decays ($\gamma \gamma \rightarrow a$) since the photon’s thermal mass kinematically forbids these in the ALP ultra-relativistic regime where thermal production predominantly occurs. Furthermore, as ALP interactions in \Eq{eq:aFFdual} are non-renormalizable, freeze-in production through scatterings is UV-dominated and sensitive to the reheat temperature $T_R$. To maintain a consistent effective field theory (EFT) description, we restrict $T_R \lesssim g_{a\gamma\gamma}^{-1}$. Conversely, $T_R$ cannot be arbitrarily low without compromising the success of the standard hot big bang model; thus, we impose the conservative lower bound $T_R \gtrsim 5\, \text{MeV}$~\cite{Kawasaki:1999na,Kawasaki:2000en,Hannestad:2004px,deSalas:2015glj,Hasegawa:2019jsa,Barbieri:2025moq}.

When the ALP constitutes the entire observed DM abundance, we focus on the mass range $20.4\,\text{eV} < m_a < 10\,\text{keV}$. This is precisely where 21 cm constraints for this decay channel in the single-component scenario are the strongest, as shown in Fig.~\ref{fig:singlecomponentphoton}. While it is in principle possible to set bounds for lower masses, the ALP lifetime scaling $\tau \propto m_a^{-3}$ leads to a rapid weakening of constraints for $m_a \lesssim 20.4\,\text{eV}$, making it unlikely that 21\,cm observations can improve upon existing complementary bounds. If the ALP coexists with a stable cold DM component, we consider even larger masses, while always respecting the inequality $m_a < T_R$. Either way, all production rates are computed in the ultra-relativistic limit for both ALPs and the charged SM particles in the plasma. For ALPs, this is justified since $T_R$ is always much larger than $m_a$. Although some charged particles may have masses comparable to or larger than $T_R$, their number densities become exponentially suppressed once they are non-relativistic, following the Maxwell-Boltzmann distribution, and thus contribute negligibly.

We consider reheat temperatures $T_R \leq 50\,\text{GeV}$, where scatterings involving external $W^\pm$ bosons are negligible due to their Boltzmann-suppressed abundances. For temperatures above the QCD confinement scale and below this upper bound, all electrically charged particles in the plasma are elementary fermions, enabling reliable perturbative calculations of ALP production rates. Below the QCD confinement scale, quarks and gluons hadronize, so we restrict production rate calculations to charged leptons only. In the intermediate temperature range, we interpolate smoothly between the quark-gluon plasma and hadronic regimes.

The thermal production rates from Primakoff (P) and fermion-antifermion annihilation (A) processes are given by~\cite{Braaten:1991dd,bol:the,cad:cos,lan:irr}
\begin{subequations}
    \begin{align}
    \label{eq:gammaP} 
    \gamma_P(T) & \simeq g_P(T) \, \frac{\alpha g_{a\gamma\gamma}^2 \zeta(3) T^6}{36\pi^2} \left[\ln \left(\frac{T^2}{m_\gamma^2}\right) + 0.8194\right], \\
    \label{eq:gammaA} 
    \gamma_A(T) & \simeq g_A(T) \, \frac{\alpha g_{a\gamma\gamma}^2 \zeta(3) T^6}{36\pi^2} \times 0.2248 \ .
    \end{align}
\end{subequations}
Here, $\alpha$ denotes the fine-structure constant, the photon thermal mass is $m_\gamma^2 = g_P(T) e^2 T^2 / 18$. The effective number of charged species contributing to each process grows with temperature, tracking the evolving particle content of the primordial plasma. This dependence is encoded in the temperature-dependent functions $g_P(T)$ and $g_A(T)$, which respectively account for the species relevant to Primakoff and annihilation processes. Their definitions and numerical values are given in App.~\ref{app:ALP} and illustrated in Fig.~\ref{fig:gPA}. Although fermion-antifermion annihilation remains subdominant relative to Primakoff scattering\footnote{More precisely, $g_A(T)/g_P(T) \lesssim 2$ for $T \lesssim 200\,\text{GeV}$.}, we include both contributions for completeness. The computation of the ALP asymptotic abundance is obtained by inserting these rates into Eq.~\eqref{eq:YXinfty}, yielding
\be
\begin{split}
Y_a^\infty \simeq & \, \int_{0}^{T_R} \frac{d T}{T} \left(1 + \frac{1}{3} \frac{d \log g_{*s}}{d \log T} \right) \frac{\gamma_{P}(T) + \gamma_{A}(T)}{s(T) H(T)}  \\ 
\simeq & \left( \frac{g_{a\gamma\gamma}}{10^{-15} \, {\rm GeV}} \right)^2 \times \left\{ \begin{array}{cccl}
1.6 \times 10^{-18} & \quad & \quad & T_R = 5 \, {\rm MeV} \\
9.0 \times 10^{-17} & \quad & \quad & T_R = 500 \, {\rm MeV} \\
5.1 \times 10^{-15} & \quad & \quad & T_R = 50 \, {\rm GeV} 
\end{array} \right. \ .
\end{split}
\ee

\begin{figure}
\centering
\includegraphics[width=0.65\linewidth]{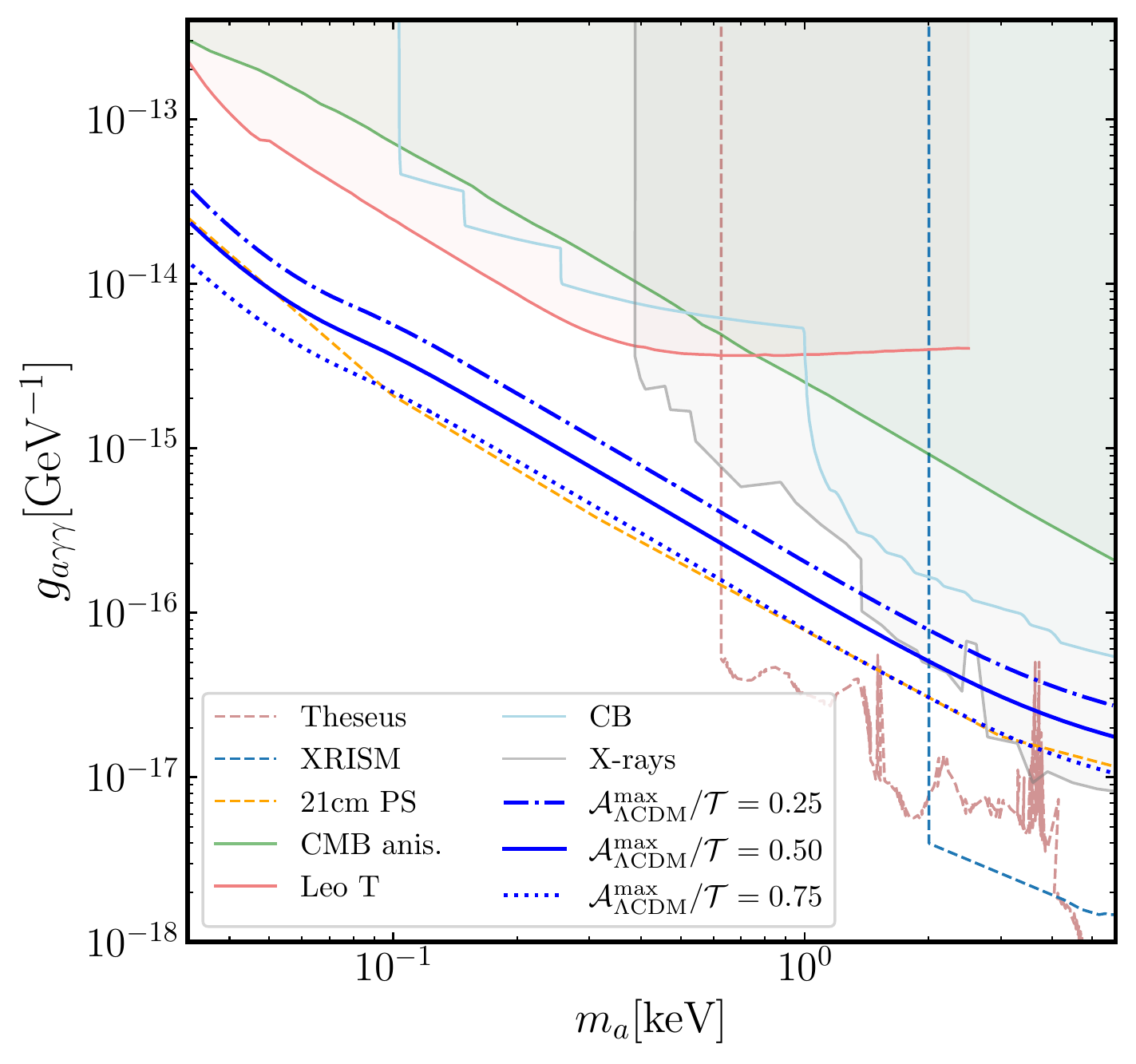}
\caption{Current constraints and future prospects for an ALP coupled to photons in the $(m_a, g_{a\gamma\gamma})$ plane. We remain agnostic about the production mechanism and assume that the ALP abundance reproduces the observed DM energy density. The blue lines show the results of this work for three representative values of $\mathcal{A}_{\Lambda \rm CDM}^{\rm max}/\mathcal{T}(z)$, assuming a fixed signal redshift $z_t = 15$. Light solid lines show existing bounds~\cite{liu:exo2,cap:cmb,wad:str,por:nov,cad:cos,fer:dod}, as specified in the legend, while dashed lines denote projected constraints~\cite{tho:the,des:lim,sun:inh}. Both are taken from \textsf{AxionLimits}~\cite{cia:caj}.}
    \label{fig:ALPbound1}
\end{figure}

Having established the theoretical framework and outlined the key elements of our analysis, we now proceed to present and examine our results. In the first scenario considered, we do not specify a particular ALP production mechanism; instead, we assume the existence of a cosmic ALP population with an abundance matching the observed DM density. Under this assumption, we impose the condition necessary to recast the projected constraints derived in Sec.~\ref{sec:DMbounds}. Specifically, this bound is implemented via the requirement $\tau_{a \rightarrow \gamma \gamma} > \tau_{\min}$, where the ALP lifetime is defined as the inverse of the decay width $\Gamma_{a \rightarrow \gamma \gamma}$ given in Eq.~\eqref{eq:Gammaaggg}. The lower limit $\tau_{\min}$ as a function of the ALP mass and for various signal amplitudes and redshifts can be read from Fig.~\ref{fig:singlecomponentphoton}. The three blue lines in Fig.~\ref{fig:ALPbound1} indicate the projected constraints derived here from the 21 cm monopole signal, corresponding to the three benchmark values of $\mathcal{A}_{\Lambda \rm CDM}^{\rm max}/\mathcal{T}(z)$ adopted in this work, assuming a fixed signal redshift $z_t = 15$. The same figure also presents current bounds and projected sensitivities from forthcoming experiments in the $(m_a, g_{a\gamma\gamma})$ parameter space. We emphasize that our results pertain to the global 21\,cm signal; complementary analyses focusing on the 21\,cm power spectrum, such as Refs.~\cite{fac:21c,sun:inh}, provide additional, independent constraints on the ALP parameter space. Together, these findings underscore the primary conclusion of this section: future 21\,cm observations hold the potential to probe regions of parameter space previously unexplored in this scenario.

\begin{figure}
\centering
\includegraphics[width=0.48\linewidth]{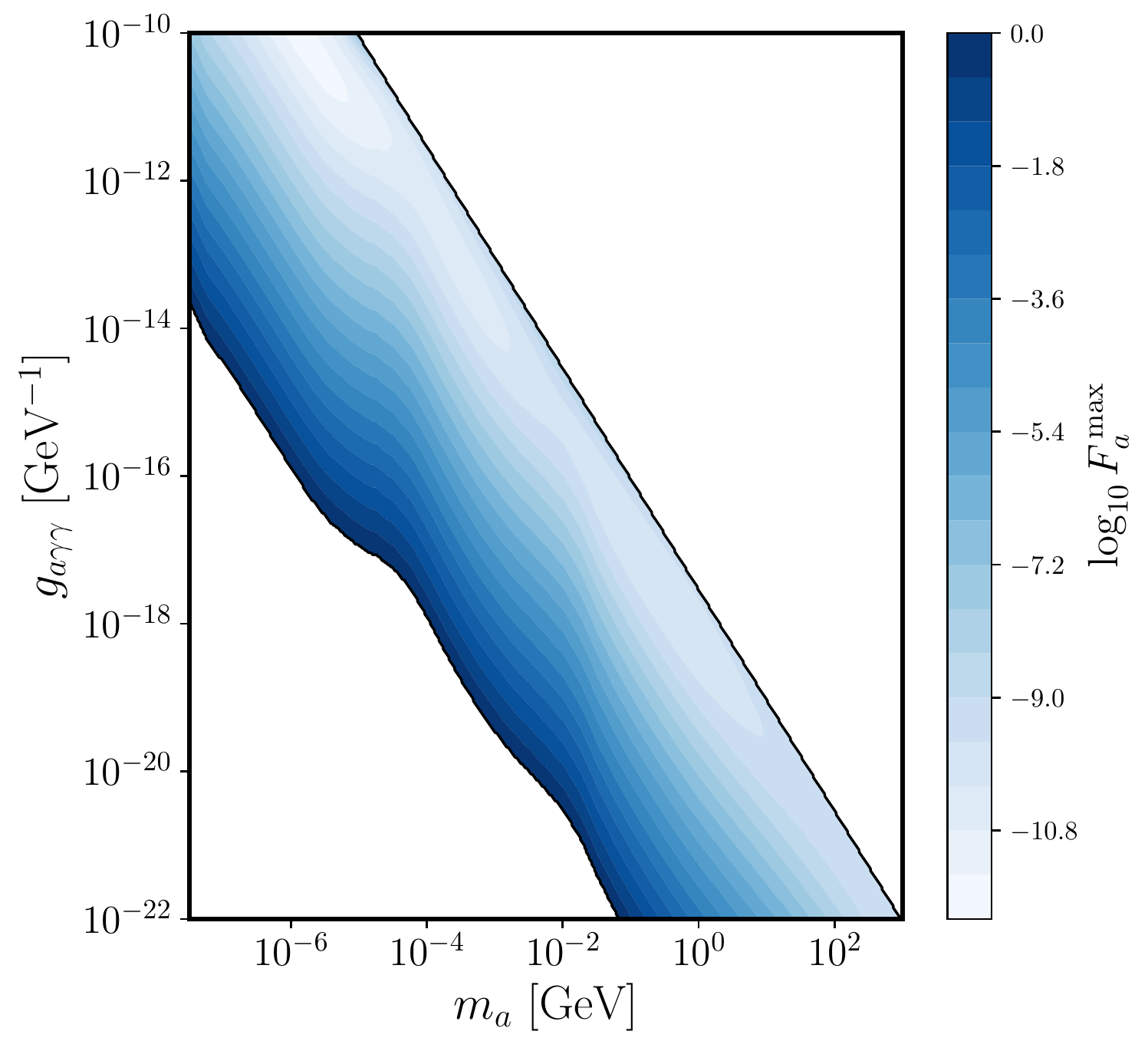} $\quad$
\includegraphics[width=0.48\linewidth]{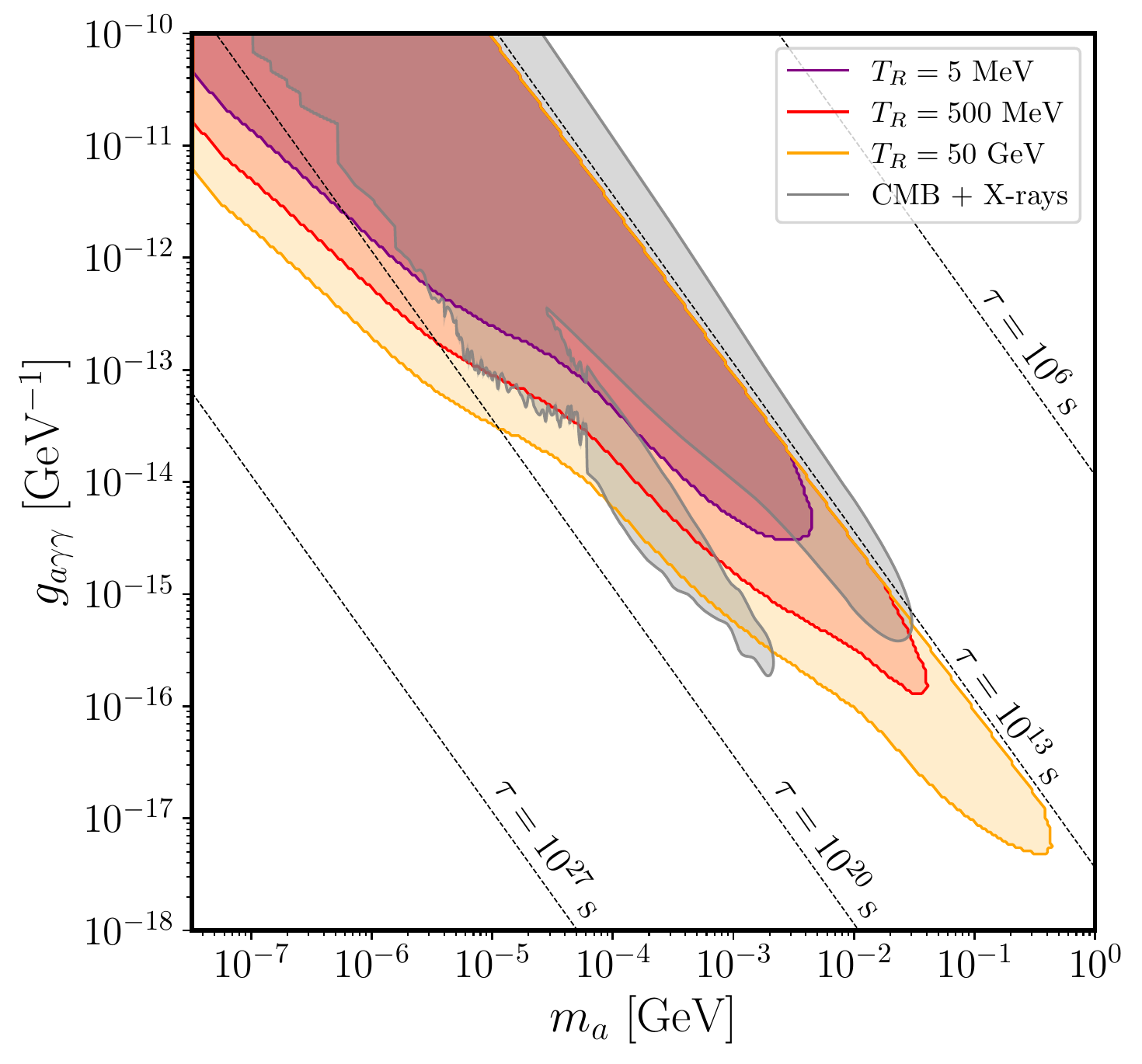}
\caption{Non-minimal scenario where the ALP is an unstable DM sub-component and the majority of DM is made of some other unspecified cold component. The results shown here hold for a 21\,cm signal amplitude $\mathcal{A}_{\Lambda \rm CDM}^{\max}/\mathcal{T}(z) = 0.50$ at the redshift value $z_t = 15$. \textbf{Left panel:} Heat map in the $(m_a, g_{a\gamma\gamma})$ plane showing the largest allowed ALP fraction $F_a^{\max}$ consistent with the global 21\,cm signal. The color scale encodes the maximum allowed ALP fraction for each point in the parameter space. \textbf{Right panel:} 21\,cm projections on the irreducible axion background for three different reheat temperatures. For this case, we evaluate the ALP abundance self-consistently for each point in the parameter space. The shaded gray regions are excluded by CMB and X-rays~\cite{lan:irr}. Solid black lines identify benchmark values for the ALP lifetime.}
    \label{fig:ALPbound2}
\end{figure}

We now turn to the non-minimal scenario, in which the CDM component consists of another species, and the ALP constitutes an additional dark degree of freedom. As before, we do not specify an explicit production mechanism; instead, we fix the ratio between the ALP mass density and the total DM density as
\begin{equation}
    F_a \equiv \frac{\rho_a}{\rho_a + \rho_{\mathrm{CDM}}} \ .
\end{equation}
The largest allowed ALP fraction, $F_a^{\max}$, at each point in the parameter space is shown in the heat map in the left panel of Fig.~\ref{fig:ALPbound2}, where the color scale encodes the maximum value consistent with the global 21\,cm signal. The results in this figure were obtained for a signal amplitude $\mathcal{A}_{\Lambda \rm CDM}^{\max}/\mathcal{T}(z) = 0.50$ and a signal redshift $z_t = 15$.

Finally, we commit to the specific production channels identified at the beginning of this subsection and compute the ALP abundance self-consistently. Using this, we recast the bounds derived in Sec.~\ref{subsec:sub-component} by imposing 
\begin{equation}
    \mathcal{F}_a(g_{a\gamma\gamma}, m_a, T_R) = \frac{m_a\, s(t_0) Y^\infty_a}{\rho_{\mathrm{DM}}(t_0)} < F_a^{\mathrm{max}} \ .
\end{equation}
Here, following the notation of Ref.~\cite{lan:irr}, $\mathcal{F}_a$ denotes the fractional DM abundance stored in ALPs produced via freeze-in, assuming they are absolutely stable. The quantity $F_a^{\mathrm{max}}$ denotes the maximal allowed mass fraction of the DM sub-component, and can be read from the heat map in the left panel of Fig.~\ref{fig:ALPbound2}. The corresponding results are shown in the right panel of Fig.~\ref{fig:ALPbound2}. As expected, 21\,cm constraints strengthen the bounds on the irreducible axion abundance for DM masses below a few keV, as well as within the narrow region left open between existing X-ray and CMB bounds. This implies that incredibly tiny populations of non-DM ALPs, with fractional abundances in the range $10^{-11} \lesssim F_a \lesssim 10^{-9}$ and lifetimes in the range $10^{13} \, \text{s} \lesssim \tau \lesssim 10^{20} \, \text{s}$, could be excluded.

Although considering only freeze-in production is a conservative approach, it is worth commenting briefly on the \emph{misalignment} production mechanism. The ALP relic abundance produced via misalignment is not irreducible, as it depends on the field's initial conditions. In turn, such initial conditions depend on whether the symmetry breaking setting the initial angle happens before or after inflation. For simplicity, let us assume that the initial misalignment angle $\theta_0$ is set during inflation. Furthermore, the fractional abundance of ALPs produced via misalignment, $\mathcal{F}_a^{(\mathrm{mis})}$, also depends on the detailed expansion history of the universe after the ALP field starts oscillating at $H \sim m_a$. It turns out that the most conservative choice, namely the one leading to the smallest possible value of $\mathcal{F}_a^{(\mathrm{mis})}$, corresponds to a period of early matter domination followed by reheating at temperature $T_R$. In this specific scenario, one can show \cite{Blinov:2019rhb} that
\begin{equation}
    \mathcal{F}_a^{(\mathrm{mis})} \simeq \left(\frac{f_a \, \theta_0}{1.3 \times 10^{15}\ \mathrm{GeV}}\right)^2 \left(\frac{T_R}{5\ \mathrm{MeV}}\right)
    \sim \left(\frac{\alpha \, \theta_0}{2 \pi g_{a\gamma\gamma}} \frac{1}{1.3 \times 10^{15}\ \mathrm{GeV}}\right)^2 \left(\frac{T_R}{5\ \mathrm{MeV}}\right),
\end{equation}
where $f_a$ is the ALP decay constant, and in the second equality we estimated $g_{a\gamma\gamma} \sim \alpha/(2 \pi f_a)$. The scaling of $\mathcal{F}_a$ (from freeze-in) and $\mathcal{F}_a^{(\mathrm{mis})}$ (from misalignment) with $g_{a\gamma\gamma}$ is different. Misalignment production is more efficient at smaller couplings since $\mathcal{F}_a^{(\mathrm{mis})} \propto g_{a\gamma\gamma}^{-2}$, while freeze-in production is more efficient at larger couplings as $\mathcal{F}_a \propto g_{a\gamma\gamma}^2$. Therefore, including misalignment would constrain a complementary portion of the parameter space.

\subsection{Pseudo-Dirac DM}
\label{sec:pseudoDirac}

In this subsection, we focus on two explicit microscopic realizations of the two-component DM framework. More specifically, we consider scenarios where the dark sector comprises two Weyl fermions, $\xi$ and $\eta$ (both left-handed), which are singlets under the SM gauge group. Being SM singlets allows us to write both Dirac and Majorana mass terms. The key assumption here is that the Majorana mass terms are naturally suppressed, resulting in mass eigenstates that are nearly degenerate. This \emph{pseudo-Dirac} limit is technically natural, as the symmetry content of the theory is enhanced when the Majorana masses vanish.

As is well known, fermion singlets cannot have renormalizable interactions with SM fields, except when coupled to the gauge-invariant combination $L H$, where $L$ and $H$ denote the lepton and Higgs doublets, respectively. However, this coupling neither stabilizes the DM ground state nor yields the desired phenomenology. We therefore do not pursue this operator further\footnote{There are several ways to forbid this interaction. For example, one can assign odd parity under a $\mathbb{Z}_2$ symmetry to the two DM states while keeping all SM fields even, or impose lepton number conservation.} and instead focus on alternative communication channels.

Broadly speaking, the general Lagrangian for this framework takes three contributions
\begin{equation}
\mathcal{L}_{\rm pseudo-Dirac} = \mathcal{L}_{\rm SM} + \mathcal{L}_{\rm dark} + \mathcal{L}_{\rm mix} \ . 
\end{equation}
The right-hand side contains the SM Lagrangian $\mathcal{L}_{\rm SM}$, operators with only dark sector degrees of freedom in $\mathcal{L}_{\rm dark}$, and operators mixing the two sectors in $\mathcal{L}_{\rm mix}$. 

Before studying specific mediation mechanisms, which corresponds to different fields and operators appearing in $\mathcal{L}_{\rm mix}$, we present a general discussion of the dark fermion mass spectrum. The most general quadratic Lagrangian we can write for two canonically-normalized left-handed Weyl fermions reads
\begin{equation}
\mathcal{L}_{\rm dark} \supset \mathcal{L}^{(\rm quadratic)}_{\rm fermions} = \xi^\dag i \bar{\sigma}^\mu \partial_\mu \xi + \eta^\dag i \bar{\sigma}^\mu \partial_\mu \eta - \left[m_D \, \xi \eta + \frac{1}{2} m_\xi \, \xi \xi + \frac{1}{2} m_\eta \, \eta \eta + {\rm h.c.}\right] \ .
\end{equation}
We assume that the Majorana terms are naturally suppressed compared to the Dirac mass, $m_{\xi, \eta} \ll m_D$, and we identify the mass eigenstates and eigenvalues in this limit
\begin{subequations}
\begin{align}
\chi_1 = & \, i\frac{\eta-\xi}{\sqrt{2}} \ , \qquad \qquad \qquad m_{\chi_1} = m_D - \frac{ m_\xi +  m_\eta}{2} \equiv m_\chi \ , \\
\chi_2 = & \, \frac{\eta +\xi}{\sqrt{2}}\ , \;\, \qquad \qquad \qquad m_{\chi_2} = m_D + \frac{ m_\xi + m_\eta}{2} \equiv m_\chi (1 + \delta) \ .
\end{align}
\label{eq:changeofbasis}
\end{subequations}

\subsubsection*{Renormalizable interactions via a vector mediator}

The first model we investigate contains only renormalizable operators in $\mathcal{L}_{\rm mix}$. In particular, we extend the SM gauge group with a new Abelian factor $U(1)_D$ with $g_d$ the corresponding gauge coupling. Consequently, the dark covariant derivative reads $D_\mu = \partial_\mu - i g_d q_d A_\mu^\prime$ with $q_d$ and $A_\mu^\prime$ the Abelian dark charge and gauge field, respectively. All SM fields are neutral under the new gauge group, but the dark fermions carry equal and opposite charges. Without any generality loss, we set their absolute value equal to one, $q_\xi = - q_\eta = 1$. This field content is not enough to ensure the desired phenomenology. For this reason, we introduce a dark Higgs $\phi$ field that acquires a $U(1)_D$-breaking vacuum expectation value (vev). Moreover, we assign the charge $q_\phi = - 2$ so we can write Yukawa operators in $\mathcal{L}_{\rm dark}$. The vev of $\phi$ generates a non-vanishing mass for the gauge boson $A_\mu^\prime$ and also Majorana mass terms. 

The dark sector Lagrangian takes the form
\begin{equation}
\mathcal{L}_{\rm dark} = \left| D_\mu \phi \right|^2 + \xi^\dag i \bar{\sigma}^\mu D_\mu \xi + \eta^\dag i \bar{\sigma}^\mu D_\mu \eta - \frac{1}{4} F^{\prime \mu\nu} F^\prime_{\mu\nu} - \left[ y_\xi \phi \xi \xi + y_\eta \phi^\dag \eta \eta + {\rm h.c.} \right] - V_d(\phi)\ .
\end{equation}
We do not need to provide the explicit expression of the scalar potential $V_d(\phi)$, all we need is that the vacuum state is for a non-vanishing value of the dark Higgs field, $\langle \phi \rangle = v_\phi / \sqrt{2}$. Without any loss of generality, the vev $v_\phi$ can be taken real and positive. We also take the mass of the dark physical Higgs boson (i.e., the radial mode) large enough that it cannot affect neither the production nor the late-universe phenomenology. Once we evaluate the above Lagrangian with the Higgs field sitting at its vev, we identify both a gauge boson mass and Majorana mass terms
\begin{equation}
m_{A^\prime} = |q_\phi| g_d v_\phi \ , \qquad \qquad \qquad (m_\xi, m_\eta) =  \left(\frac{y_\xi v_\phi}{\sqrt{2}} , \frac{y_\eta v_\phi}{\sqrt{2}} \right)\ .
\end{equation}

A new Abelian group allows for renormalizable interactions between the dark and the visible sectors via the kinetic mixing operator
\begin{equation}
\mathcal{L}^{(1)}_{\rm mix} = - \frac{\epsilon}{2} F^\prime_{\mu\nu} B^{\mu\nu} \ .
\label{eq:Lmix1}
\end{equation}
Here, $F^\prime_{\mu\nu} = \partial_\mu A^\prime_\nu - \partial_\nu A^\prime_\mu$ is the dark Abelian field strength, and we write down the theory in the unbroken electroweak phase so the kinetic mixing is with the hypercharge field strength $B_{\mu\nu}$. This operator induces non-canonical kinetic terms for the gauge bosons, and we need to identify the appropriate degrees of freedom. The following change of variables 
\begin{equation}
B_\mu \; \rightarrow \; B_\mu - \frac{\epsilon}{\sqrt{1-\epsilon^2}} A^\prime_\mu \ , \qquad \qquad \qquad \qquad
A^\prime_\mu \; \rightarrow \; \frac{1}{\sqrt{1-\epsilon^2}} A^\prime_\mu \ ,
\end{equation}
provides a canonical Lagrangian for any finite value of the coupling $\epsilon$. Experimental constraints impose $\epsilon \ll 1$ so we keep only the linear terms and neglect $\mathcal{O}(\epsilon^2)$ contributions. In particular, the mass of the dark gauge boson is not affected by this change of field basis. 

Keeping only linear contributions in the kinetic mixing parameter, the net effect of this field redefinition is to induce an additional $\epsilon$-suppressed interaction between the hypercharge current $J^\mu_Y$ and the dark photon $A_\mu^\prime$. After performing this operation, we find the following set of gauge interactions for physical mass eigenstates
\begin{equation}
\mathcal{L}_{\rm interactions} = g_Y ( B_\mu - \epsilon A^\prime_\mu) J^\mu_Y + i \, g_d \, A_\mu^\prime \left( \chi_2^\dag \bar{\sigma}^\mu \chi_1 - \chi_1^\dag \bar{\sigma}^\mu \chi_2 \right) \ . 
\label{eq:Lint521}
\end{equation}
where $g_Y$ is the hypercharge coupling constant.

The phenomenology of this scenario has been extensively studied. Regarding production mechanisms, both thermal freeze-out~\cite{gon:cos} and freeze-in~\cite{hee:ine} have been examined in the literature. Each can account for the observed DM relic abundance, though they operate in distinct regions of model parameter space. Freeze-out is associated with larger annihilation cross sections and is already subject to stringent constraints from current and projected experimental data. In contrast, the freeze-in regime is less constrained. In this work, we focus on the freeze-in scenario, specifically on the parameter space identified in \cite{hee:ine}, where the observed DM abundance is achieved through non-thermal production. Due to its feeble couplings, the excited state $\chi_2$ is not efficiently depleted immediately after its production and remains long-lived on cosmological timescales. As a result, it is a suitable candidate for constraints from 21 cm cosmology.

The overall mass scale $m_\chi$ cannot be too light, as the warmness of DM produced via freeze-in would suppress cosmological perturbations at small scales. This sets a lower bound on $m_\chi$ at the level of tens of keV~\cite{Dvorkin:2019zdi,Dvorkin:2020xga,DEramo:2020gpr,DEramo:2025jsb}. Following Ref.~\cite{hee:ine}, we focus on the mediator mass range $10\,\text{MeV} \lesssim m_{A'} \lesssim 10\,\text{GeV}$, which is the target of current and upcoming experimental searches. To ensure that the mediator mass is negligible during freeze-in, we consider the limit $m_{A'} \ll m_{\chi}$. We explore the DM mass range $30\,\text{MeV} \lesssim m_{\chi} \lesssim 1\,\text{TeV}$; below this range, collider bounds become very stringent, while above it, the relic abundance computation in Ref.~\cite{hee:ine} becomes unreliable, as it was performed entirely in the broken electroweak phase.
 
We specialize to the regime of \emph{visible freeze-in}, where a single combination of couplings is responsible for both generating the correct DM relic abundance and inducing observable late-time signatures. This scenario is realized by requiring $g_d \lesssim g_Y \epsilon$, which ensures that DM is produced directly from the SM thermal bath rather than from a dark sector bath. The coupling combination needed to reproduce the observed relic abundance is $g_d \epsilon \sim 10^{-11}$.

Finally, we primarily focus on mass splittings in the range $2m_e/m_{\chi} < \delta < m_{A^\prime}/m_{\chi}$. This choice forbids the two-body decay $\chi_2 \to \chi_1 A^\prime$, thereby ensuring that $\chi_2$ remains long-lived, while allowing for the observable three-body decay $\chi_2 \to \chi_1 e^+ e^-$. For smaller mass splittings, $\delta < 2m_e/m_{\chi}$, the only available SM decay channels are the loop-induced processes $\chi_2 \to \chi_1 \gamma\gamma$ and $\chi_2 \to \chi_1 \bar{\nu} \nu$, both of which are highly suppressed.

The decay width for the visible channel within the parameter region we focus on, up to corrections due to the finite electron mass, results in~\cite{hee:ine}
\begin{equation}
    \Gamma_{\chi_2\to\chi_1 e^+ e^-}\simeq \frac{(g_d \epsilon)^2(e \cos\theta_w)^2}{60\pi^3}\frac{m_{\chi}^5}{m_{A'}^4}\delta^5 \ ,
\end{equation}
where $e$ is the electron charge and $\theta_w$ the weak mixing angle. As remarked above, we need the decay rate to be larger than $1/\tau_{CMB}$ in order to be able to set bounds on this model. This constraints provides a lower value for the mass splitting
\begin{equation}
    \delta \gtrsim \left(\frac{60\pi^3}{(g_d \epsilon)^2(e \cos\theta_w)^2\tau_{CMB}}\frac{m_{A'}^4}{m_{\chi}^5}\right)^{1/5}\simeq 2\cdot 10^{-4} \left(\frac{10^{-11}}{g_d \epsilon}\right)^2\left(\frac{m_{A'}}{1\text{ GeV}}\right)^{4/5} \left(\frac{35\text{ GeV}}{m_{\chi}}\right).
\end{equation}
We consider the scenario in which the combination of the produced $\chi_1$ and $\chi_2$ saturates the DM relic abundance. Furthermore, the freeze-in production via the interactions in \Eq{eq:Lint521} leads to an initial population of dark particles democratically split into $\chi_1$ and $\chi_2$. This is also the scenario considered in Ref.~\cite{hee:ine}. The consequent scaling of our bound results in
\begin{equation}
    m_{A'}>\left[\frac{(g_d \epsilon)^2 (e \cos\theta_w)^2}{60\pi^3}\right]^{1/4}\left(\delta  m_{\chi}\right)^{5/4} \Bar{\tau}^{1/4} \gtrsim 50 \text{ GeV},
\end{equation}
where the limiting value $\Bar{\tau}$ can be read from Fig. \ref{fig:singlecomponentsboundelectrons}. In the last inequality we have set $\Bar{\tau}=10^{25}$ s, $g_d \epsilon =10^{-12}$ and $\delta m_{\chi}=10^{-3}$ GeV to get the most conservative (i.e. smaller) value for our bound. Referring directly to Fig.~$5$ of Ref.~\cite{hee:ine}, 21 cm observations would be able to probe the unconstrained upper left portion of the $(\delta m_{\chi}, m_{A'})$ parameter space for this model.

\subsubsection*{Higher-dimensional interactions via electromagnetic dipoles}

The framework discussed above represents one of the most compelling scenarios for nearly degenerate two-component DM, particularly well suited for constraints from 21 cm cosmology. We now turn to another well-motivated setup in which interactions between the dark and visible sectors arise from non-renormalizable operators. Our focus remains on pseudo-Dirac DM, which retains the same mass spectrum as in the previous case.

In contrast to the previous scenario, we now work in the heavy mediator limit and introduce an EFT framework where the terms in $\mathcal{L}_{\rm mix}$ correspond to contact interactions between the dark fermions $\xi$ and $\eta$ and the SM fields. In this subsection, we focus on the case of an \emph{inelastic dipole}, previously explored in Ref.~\cite{der:dar} in the context of a novel mechanism for generating X-ray lines. For dipole operators, the DM interactions are conveniently formulated in terms of the Dirac spinor $\psi_D = (\xi \;\; \eta^\dag)^T$, expressed in the Weyl basis for the Dirac matrices.

We begin by writing the interactions for this Dirac field in the electroweak unbroken phase, where the dipole couples to the hypercharge gauge field strength $B_{\mu\nu}$
\begin{equation}
\mathcal{L}^{(2)}_{\rm mix} = - \frac{1}{2 \cos\theta_w \Lambda} 
\overline{\psi_D} \, \sigma^{\mu\nu} \left( c_M + c_E \, i \gamma^5 \right) \psi_D \, B_{\mu\nu} \ .
\label{eq:Lmix2}
\end{equation}
Here, $\sigma^{\mu\nu} \equiv (i / 2) [\gamma^\mu, \gamma^\nu]$, with $\gamma^\mu$ the Dirac matrices. The weak mixing angle $\theta_w$ appears in the overall normalization to match the definition of the electromagnetic dipoles given in Ref.~\cite{der:dar}. The cutoff scale $\Lambda$ is interpreted as the mass of the heavy mediator, while $c_M$ and $c_E$ denote the coefficients of the magnetic and electric dipole moments, respectively.
 
We can expand \Eq{eq:Lmix1} into Weyl components and apply the field redefinition from \Eq{eq:changeofbasis} to identify the interactions of the mass eigenstates $\chi_1$ and $\chi_2$. For convenience, we introduce the corresponding four-component Majorana spinors $\psi_{i} = (\chi_i \;\; \chi_i^\dag)^T$ with $i = \{1,2\}$. In terms of these fields, the dipole interactions in the unbroken phase take the form
\begin{equation}
\mathcal{L}^{(2)}_{\rm mix} = - \frac{i}{2 \cos\theta_w \Lambda} 
\overline{\psi_2} \, \sigma^{\mu\nu} \left( c_M + c_E \, i \gamma^5 \right) \psi_1 \, B_{\mu\nu} \, .
\label{dipole moment Lagrangian above EWSB}
\end{equation}
The potential signal arises from the decay of $\chi_2$ into its lighter partner $\chi_1$ and visible particles. To compute the decay rate, we must consider the Lagrangian in the broken phase
\begin{equation}
\mathcal{L}^{(2)}_{\rm mix} = - \frac{i}{2 \Lambda} 
\overline{\psi_2} \, \sigma^{\mu\nu} \left( c_M + c_E \, i \gamma^5 \right) \psi_1 \, F_{\mu\nu} + \tan\theta_w \frac{i}{2 \Lambda} 
\overline{\psi_2} \, \sigma^{\mu\nu} \left( c_M + c_E \, i \gamma^5 \right) \psi_1 \, Z_{\mu\nu} \, ,
\label{dipole moment Lagrangian below EWSB}
\end{equation}
where $F_{\mu\nu}$ and $Z_{\mu\nu}$ are the photon and $Z$ boson field strengths, respectively, arising from the basis rotation $B_{\mu\nu} \to \cos\theta_w F_{\mu\nu} - \sin\theta_w Z_{\mu\nu}$. For the mass splitting range considered here, the decay $\chi_2 \to \chi_1 Z$ is kinematically forbidden. However, the operator proportional to $F_{\mu\nu}$ induces the kinematically allowed decay $\chi_2 \to \chi_1 \gamma$, with a partial width given by
\begin{equation}
\Gamma_{\chi_2 \to \chi_1 \gamma} = \frac{c_M^2 + c_E^2}{8\pi \Lambda^2} \frac{(m_{\chi_2}^2 - m_{\chi_1}^2)^3}{m_{\chi_2}^3} \simeq \frac{c_M^2 + c_E^2}{\pi \Lambda^2} \, m_{\chi}^3 \, \delta^3  \, ,
\end{equation}
where in the last expression we expanded in the small parameter $\delta= (m_{\chi_2} - m_{\chi_1}) / m_{\chi_1} \ll 1$.

The machinery developed in this work can be applied to this scenario as long as two conditions are satisfied: $\Gamma_{\chi_2 \to \chi_1 \gamma} \, \tau_{\rm CMB} < 1$ and $\delta m_{\chi} > 2\mathcal{R} \simeq 20.4 \, {\rm eV}$. These requirement allow us to immediately identify some key features of the analysis we are about to perform. Assuming unit values for the dimensionless Wilson coefficients ($c_{M,E}=1$), these inequalities combine to yield a lower bound on the cutoff scale, $\Lambda \gtrsim 10^7$ GeV. This is significantly larger than the value required to reproduce the DM relic abundance via thermal freeze-out, which is approximately $200$ GeV~\cite{der:dar}. For this reason, we focus on non-thermal production via freeze-in. It is instructive to exploit the connection between the minimal energy deposition relevant for our study, $\delta m_{\chi}$, and the corresponding minimal cutoff scale, $\Lambda_{\rm min}$, that satisfies the constraints. These two quantities are related through $\Lambda_{\rm min} \propto \delta^3 m_{\chi}^3$. If we consider $\delta m_{\chi}\gtrsim 1$ GeV, this condition would require $\Lambda \gtrsim M_{\rm Pl}$. The second key implication is that we must consider energy depositions below 1 GeV in order for our EFT approach to remain valid and consistent. The conclusions just derived are valid under the assumption that the dimensionless Wilson coefficients are set to unit values. If they are not, the bounds on the minimal cutoff scale and energy deposition can be trivially rescaled since the decay width depends only on the specific combination
\begin{equation}
\Lambda_{\rm dip} \equiv \frac{\Lambda}{\sqrt{c_M^2 + c_E^2}} \ .
\end{equation}

As far as DM production in the early universe is concerned, the small mass splitting between the two states is negligible at the production epoch. The relic density calculation can be carried out in the so-called Dirac limit with the DM relic density depending only $m_\chi$ and $\Lambda_{\rm dip}$. As for ALP freeze-in production in Sec.~\ref{subsec:ALP}, we are dealing with non-renormalizable operators and therefore with UV-dominated freeze-in with the relic density depending on the reheating temperature $T_R$. This introduces an additional constraint on the cutoff scale $\Lambda$, since our EFT framework can reliably describe the dynamics only up to temperatures $T \lesssim \Lambda$. At higher temperatures, the heavy degrees of freedom that were integrated out to obtain the effective Lagrangian in Eq.~\eqref{eq:Lmix2} become dynamical.

A first consistency check is to verify that the freeze-in production satisfies this condition. To be within the freeze-in regime, the interaction rate, $\Gamma_{int}(T) \sim T^3/\Lambda_{\rm dip}^2$, needs to be much smaller than the expansion rate, $H(T) \sim T^2/M_{\rm Pl}$, for all the cosmological evolution. This gives us an upper bound for the reheating temperature
\begin{equation}
    \frac{\Gamma_{int}(T)}{H(T)} \sim \frac{T M_{\rm Pl}}{\Lambda_{\rm dip}^2}<1 \quad\Rightarrow \quad T_R<\frac{\Lambda_{\rm dip}^2}{M_{\rm Pl}} \ .
\end{equation}
Thus, freeze-in production takes plane within the EFT validity regime as long as $\Lambda\lesssim M_{\rm Pl}$. Furthermore, BBN requires $T_R\gtrsim 5$ MeV, which in turn implies $\Lambda_{\rm dip} \gtrsim 10^9$ GeV. 

We focus on reheating temperatures above the weak scale, $T_R \gtrsim 250 \, {\rm GeV}$, which justifies performing the calculation in the unbroken electroweak phase, as freeze-in is UV dominated. In this regime, DM production proceeds via three channels: $f\Bar{f} \to \chi\Bar{\chi}$, $H H^\dagger \to \chi\Bar{\chi}$, and $B B \to \chi\Bar{\chi}$. The last process is negligible, being induced by a double dipole insertion and suppressed by $\mathcal{O}(1/\Lambda_{\rm dip}^4)$. We therefore retain only the contributions from fermion-antifermion and Higgs doublet annihilations, both scaling as $\mathcal{O}(1/\Lambda_{\rm dip}^2)$.

The freeze-in production can be computed using the formalism outlined in App.~\ref{app:BE}. In particular, the asymptotic comoving number density, $Y_{\chi}^\infty \equiv Y_{\chi_1}^\infty + Y_{\chi_2}^\infty$, is given by \Eq{eq:YXinfty}. The integrand of that expression encodes both the cosmological background and the total collision operator, summed over all relevant production processes. The general form of the collision operator for a generic binary scattering process is given in \Eq{eq:scatteringrate}. We now evaluate its explicit expression for the DM production channels considered here. Given the high reheating temperature, all particle masses can be neglected: SM particles are massless in the unbroken phase,\footnote{Thermal masses are negligible for the processes considered in this dipole case.} and the DM mass is always assumed to be smaller than $T_R$.

We find it convenient to classify fermion-antifermion annihilations by identifying the SM Weyl fields in the initial state. Denoting by $f$ a generic SM fermion with well-defined gauge quantum numbers, the production cross section reads
\begin{equation}
    \sigma_{f\Bar{f} \to \chi\Bar{\chi}} = \frac{1}{\mathcal{C}_f \, \mathcal{W}_f} \frac{g_Y^2 Y_f^2}{12 \pi \cos^2\theta_w \Lambda_{\rm dip}^2}  \ .
\end{equation}
Here, $g_Y$ is the hypercharge coupling, and $Y_f$ is the hypercharge of the Weyl fermion $f$. The factors $\mathcal{C}_f$ and $\mathcal{W}_f$ denote the number of color and weak-isospin components, respectively (i.e., $\mathcal{C}_f = 3$ for quarks and $1$ for leptons; $\mathcal{W}_f = 2$ for weak doublets and $1$ for singlets). Plugging this expression into the general formula for the collision rate and summing over all SM fermions, we find the total production rate for this contribution
\begin{equation}
 \gamma_f(T) = \left(\sum_f \mathcal{C}_f^2 \mathcal{W}_f^2 \, \sigma_{f\Bar{f} \to \chi\Bar{\chi}} \right) \frac{T}{32 \pi^4} \int_{0}^\infty ds\, s^{3/2} K_1\left(\frac{\sqrt{s}}{T}\right) = 
 \frac{g_Y^2 T^6 \, \sum_f \mathcal{C}_f \mathcal{W}_f Y_f^2}{12 \pi^5 \cos^2\theta_w \Lambda_{\rm dip}^2} \ .
\end{equation}
The fact that the cross section is independent of the center-of-mass energy $s$ allows us to factor it out of the integral, and the remaining integrand can be evaluated using the identity $\int_{0}^\infty ds\, s^{3/2} K_1(\sqrt{s}/T) = 32 \, T^5$. The sum over SM fermions is easily performed by accounting for their multiplicities and hypercharges
\begin{equation}
    \sum_f \mathcal{C}_f \mathcal{W}_f Y_f^2 = 3 \times \left[ 3 \left(2\left(\tfrac{1}{6}\right)^2 + \left(\tfrac{1}{3}\right)^2 + \left(\tfrac{2}{3}\right)^2 \right) + \left( 2 \left(\tfrac{1}{2}\right)^2 + 1 \right) \right] = 10 \ ,
\end{equation}
where the bracketed sum gives the contribution from one generation, and the factor of 3 accounts for all of them.

We now switch to the production processes where DM pairs get generated by annihilations of the components of the SM Higgs doublet. Counting the weak-isospin states as we have just done, we find
\begin{equation}
    \sigma_{HH^\dagger \to \chi\Bar{\chi}} = \frac{1}{\mathcal{W}_H} \frac{g_Y^2 Y_H^2}{ 24 \pi \cos^2\theta_w \Lambda_{\rm dip}^2} \ .
\end{equation}
Here, the SM Higgs doublet has hypercharge $Y_H=1/2$ and it has $\mathcal{W}_H=2$ complex components. The resulting production rate can be computed as done before for fermions
\begin{equation}
 \gamma_H(T) = \left(\mathcal{W}_H^2 \, \sigma_{HH^\dagger \to \chi\Bar{\chi}} \right) \frac{T}{32 \pi^4} \int_{0}^\infty ds\, s^{3/2} K_1\left(\frac{\sqrt{s}}{T}\right) = \frac{g_Y^2 \, T^6}{48 \pi^5 \cos^2\theta_w \Lambda_{\rm dip}^2}   \ .
\end{equation}
The total production rate is therefore
\begin{equation}
\gamma_{\rm dip}(T) = \gamma_f(T)+\gamma_H(T)=\frac{41 g_Y^2 \, T^6}{48 \pi^5 \cos^2\theta_w \Lambda_{\rm dip}^2} \ .
\end{equation}

We input this rate into the Boltzmann equation and use Eq.~\eqref{eq:YXinfty} to compute the DM abundance produced via freeze-in. Given the large $T_R$ values under consideration, we neglect the term involving the temperature derivative of the entropic degrees of freedom. This yields 
\begin{equation}
\begin{split}
Y_\chi^\infty \simeq & \, 2 \times \int_{0}^{T_R} \frac{d T}{T} \frac{\gamma_{\rm dip}(T)}{s(T) H(T)} = \frac{1845 \sqrt{10}}{16 \pi^8} \frac{g_Y^2}{\cos^2\theta_w \Lambda_{\rm dip}^2} \frac{M_{Pl}}{\sqrt{g_*}\,g_{*s}} T_R  \\ & \simeq 1.4 \times 10^{-11}
\left(\frac{10^{15}\text{ GeV}}{\Lambda_{\rm dip}}\right)^2\left(\frac{106.75^{3/2}}{\sqrt{g_*}g_{*s}}\right)\left(\frac{T_R}{10^6\text{ GeV}}\right)\ .
\end{split}
\label{eq:Ychidip}
\end{equation}

\begin{figure}
    \centering
    \includegraphics[width=0.65\linewidth]{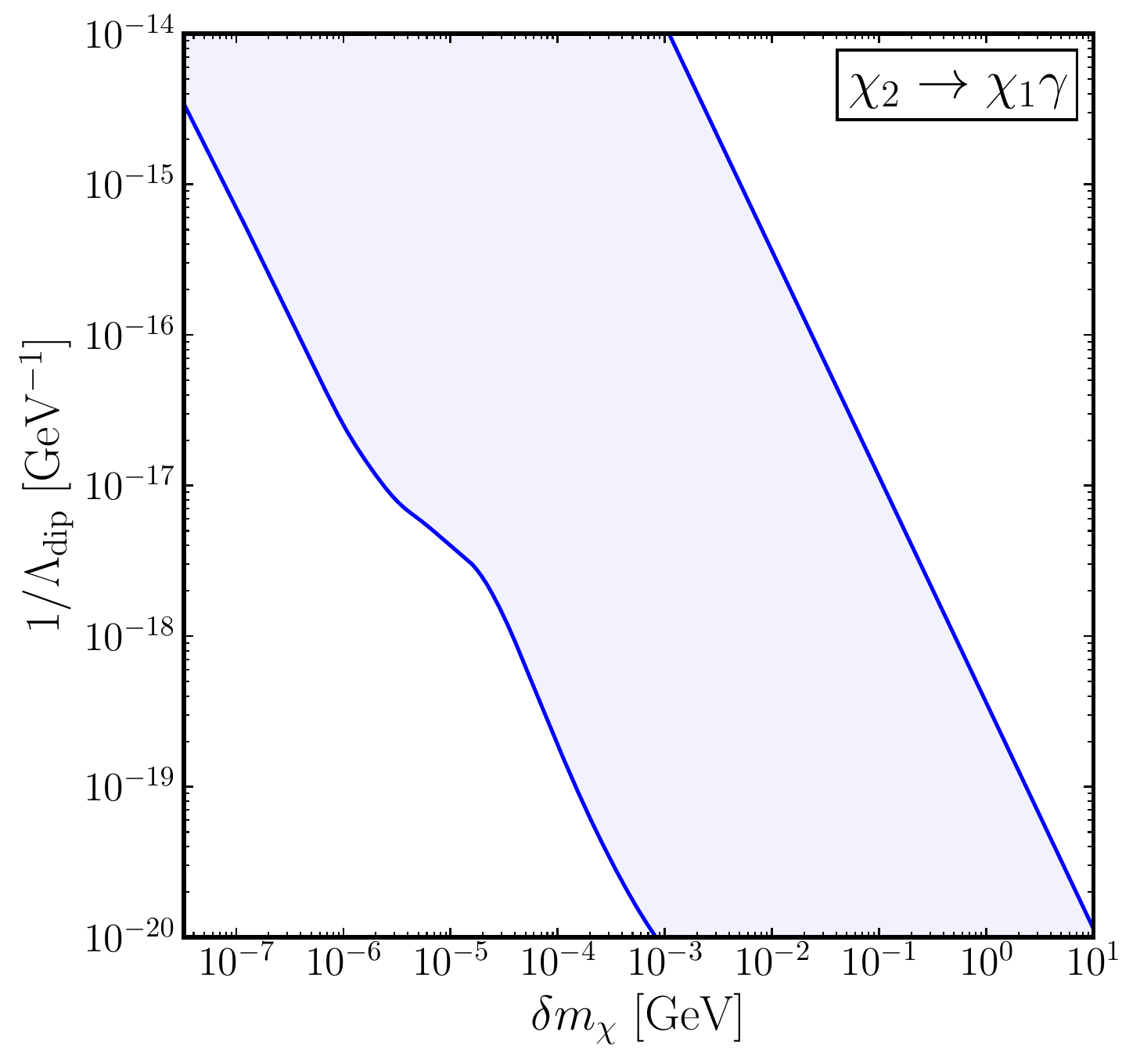}
    \caption{Excluded parameter space region of the two-component DM with electromagnetic dipole moments assuming a 21 cm monopole observation characterized by $z_t=15$ and $\mathcal{T}/\mathcal{A}_{\Lambda\rm CDM}^{\rm max}=0.50$. Here, we assume an initial fraction $F_{\chi_2}=1/2$. The bounds are represented in the space spanned by the dimensionful coupling $1/\Lambda_{\rm dip}$ and by the energy injected into the IGM per decay event $\delta m_\chi$.}
    \label{fig: DM dipole bounds}
\end{figure}

\begin{figure}
    \centering
    \includegraphics[width=0.48\linewidth]{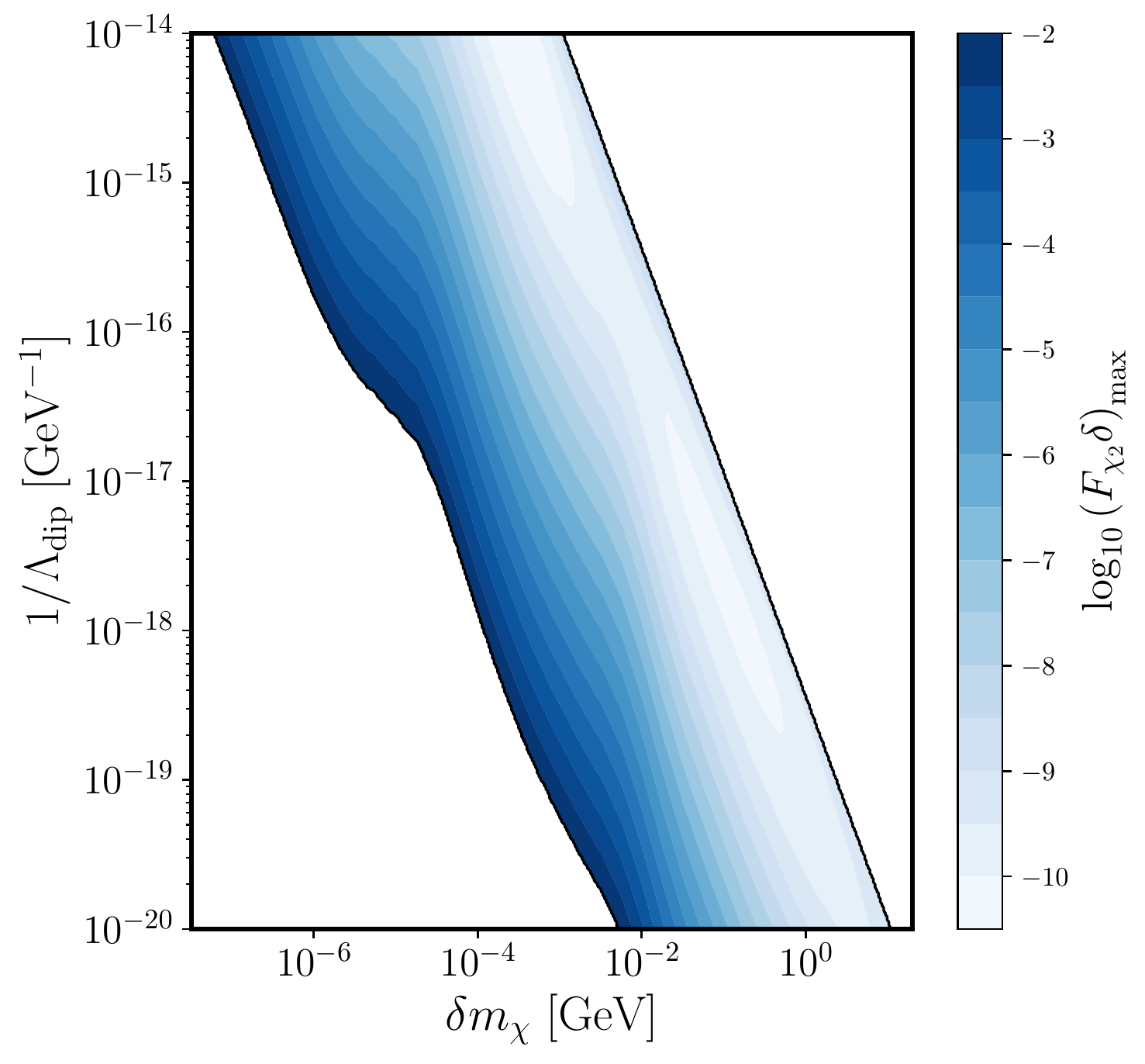}
    \includegraphics[width=0.48\linewidth]{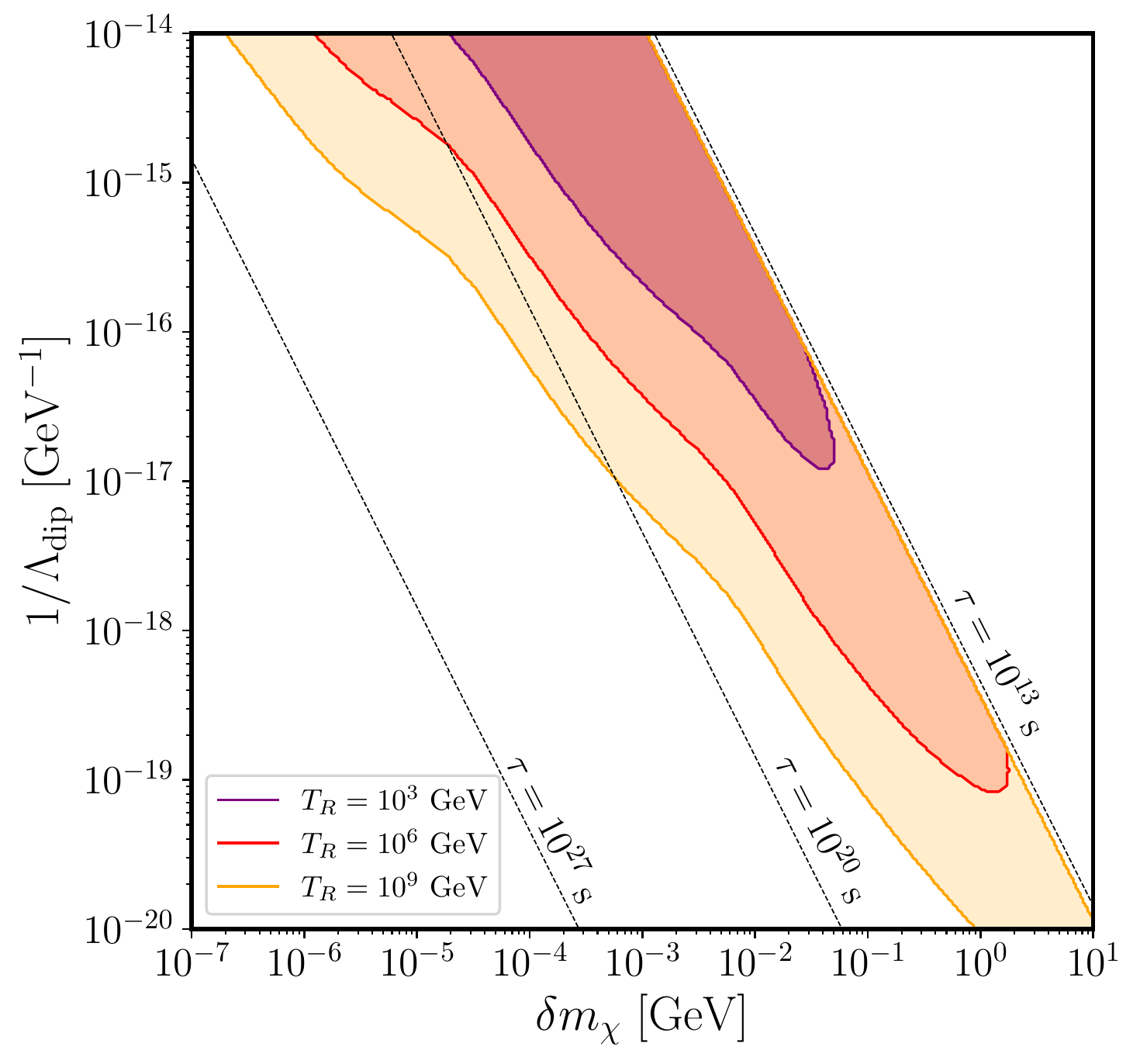}
    \caption{
Bounds on nearly degenerate two-component DM with electromagnetic dipole moments, assuming $\mathcal{T}/\mathcal{A}_{\Lambda\rm CDM}^{\rm max}=0.50$ and $z_t=15$. \textbf{Left panel}: heat map for the maximal values of $F_{\chi_2}\delta$ in the plane spanned by the dimensionful coupling $1/\Lambda_{\rm dip}$ and the energy injected into the IGM per decay event $\delta m_\chi$. \textbf{Right panel}: 21 cm bounds on the same model assuming pure freeze-in production. Colored regions would be excluded and different colors correspond to different reheat temperatures.}
    \label{fig: DM dipole heat maps}
\end{figure}

We present our bounds for this model following the same approach as in Sec.~\ref{subsec:ALP}. First, we consider the scenario where $\chi_1$ and $\chi_2$ together make up the entire DM relic abundance, and we determine how 21 cm cosmology constraints shape the allowed $(\delta m_{\chi}, \Lambda_{\rm dip})$ parameter space. This analysis requires tuning the reheating temperature $T_R$ for each value of $\delta m_{\chi}$ so that the freeze-in abundance matches the observed DM density, or alternatively, assuming the presence of some mechanism that ensures the correct relic abundance. Additionally, it depends on how the total DM energy density is divided between the two states. The resulting constraints for the case $F_{\chi_2} = 1/2$ are shown in Fig.~\ref{fig: DM dipole bounds}. We then explore what is the maximal fraction $F_{\chi_2}$ compatible with an observation of the 21 cm monopole signal. In the left panel of Fig.~\ref{fig: DM dipole heat maps}, we remain agnostic about the production mechanism and compute the maximal value for the combination $F_{\chi_2} \delta$ that would be allowed by 21 cm observations across the entire $(\delta m_{\chi}, \Lambda_{\rm dip})$ parameter space. Finally, in the right panel of Fig.~\ref{fig: DM dipole heat maps}, we assume freeze-in production and evaluate the relic density self-consistently. In particular, we recast the model-independent bounds derived in Sec.~\ref{subsec:nearlydegenerate} by requiring
\begin{equation}
    \mathcal{F}_{\chi_{2}}\left(\delta m_{\chi}, \Lambda_{\rm dip}, T_R\right)
    = \frac{1}{2} \frac{m_{\chi} \, s(t_0) Y_{\chi}^\infty}{\rho_{\rm DM}(t_0)}
    < \frac{\left(F_{\chi_2} \delta\right)_{\rm max}}{\delta},
\end{equation}
where $Y_{\chi}^\infty$ is the total freeze-in abundance computed as prescribed by Eq.~\eqref{eq:Ychidip}, and the overall factor of $1/2$ keeps into account that the two states are produced democratically.

Providing a comprehensive list of complementary probes that can constrain this scenario lies beyond the scope of this work. Our aim here is to illustrate how the bounds we derive can be applied across a variety of models. Nonetheless, in the relevant range of masses and lifetimes, the most important complementary constraints to consider are those from CMB observations and X-ray searches, as shown in the right panel of Fig.~\ref{fig:ALPbound2}. The hierarchy among these constraints is expected to remain unchanged, with 21 cm observations providing the most sensitive probe in the regions $2\mathcal{R} \lesssim \delta m_\chi \lesssim 20~\text{keV}$ and $10^{-1}~\text{MeV} \lesssim \delta m_\chi \lesssim 10~\text{MeV}$.

\section{Conclusions}
\label{sec:summary}

The consolidation of the $\Lambda$CDM model over the past decades has been made possible by key observational pillars, including the CMB and measurements of the local universe. These are highly complementary, as they probe different epochs and length scales of cosmic history. In recent years, 21 cm cosmology has entered a new era of precision, offering unprecedented opportunities to probe the evolution of our universe. In addition to illuminating standard processes such as primordial star formation and reionization, it also provides novel avenues for investigating the nature of DM. Indeed, the growing body of literature in this area explores how various dark sectors could influence the evolution of the cosmic 21 cm signal, thereby providing new constraints through future observations.

This work provided a comprehensive study of the 21 cm monopole signal, initially exploring it in a model-independent framework and subsequently examining the signatures of non-minimal dark sectors within specific microscopic models. In doing so, it established a robust foundation for understanding how different dark sectors can impact the 21 cm signal, offering new insights into the interplay between cosmology and DM. Before exploiting non-minimality for the dark sector, we revisited in Sec.~\ref{sec:DMbounds} the projected sensitivity to DM decays. More specifically, we assumed that DM consists of a single particle species with a very long, yet finite, lifetime. Including the back-reaction due to DM energy injections enhances the projected sensitivities by a factor of a few, depending on the threshold amplitude employed and the redshift of the trough. Nevertheless, the order of magnitude and qualitative behavior of the constraints remain consistent with previous studies, as expected. The dependence of the projected bounds on $\mathcal{T}/\mathcal{A}_{\Lambda \rm CDM}^{\rm max}$ and $z_t$ is physically well understood, given the qualitative evolution of the IGM temperature and ionization fraction. A final extension of earlier work, for the decay mode to a photon pair, is the set of constraints for DM mass values in the range $20.4\ \text{eV}\lesssim m_\chi\lesssim 10\ \text{keV}$. Our projected constraints within this mass window would be the most stringent available, excluding lifetimes $\tau\lesssim 10^{28}$ s. This result is in agreement with a recent study that performed a sensitivity analysis to DM decays, considering the 21 cm power spectrum as the observable \cite{sun:inh}.

We ventured into non-minimality starting from Sec.~\ref{sec:non-minimal}, where we conducted a model-independent study. First, we considered a scenario with a metastable DM sub-component. In this case, depending on the fractional abundance of the metastable species, the allowed lifetimes can be much shorter than the age of the universe, thereby modifying the power injected into the IGM per unit volume. We derived prospective constraints on the full model space spanned by $\left(m_\chi, \tau, F_\chi\right)$, specifically computing the maximal fractional abundance of the metastable species that can be present for a given mass and lifetime. In this scenario, 21 cm cosmology could probe previously unexplored regions of parameter space. Building on these results, we next considered a nearly degenerate two-component DM model, which commonly appears in extensions of the SM, such as inelastic DM. We investigated a phenomenological framework where the heavier state decays to the lighter one, emitting one or two SM particles. If the relative mass splitting between the two dark states is small, $\delta \lesssim 10^{-2}$, then the power injected per unit volume is closely related to that in the previous scenario.

Finally, we demonstrated the broad applicability of the model-independent results with explicit Lagrangian realizations of non-minimal dark sectors in Sec.~\ref{sec:Lag}. First, we considered an ALP coupled to photons via a dimension-5 contact interactions. By recasting the lifetime constraints from the single-component DM scenario in the mass range where they are most competitive ($20.4\ \text{eV}\lesssim m_\chi\lesssim 10\ \text{keV}$), we derived the limiting values for the coupling $g_{a\gamma\gamma}$ as a function of the ALP mass. As a representative example, we find $g_{a\gamma\gamma} \lesssim 3 \times 10^{-15} \ \text{GeV}^{-1}$ for $m_a = 100$ eV, which is roughly one order of magnitude stronger than existing bounds. Next, by recasting the constraints on the metastable DM sub-component, we derived the maximal allowed fractional abundance of an ALP coupled to photons, given a specified mass and coupling. With these results, we were able to compute the projected 21 cm constraints on the irreducible ALP abundance, as defined in \cite{lan:irr}. In this case, our bounds would surpass current limits. Second, we examined the pseudo-Dirac DM framework, considering both a realization with renormalizable interactions via a vector mediator, and another with higher-dimensional interactions through electromagnetic dipoles. In both cases, our results proved to be particularly effective in constraining the model parameter space.

The quest to unravel the composition of the dark universe is more open now than ever before. While assuming minimality, with only one new species to account for DM, certainly has its advantages, it could lead us to overlook key phenomenological signatures. Given the complexity of the visible universe, with the SM gauge group being the direct product of several distinct contributions, and matter that confines while being charged under multiple representations, it is worth considering the hypothesis of non-minimal dark sectors. In this work, we took a small step toward this direction by investigating dark sectors composed of two components. Typically, one component is stable, ensuring the persistence of DM to the present day, while the other is metastable and can decay, leaving cosmological imprints. Our study focused on the 21 cm monopole signal, and our findings point to several promising avenues for future research. These include extending similar analyses to the 21 cm power spectrum and placing constraints on other irreducible abundances, such as axion-like particles coupled exclusively to electrons, positrons, or gluons. Overall, this work provides a comprehensive framework for probing non-minimal dark sectors, underscoring the potential of 21 cm cosmology to uncover their signatures. Continuing these investigations in future studies will be essential for deepening our understanding of the dark universe.

\acknowledgments
We thank Wenzer Qin for correspondence that clarified some issues with the \textsf{DarkHistory} code.
F.C. is supported by the U.S. Department of Energy, Office of Science, Office of High Energy Physics under Award Number DE-SC0011632, and by the Walter Burke Institute for Theoretical Physics. F.D. is supported by Istituto Nazionale di Fisica Nucleare (INFN) through the Theoretical Astroparticle Physics (TAsP) project, and in part by the Italian MUR Departments of Excellence grant 2023-2027 ``Quantum Frontiers''. CloudVeneto is acknowledged for the use of computing and storage facilities.

\appendix

\section{Ionization and Thermal History}
\label{app:ion&therhistory}

In this appendix, we review the formalism needed to study the evolution of the gas temperature $T_g$ and the ionization fraction $x_e \equiv n_e/n_H$, i.e., the ratio between the number densities of free electrons and hydrogen (both neutral and ionized), during the cosmic dark ages. 

In standard cosmology, the gas temperature evolution is obtained by imposing energy conservation in a comoving volume, while the ionization history is, to a good approximation, described by the Peebles model~\cite{pee:rec}, also known as the Three-Level Atom (TLA) model. The evolution equations read~\cite{liu:imp}
\begin{subequations}
\begin{align}
\label{stardard Tg evolution}\dot{T}_g & = \dot{T}_g^{(0)} \equiv -2 H T_g + \Gamma_C (T_\gamma - T_g),\\
\label{stardard XHII evolution} \dot{x}_{H {\rm II}} & = \dot{x}_{H {\rm II}}^{(0)} \equiv -\mathcal{P}_2 \left[ \alpha_B n_e x_{H {\rm II}} - 4\beta_H e^{-E_\alpha/T_\gamma}(1 - x_{H {\rm II}}) \right] \ .
\end{align}
\end{subequations}
Here, $H$ is the Hubble parameter, $\mathcal{P}_2$ is the Peebles coefficient (i.e., the probability that a hydrogen atom in the $n = 2$ state decays to the ground state before being ionized), $\alpha_B$ is the case-B recombination coefficient, $\beta_H$ is the effective photoionization rate, and $E_\alpha = 10.2~\text{eV}$ is the energy of the Ly$\alpha$ transition. The Compton scattering rate $\Gamma_C$ is given by
\begin{equation}
\Gamma_C = \frac{x_e}{1 + f_{\rm He} + x_e} \frac{8\sigma_T a_r T_\gamma^4}{3 m_e},
\end{equation}
where $f_{\rm He} = n_{\rm He}/n_H$ is the helium-to-hydrogen number ratio, $\sigma_T$ is the Thomson cross section, $a_r$ is the radiation constant, and $m_e$ is the electron mass.

The ionized helium fraction $x_{\rm He\,II}$ can also be tracked using the TLA model, leading to an equation analogous to Eq.~\eqref{stardard XHII evolution}~\cite{liu:cod, won:how}. A more accurate description of the ionization history, which includes higher-$n$ excitations and de-excitations of both hydrogen and helium, is provided by the Multi-Level Atom (MLA) formalism~\cite{sea:how, chl:cos, chl:rec, gri:cos}. The MLA approach is essential for properly modeling the spectrum of low-energy ($E < 10.2~\text{eV}$) photons. While our numerical analysis also incorporates this more refined treatment, in this appendix we restrict the discussion to the TLA model for clarity of presentation.

When considering the presence of some exotic source injecting energy into the IGM, Eqs. \eqref{stardard Tg evolution} and \eqref{stardard XHII evolution} have to be modified in order to account for this additional contribution. In this context is important to notice that the power injected into the universe at some redshift z is not entirely deposited instantaneously in the IGM, much of the energy deposition is delayed \cite{sla:ind1, sla:ind2}. It is customary to assume that the deposited power is directly proportional to the injected power, through a redshift dependent efficiency factor $f(z)$, 
\begin{equation}
\left(\frac{dE}{dV dt}\right)_{dep}=f(z)\left(\frac{dE}{dV dt}\right)_{inj}.
\end{equation} 
$f(z)$ is the ratio between the deposited power and the injected one at the same redshift $z$, even if much of the energy deposition is delayed, so that the energy deposited at some redshift $z$ originates from energy injections at earlier time.  The overall energy deposited is split in four main channels \cite{sla:ind2}
\begin{enumerate}
\item \emph{ionizations}, either of hydrogen or helium atoms;
\item \emph{excitations},  e.g. through the production of Ly$\alpha$ photons;
\item \emph{heating}, produced mainly by Coulomb scattering;
\item \emph{low-energy photons},  unable to ionize the gas, but contributing to higher-$n$ excitations and CMB spectral distortions.
\end{enumerate}
The relative importance of each channel depends on the final products of the specific process we are considering and their energies. We leave these dependencies implicit in our notation following previous literature. Various approaches have been proposed to establish the fractions of deposited power in each channel, such as the SSCK prescription \cite{shu:the, che:par} and the $3$ keV prescription \cite{gal:sys}. Nevertheless, it has been shown that such prescriptions are not accurate for a generic energy injection \cite{sla:ind1, sla:ind2}. Therefore, a more refined treatment of energy deposition is required, namely computing redshift dependent efficiency factors $f^c(z)$ quantifying deposited power pertaining to each channel $c$. Specifically, these coefficients are defined as
\begin{equation}
\left(\frac{dE}{dVdt}\right)_{dep}^c=f^c(z)\left(\frac{dE}{dV dt}\right)_{inj},
\label{eq:fczdef}
\end{equation}
where $c=i, e, h, l$ to denote ionizations, excitations, heating and low-energy photons, respectively. In terms of the $f^c(z)$ factors, it is straightforward to extend the evolution equations to account for some exotic energy injection.  The gas temperature evolution is modified only by the power deposited proceeding into heating, instead, in the TLA approximation, both the power deposited into ionizations and excitations contribute to the ionization evolution. These additional contributions read
\begin{subequations}
\begin{align}
\dot{T}_g^{(\chi)}&=\frac{2f^h(z)}{3(1+f_{He}+x_e)n_H}\left(\frac{dE}{dV dt}\right)_{inj},\\
\dot{x}_{H{\rm II}}^{(\chi)}&=\left[\frac{f^i(z)}{B_1}+\frac{(1-\mathcal{P}_2)f^e(z)}{E_\alpha n_H}\right]\left(\frac{dE}{dV dt}\right)_{inj},
\end{align}
\end{subequations}
where $B_1=13.6$ eV is the hydrogen binding energy.  The first term in square brackets counts the number of ionization per unit time and unit physical volume from the hydrogen ground state, whereas the second term counts the number of ionization from the $n=2$ level always per unit time and unit physical volume. Notice that all this discussion assumes homogeneous energy deposition.

To summarize, one has to solve the system
\begin{equation}
\begin{cases}
\dot{T}_g=\dot{T}_g^{(0)}+\dot{T}_g^{(\chi)}\\
\dot{x}_{H{\rm II}}=\dot{x}_{H{\rm II}}^{(0)}+\dot{x}_{H{\rm II}}^{(\chi)}.
\end{cases}
\end{equation}
As far as the computation of the $f^c(z)$ factors and the integration of the latter system of differential equations, we rely on the publicly available Python package \textsf{DarkHistory} (\textsf{DH}) \cite{liu:cod, liu:exo1}. This code, unlike previous ones, accounts for the back-reaction effect in the computation of efficiency factors, namely it accounts for the modification in the $T_g$ and $x_{H{\rm II}}$ evolution due to the cooling of the injected particles. Moreover, in the new version recently released \cite{liu:exo1}, excitation of hydrogen and helium atoms are described with the MLA method. This progress together with an improved treatment of low-energy electrons, allow to track accurately the spectrum of soft photons and to compute thermal and ionization histories from ideally arbitrarily soft injections.\footnote{In order to ensure reproducibility, we report in the following the options we have set in the code. When using the function $\mathtt{main.evolve}$ of \textsf{DHv1.0}: $\mathtt{helium\_TLA = True}$, $\mathtt{rtol = 1e{-}6}$, $\mathtt{coarsen\_factor = 12}$, $\mathtt{backreaction = True}$, $\mathtt{elec\_method = new}$. 
When using \textsf{DHv2.0} instead, besides the above options we have also set: $\mathtt{distort = True}$, $\mathtt{nmax = 200}$, $\mathtt{fexc\_switch = True}$, $\mathtt{reprocess\_distortion = True}$, $\mathtt{iterations = 5}$. All the unspecified options are left to their default values.}

\section{Additional Results}
\label{app: additional results}

\begin{figure}
    \centering
    \includegraphics[width=.24\linewidth]{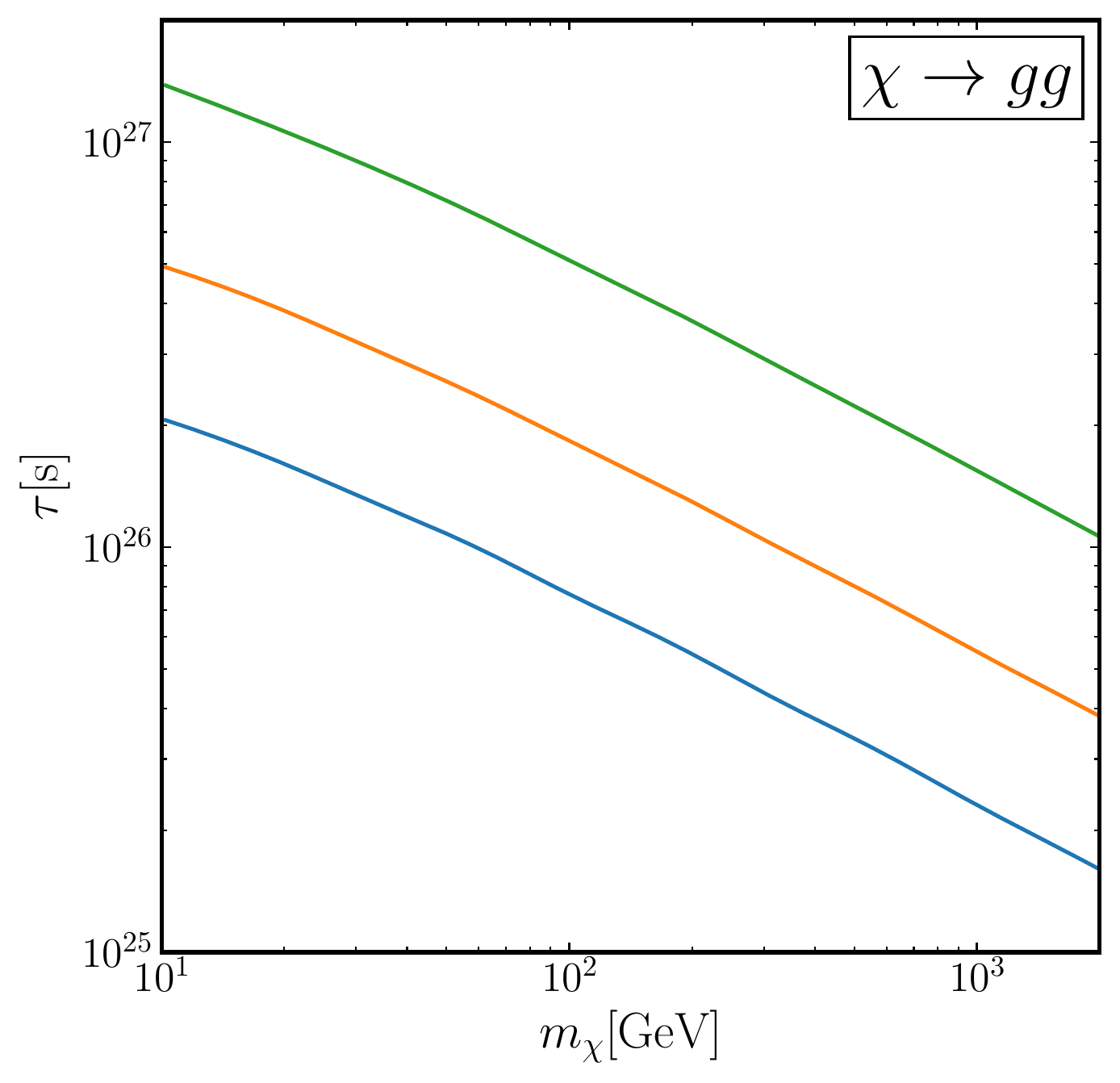}
    \includegraphics[width=.24\linewidth]{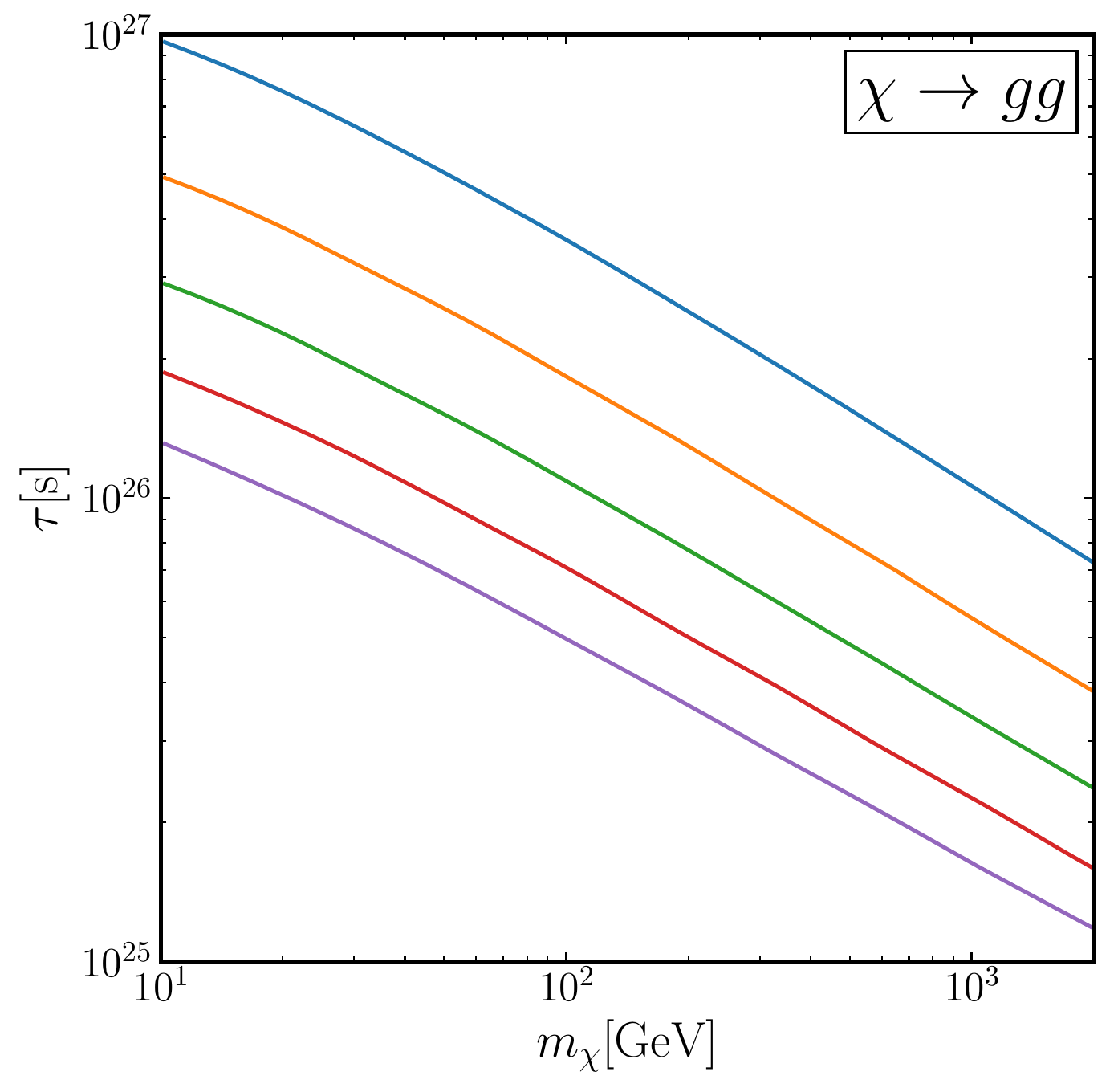}
    \includegraphics[width=.24\linewidth]{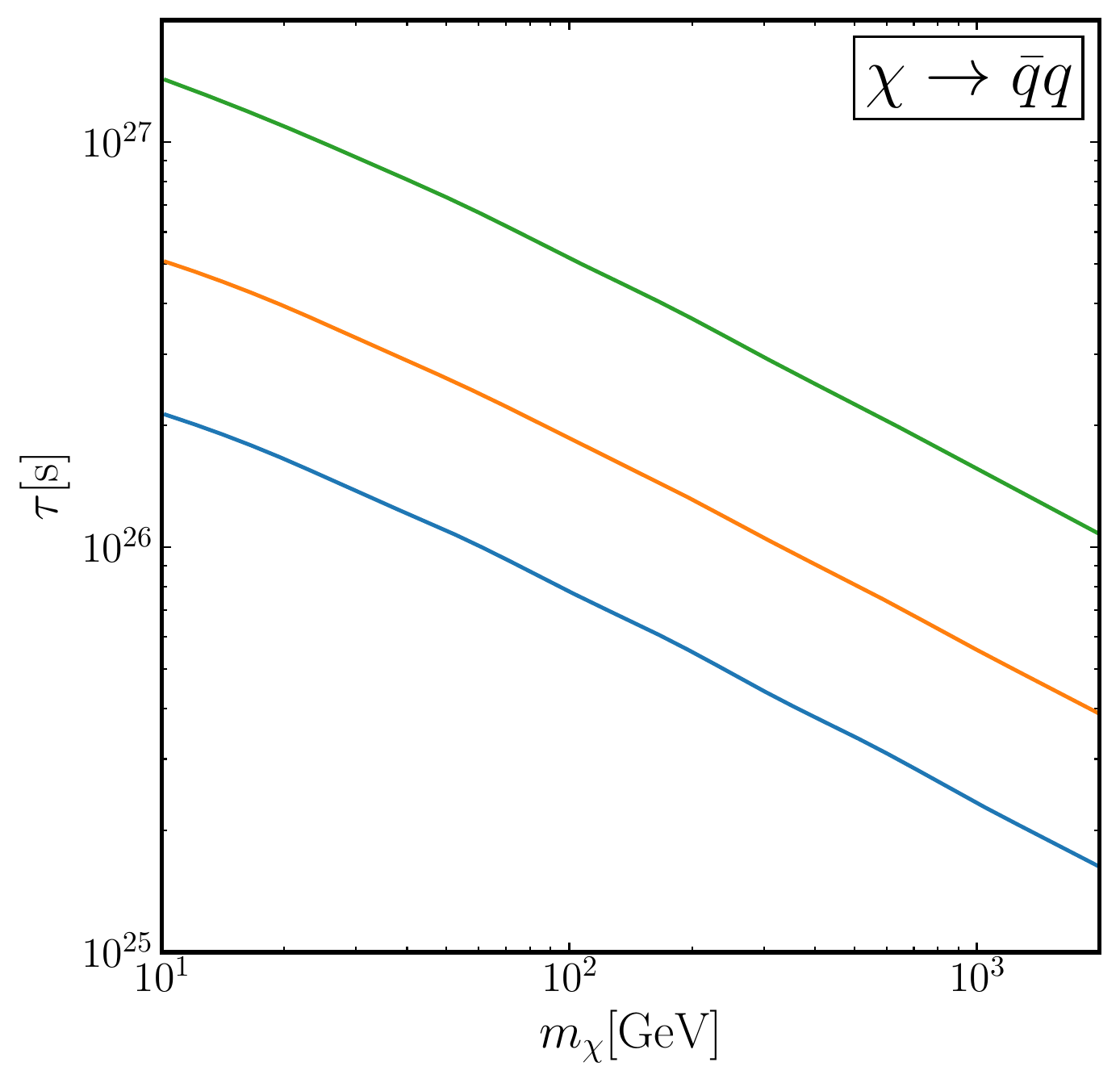}
    \includegraphics[width=.24\linewidth]{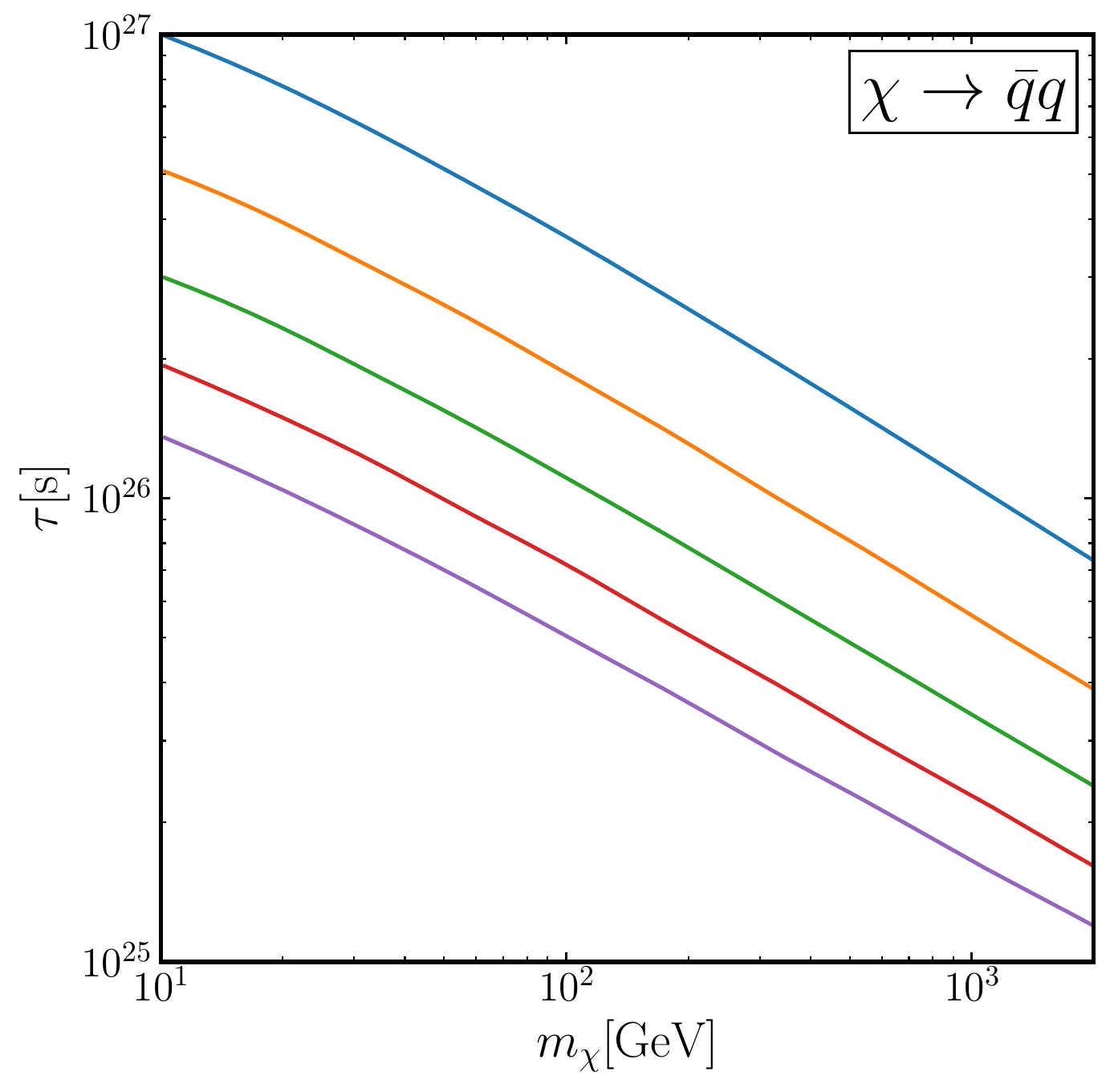}\\
    \includegraphics[width=.24\linewidth]{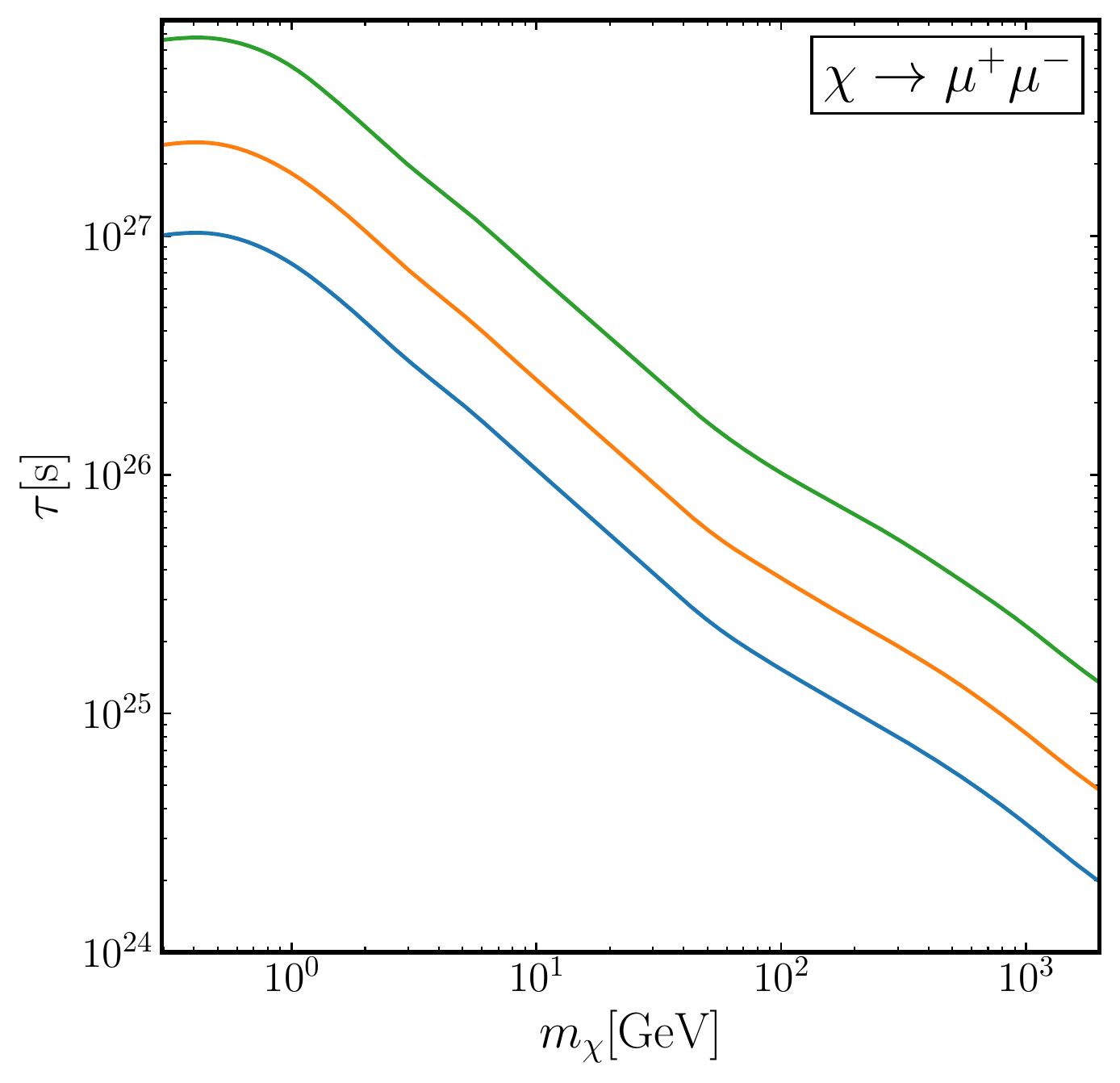}
    \includegraphics[width=.24\linewidth]{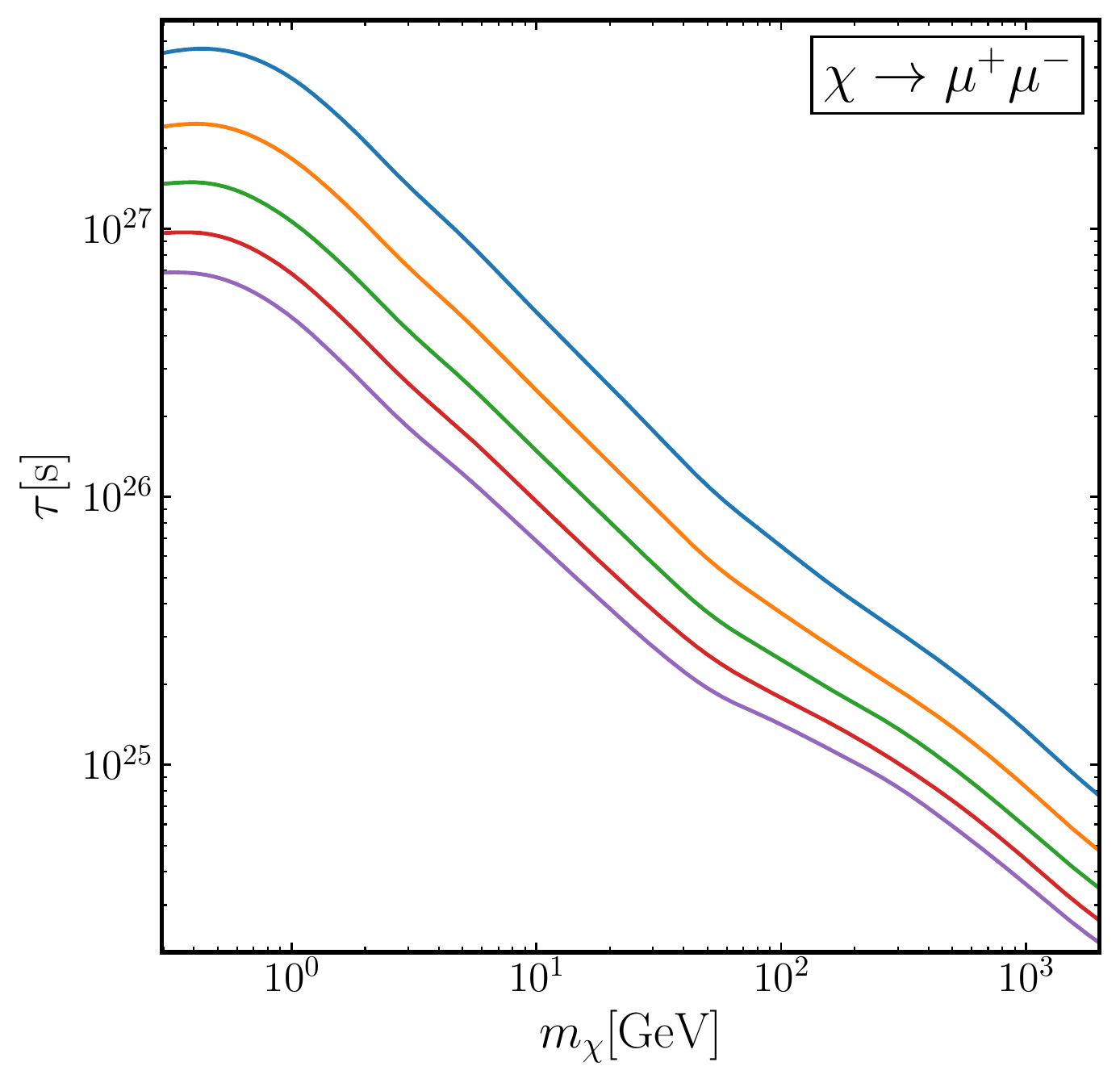}
    \includegraphics[width=.24\linewidth]{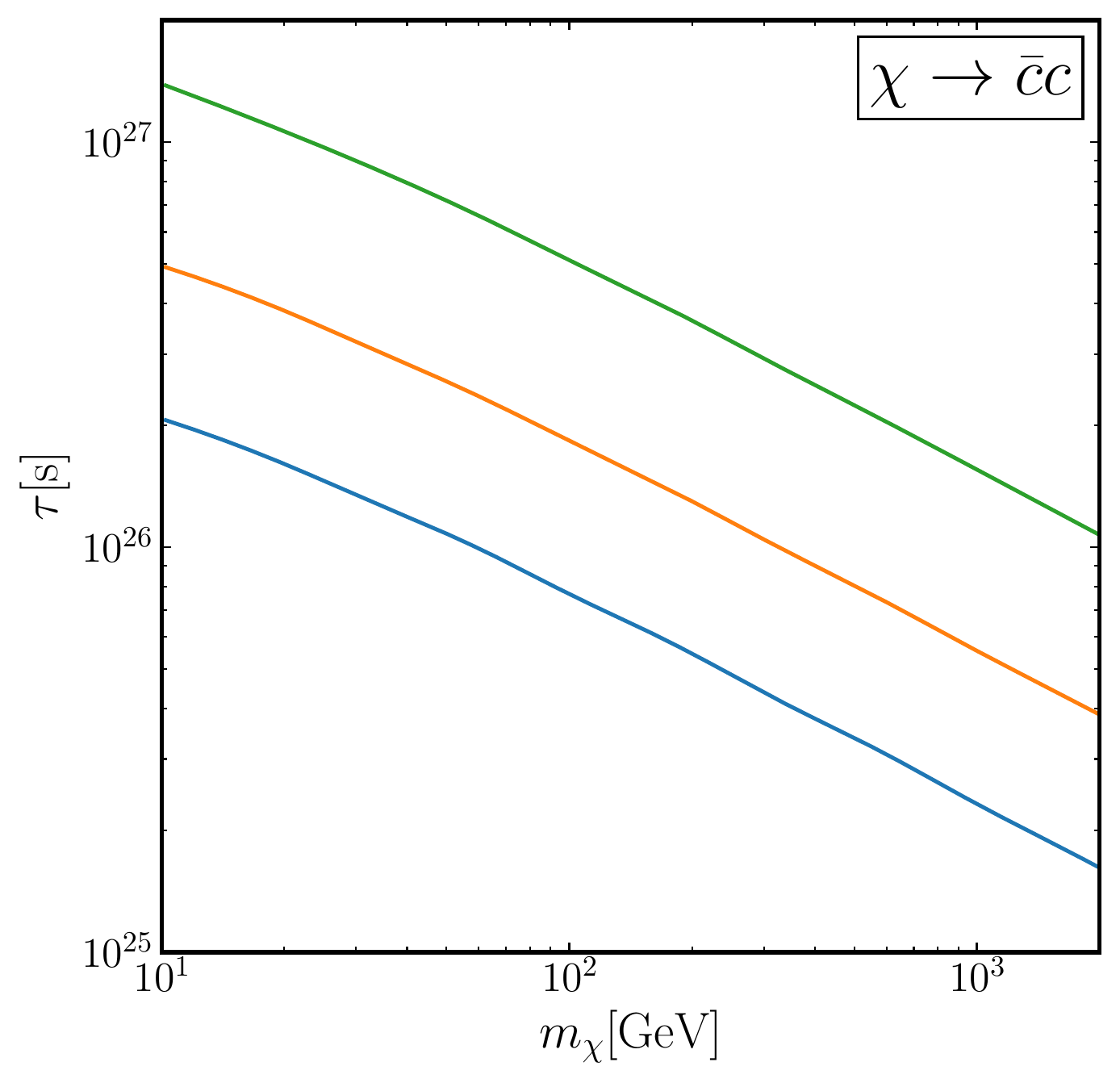}
    \includegraphics[width=.24\linewidth]{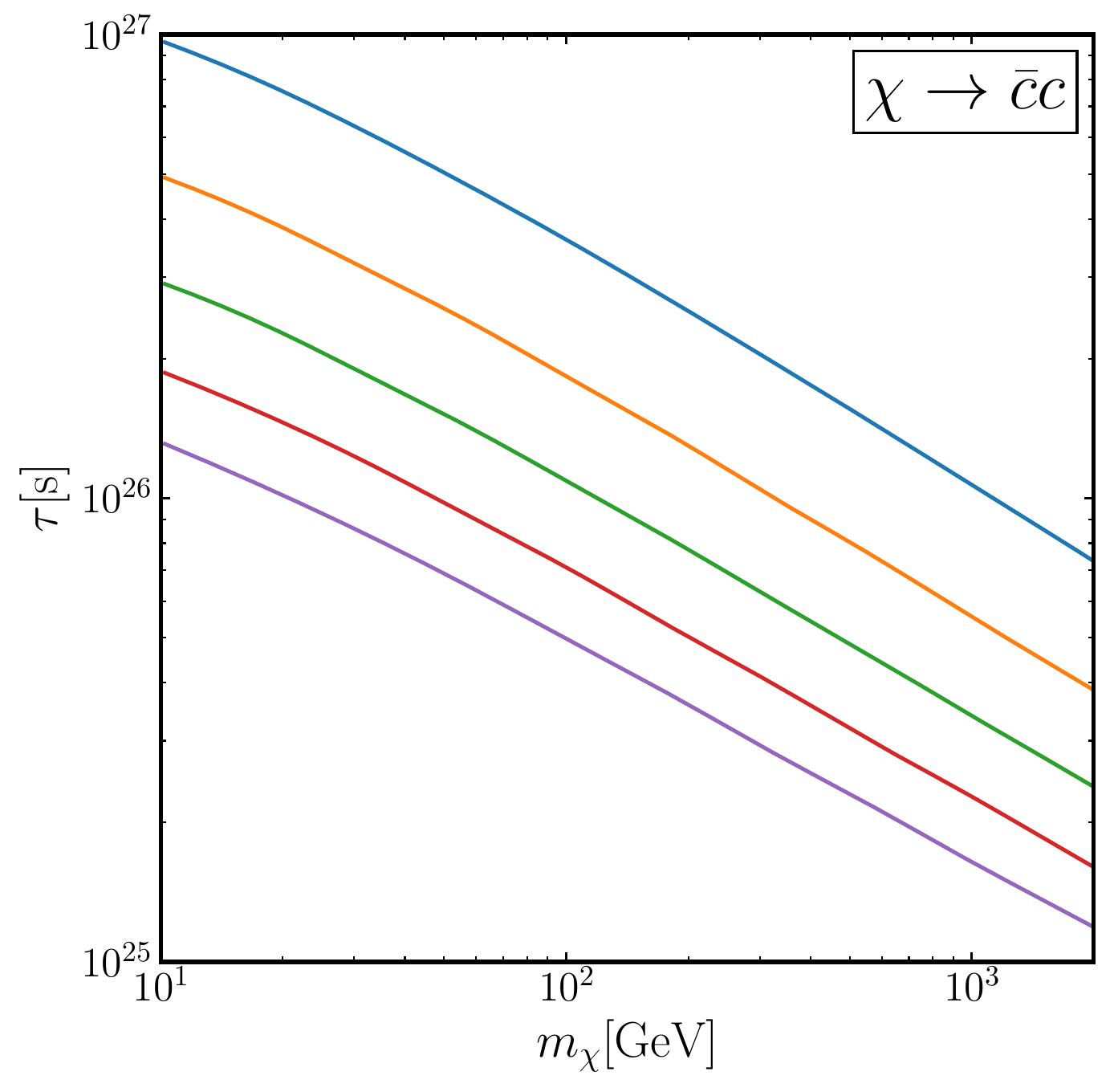}\\
    \includegraphics[width=.24\linewidth]{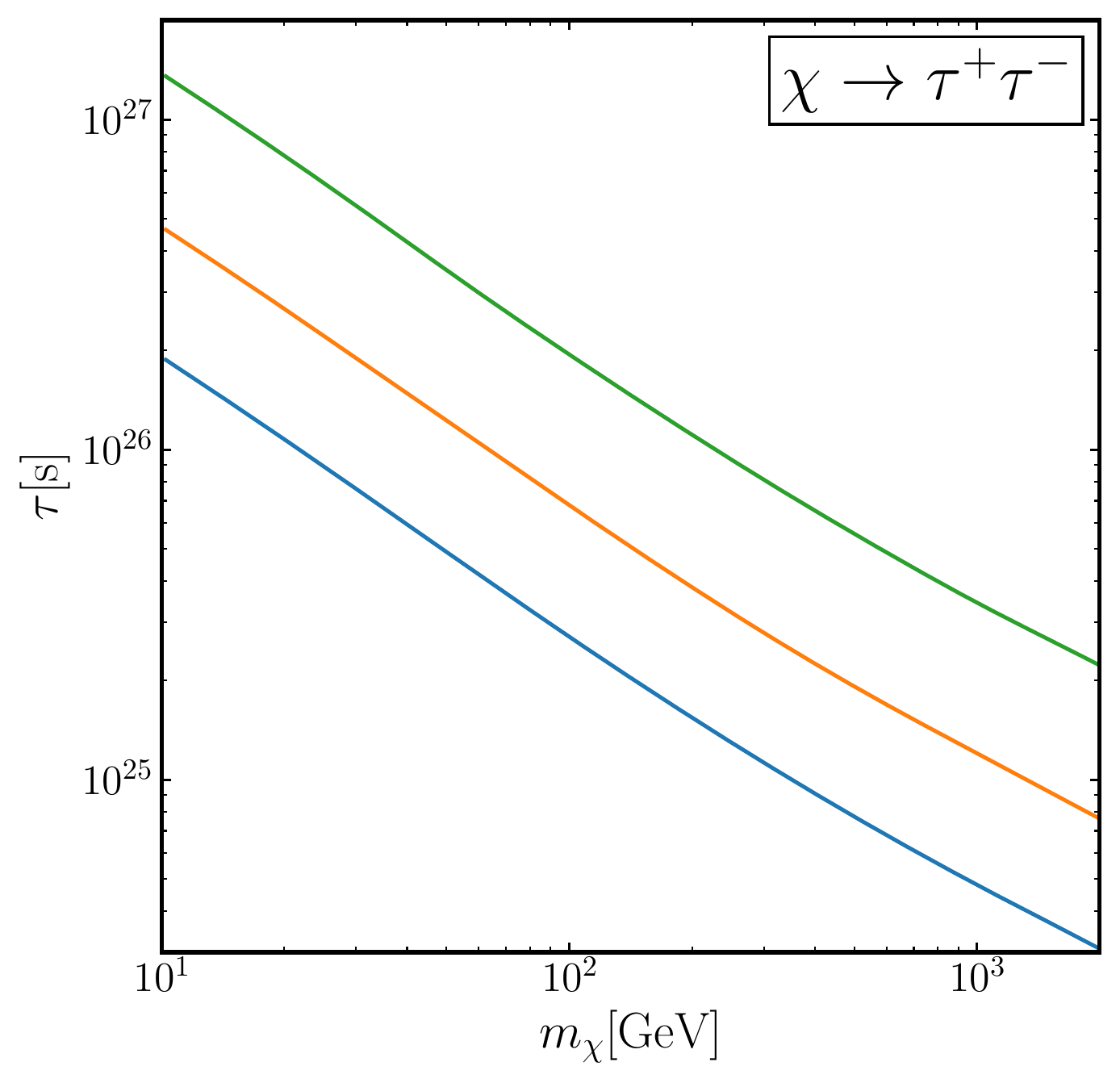}
    \includegraphics[width=.24\linewidth]{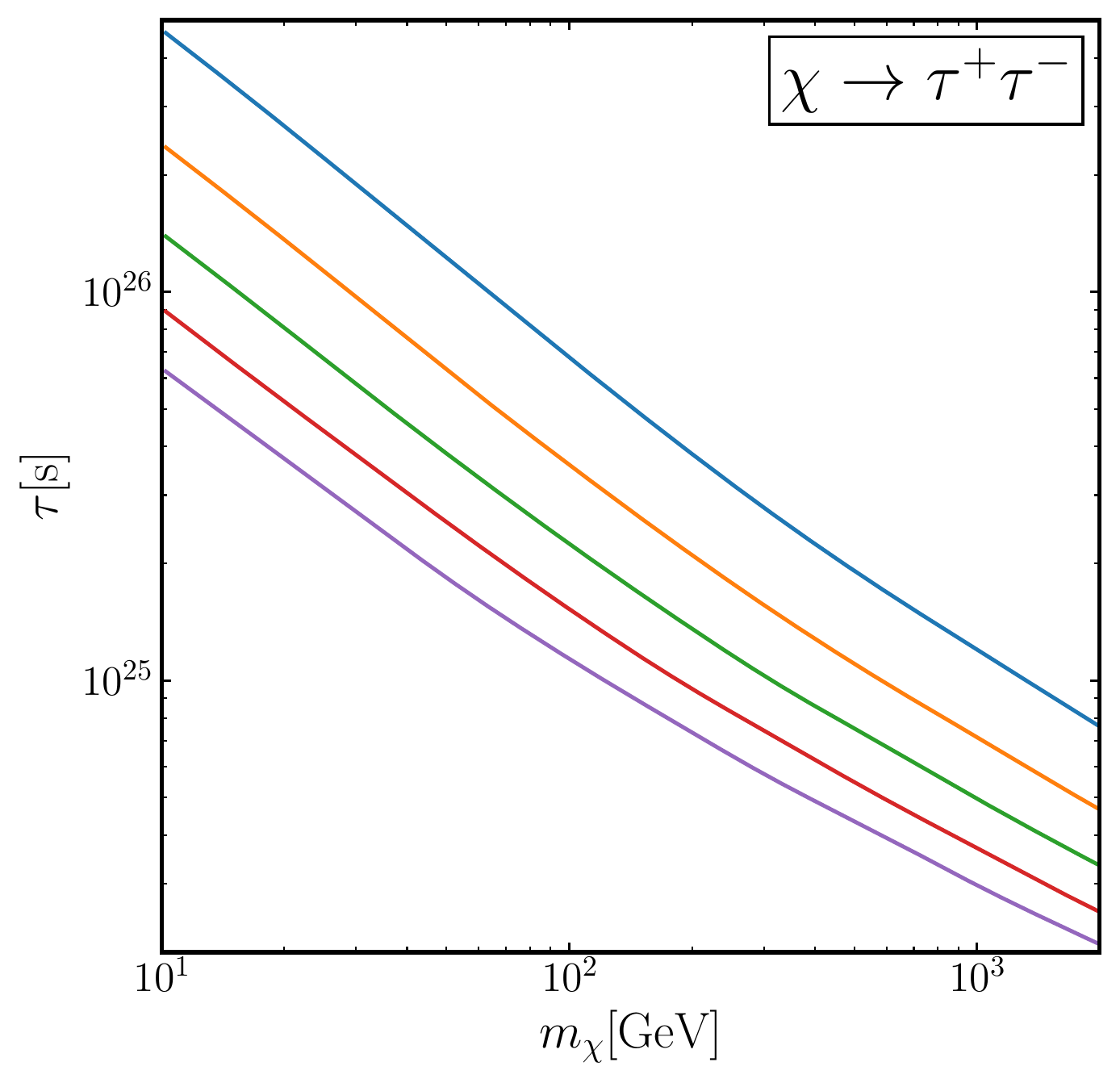}
    \includegraphics[width=.24\linewidth]{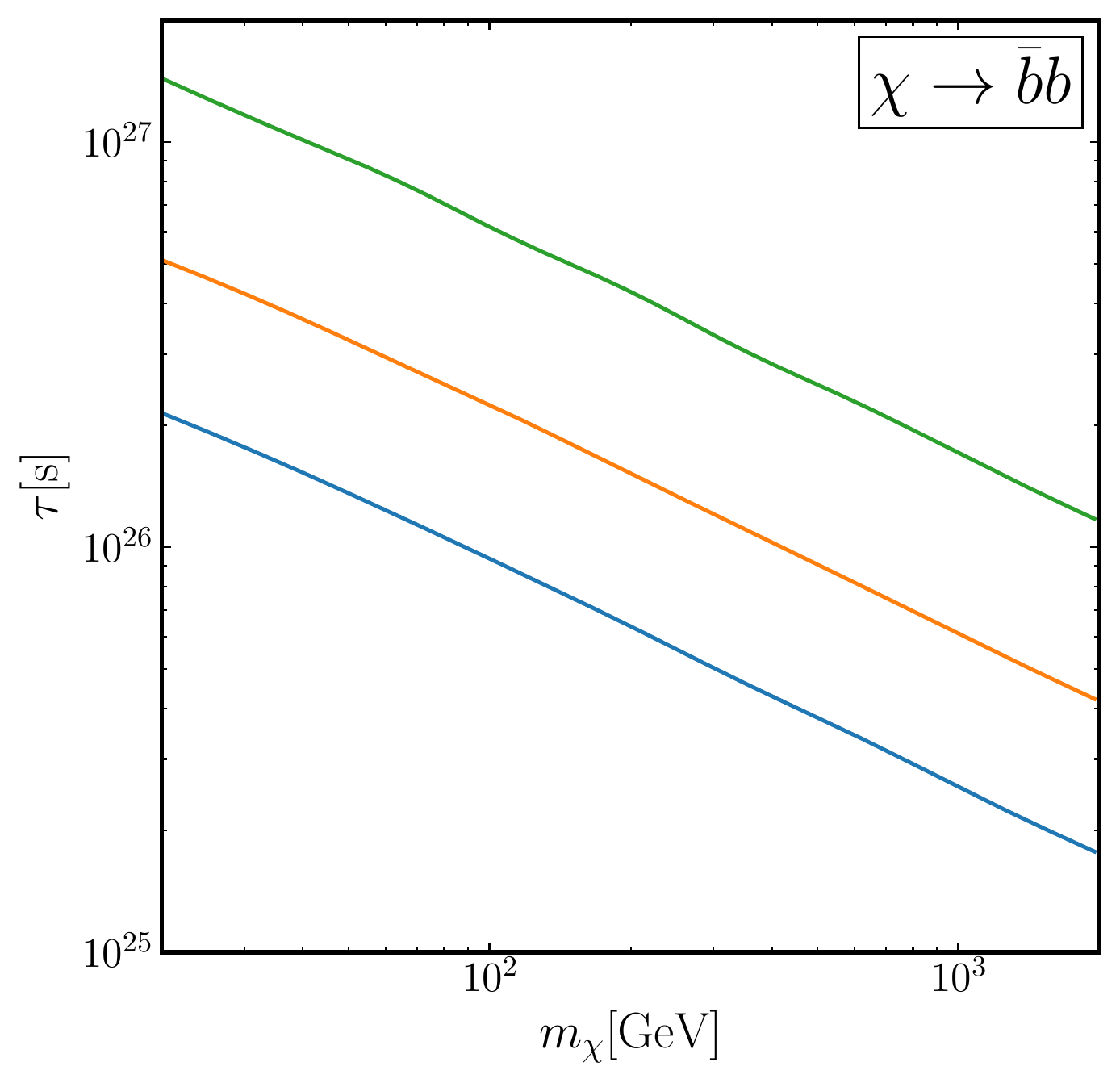}
    \includegraphics[width=.24\linewidth]{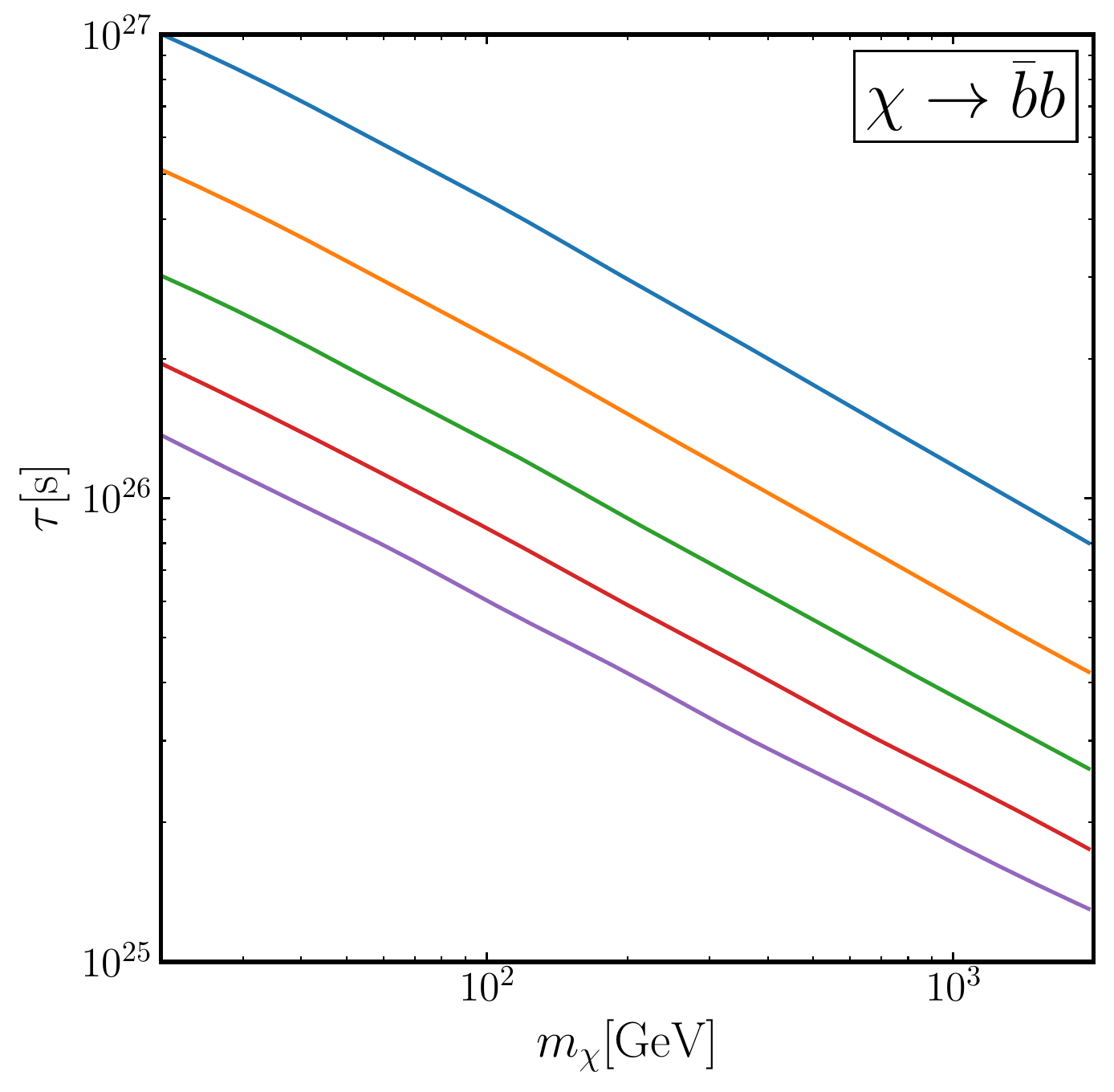}\\
    \includegraphics[width=.24\linewidth]{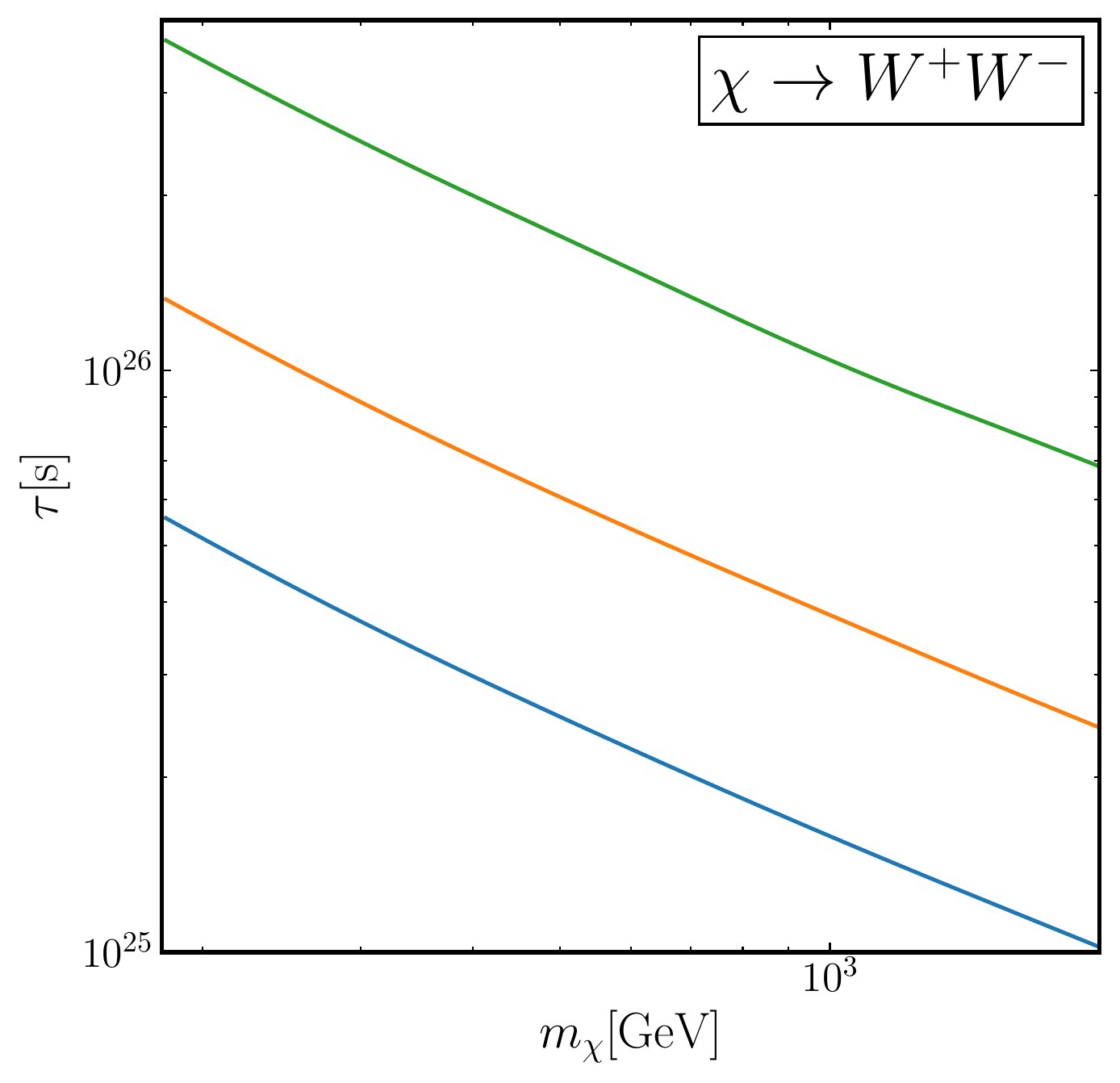}
    \includegraphics[width=.24\linewidth]{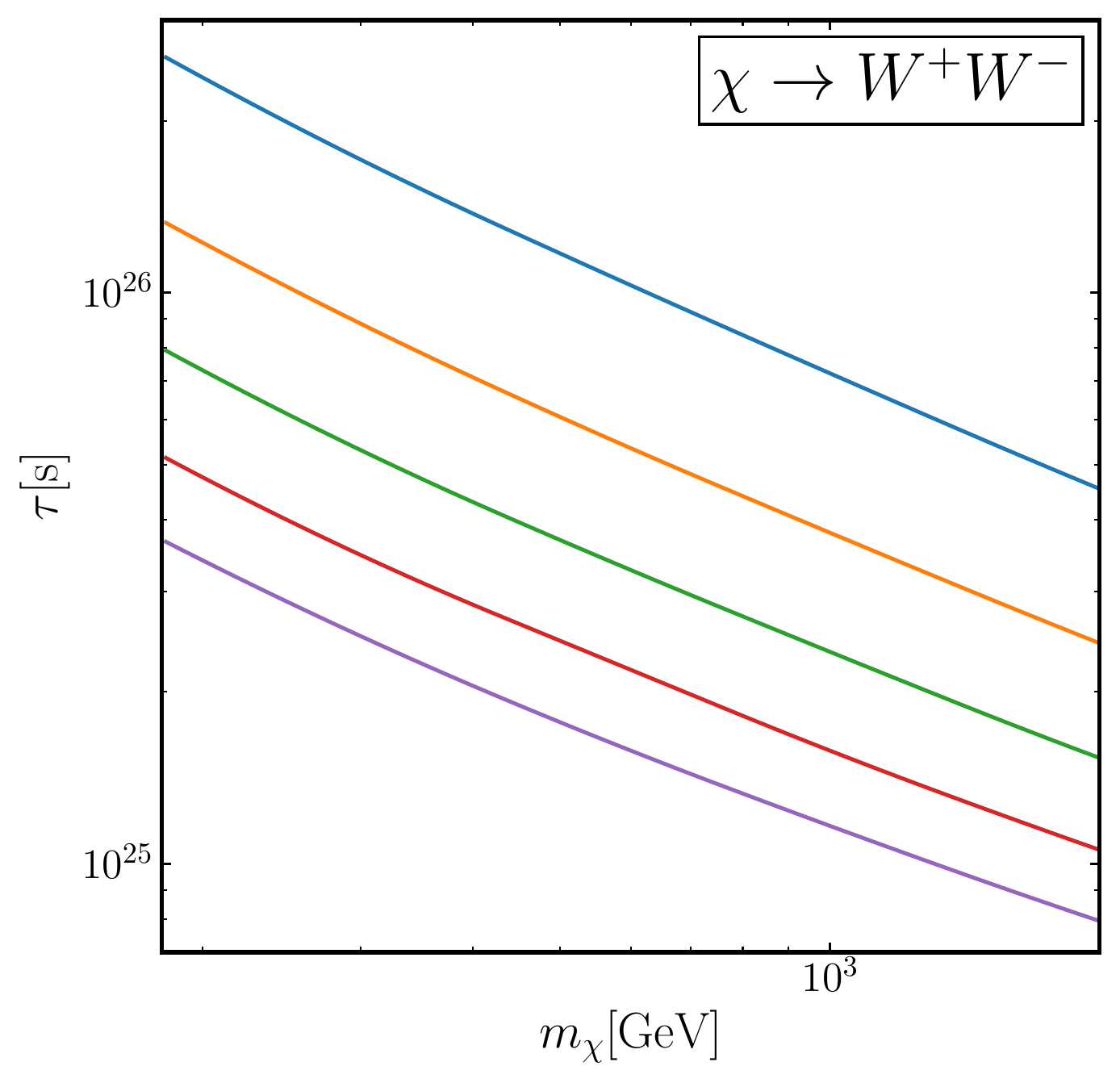}
    \includegraphics[width=.24\linewidth]{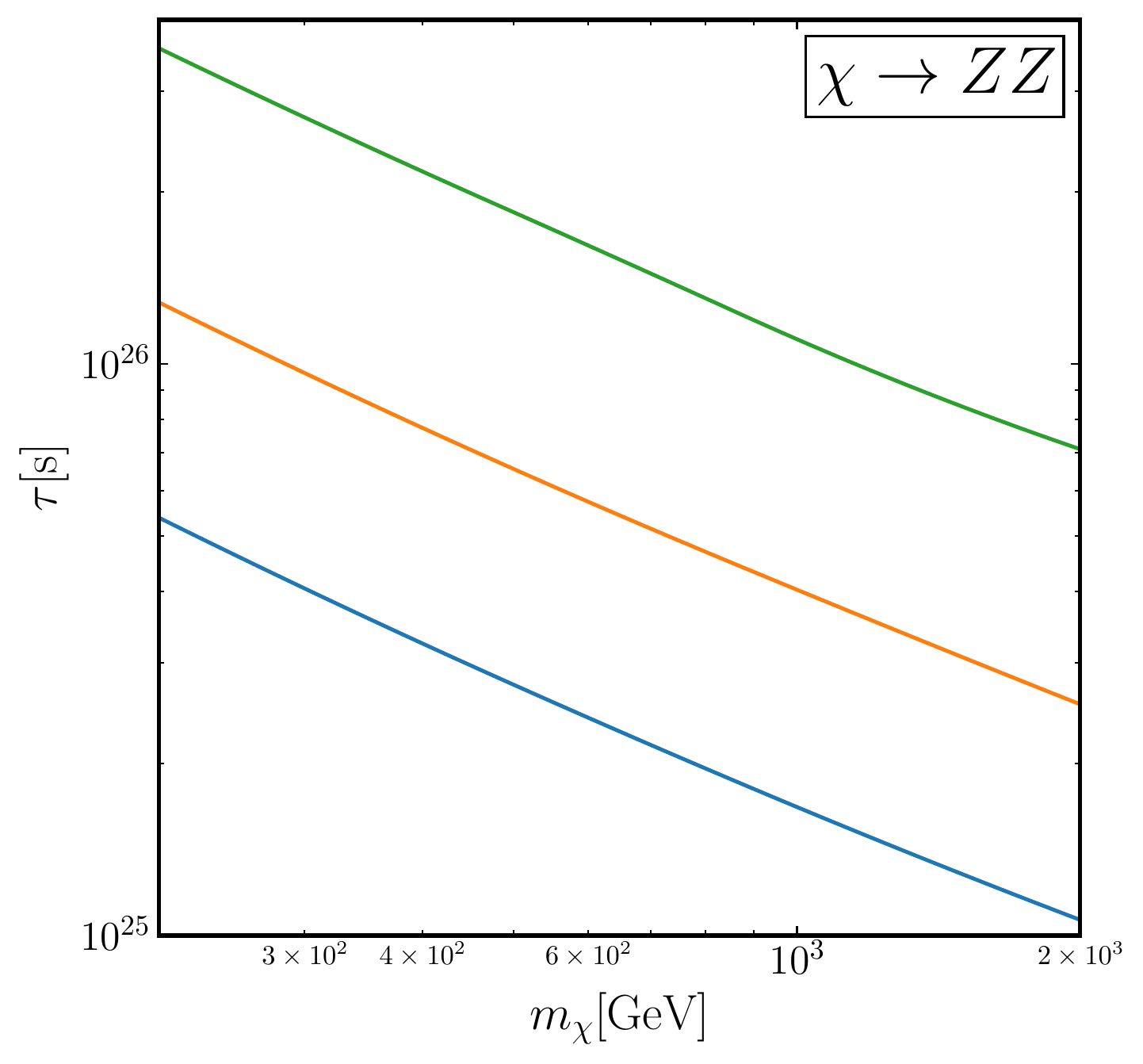}
    \includegraphics[width=.24\linewidth]{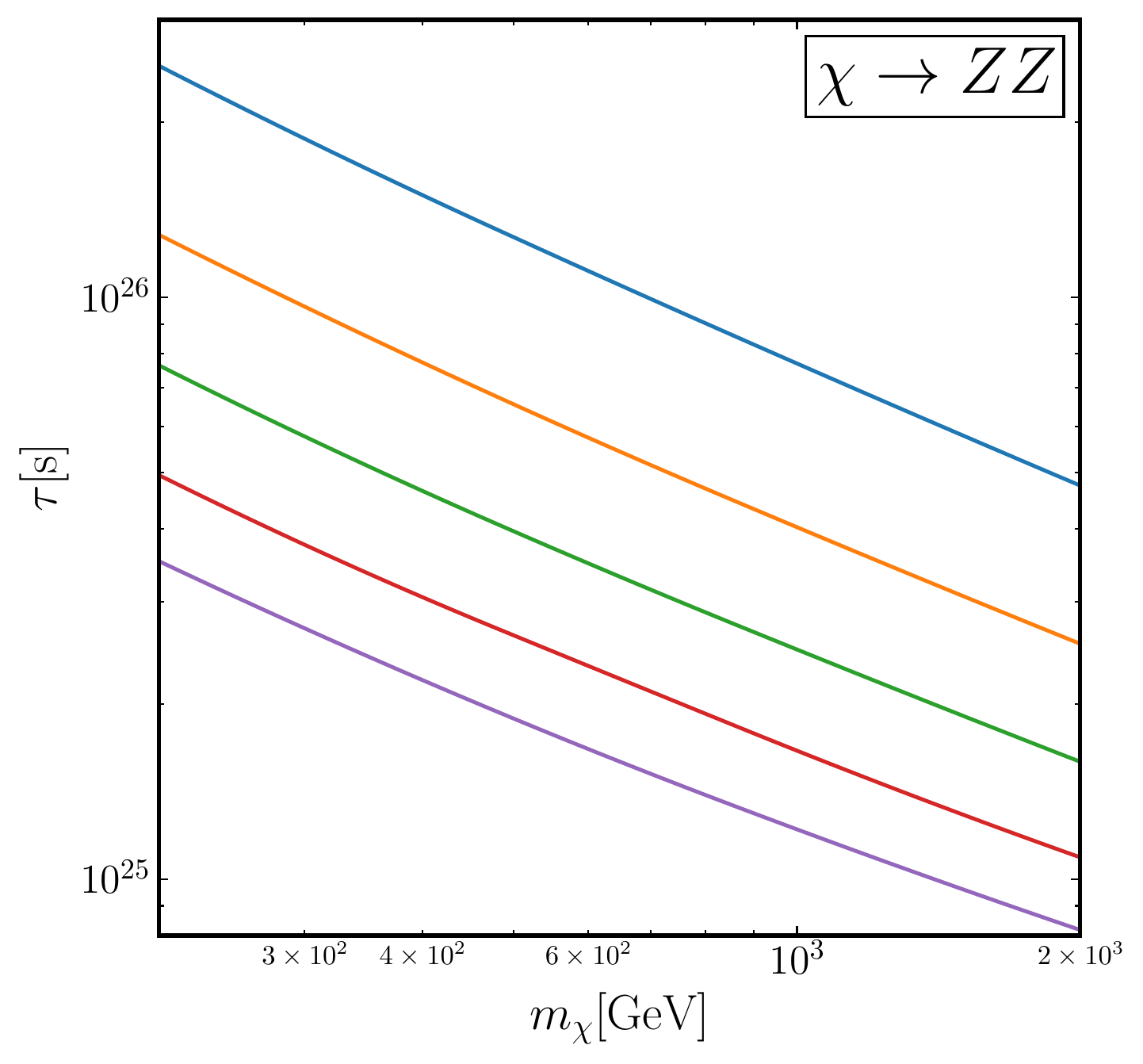}\\
    \includegraphics[width=.24\linewidth]{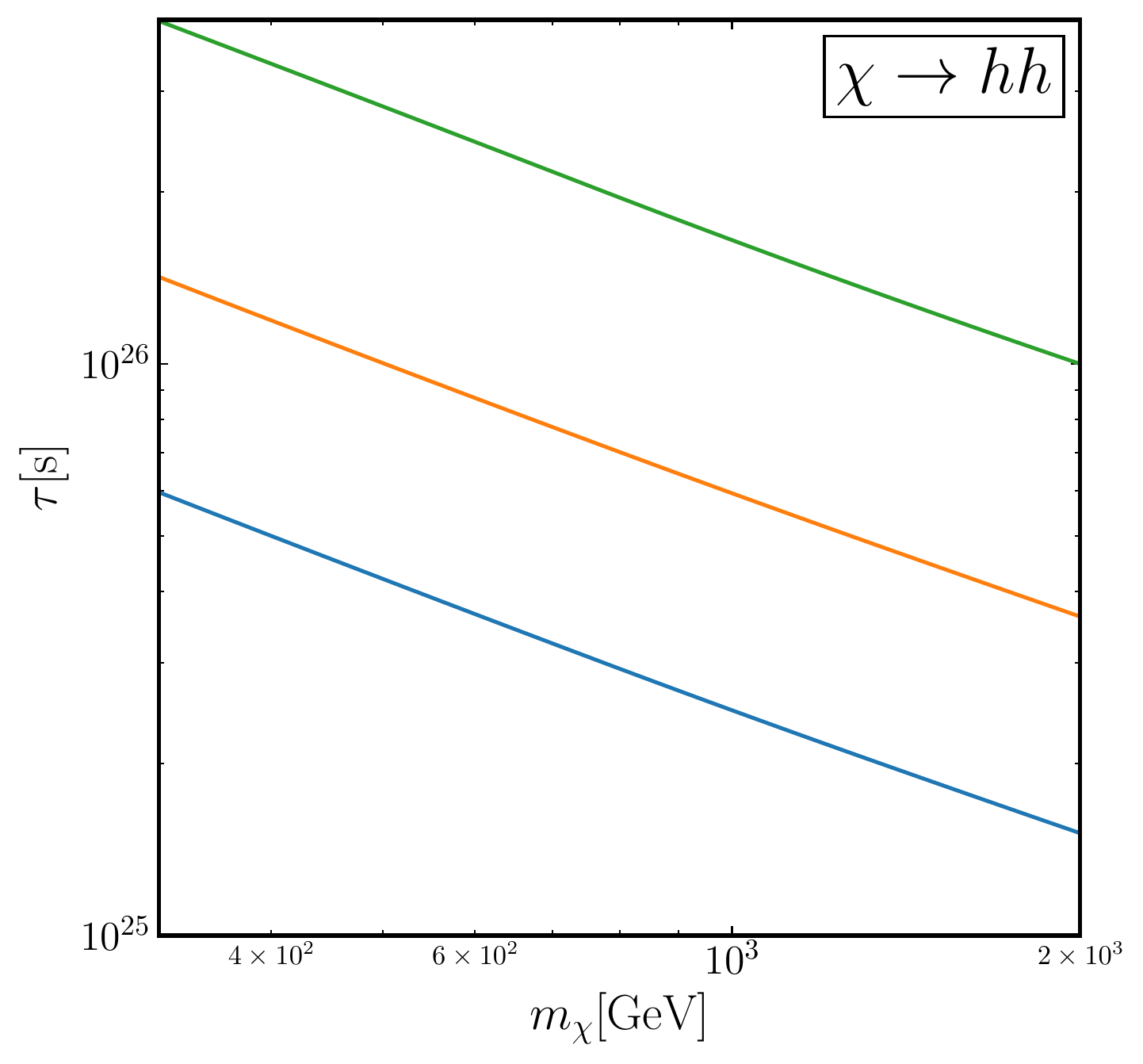}
    \includegraphics[width=.24\linewidth]{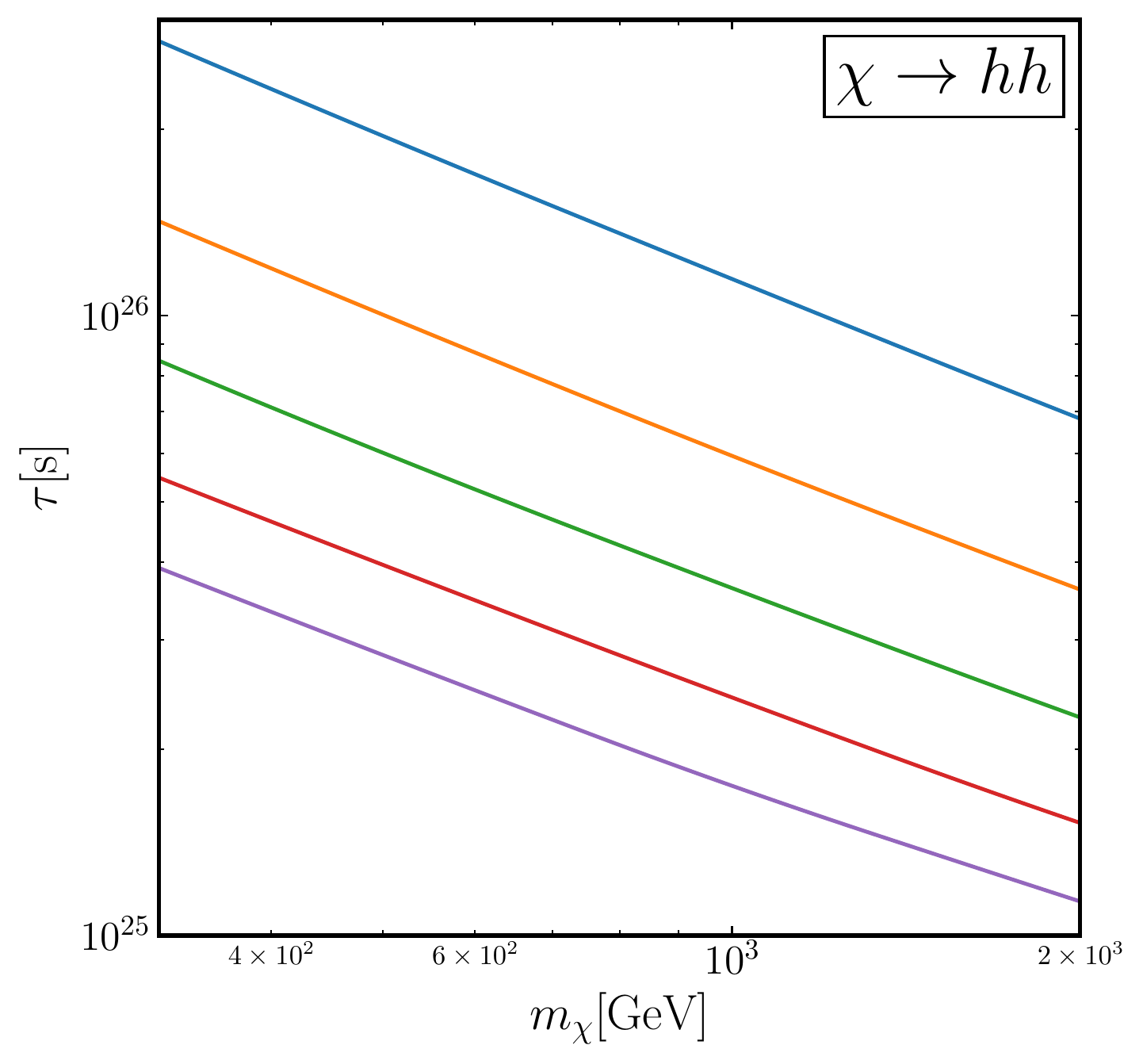}
    \includegraphics[width=.24\linewidth]{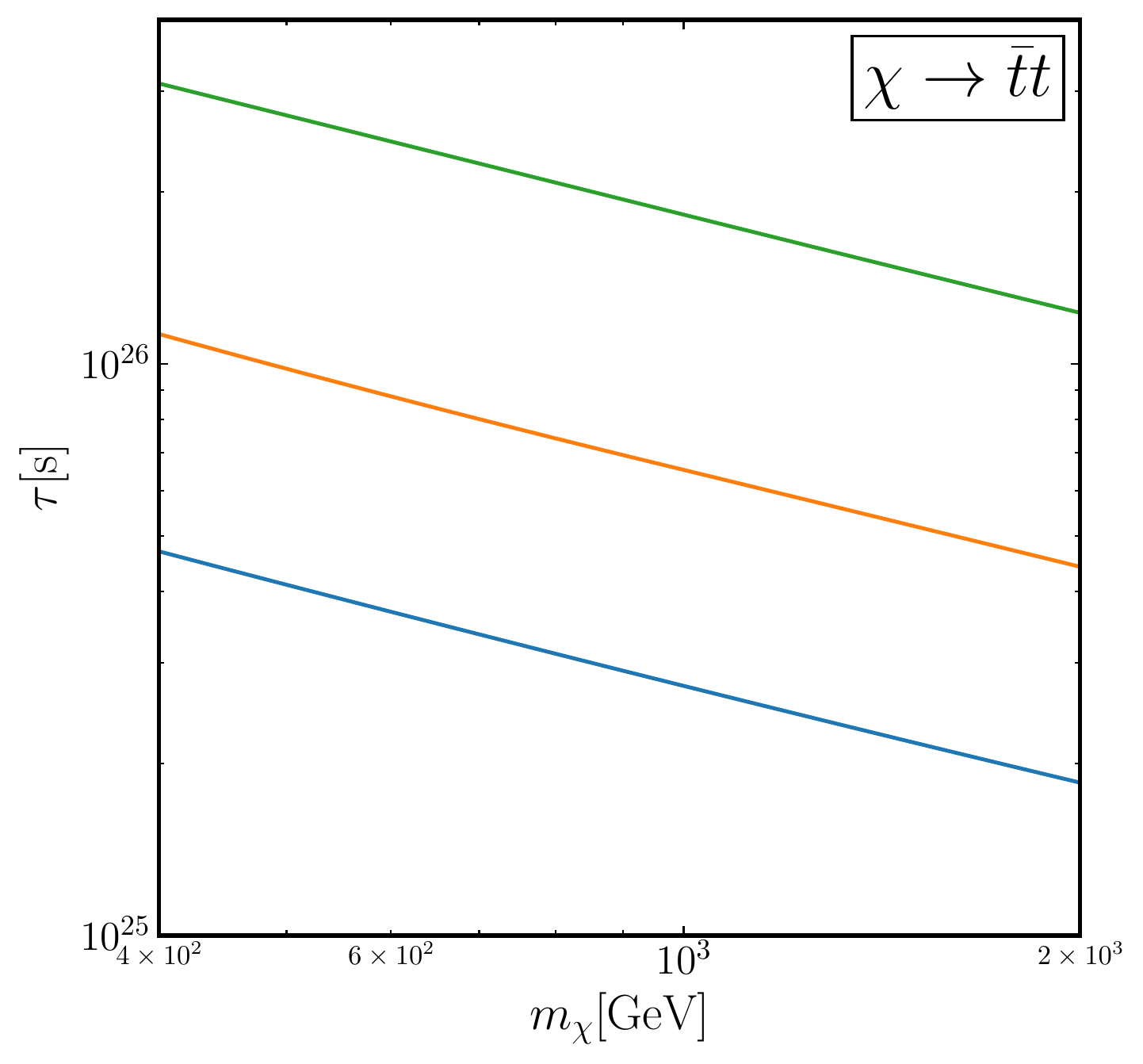}
    \includegraphics[width=.24\linewidth]{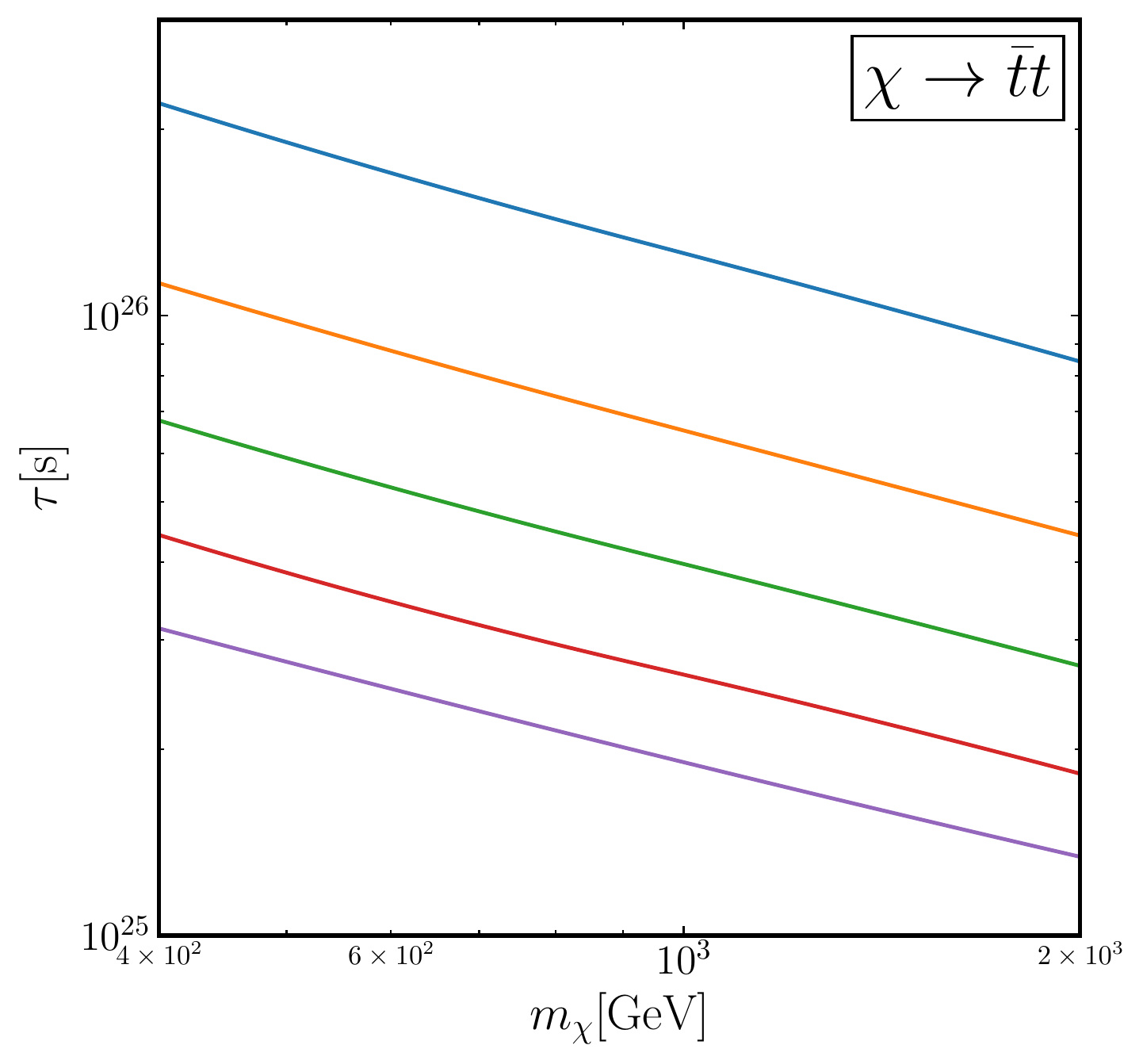}
    \caption{Bounds on the DM lifetime $\tau$ as a function of the DM mass $m_\chi$ for the single-component DM scenario, covering all decay channels except $\chi \to \gamma\gamma$ and $\chi \to e^+ e^-$, which are discussed in the main text. We show the dependence of the bounds on the amplitude threshold $\mathcal{T}/\mathcal{A}_{\Lambda\rm CDM}^{\rm max}$ and on the redshift of the signal $z_t$. The decay channel considered is indicated in the label within each panel. For readability, legends are omitted; the color coding matches that used in Figs.~\ref{fig:singlecomponentphoton} and~\ref{fig:singlecomponentsboundelectrons}. \textbf{Odd columns:} signal redshift fixed to $z_t = 15$. \textbf{Even columns:} amplitude threshold fixed to $\mathcal{T}/\mathcal{A}_{\Lambda\rm CDM}^{\rm max} = 0.50$.}
    \label{fig:singlecomponentsboundsAPP}
\end{figure}

\begin{figure}
    \centering
    \includegraphics[width=.32\linewidth]{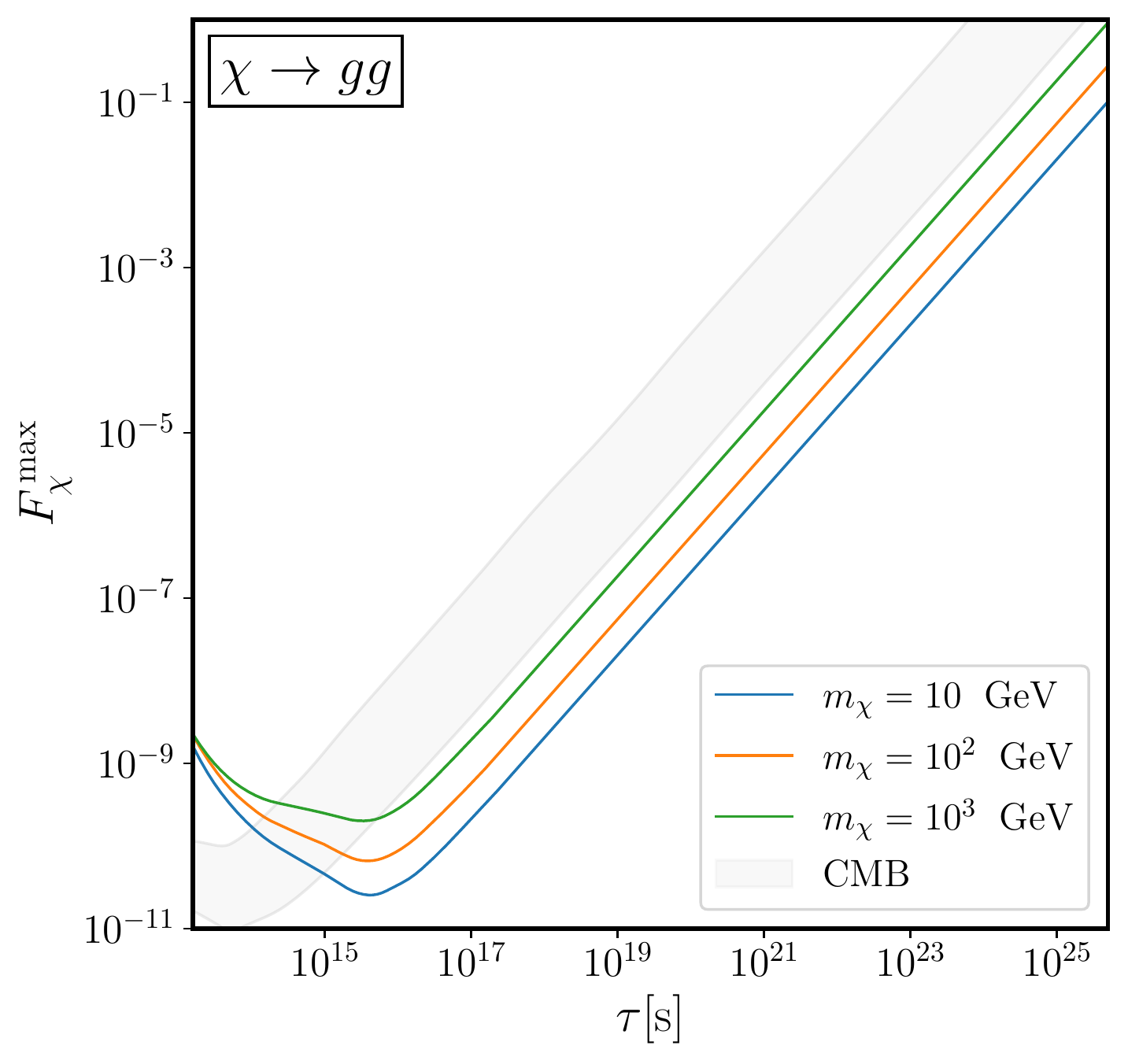}
    \includegraphics[width=.32\linewidth]{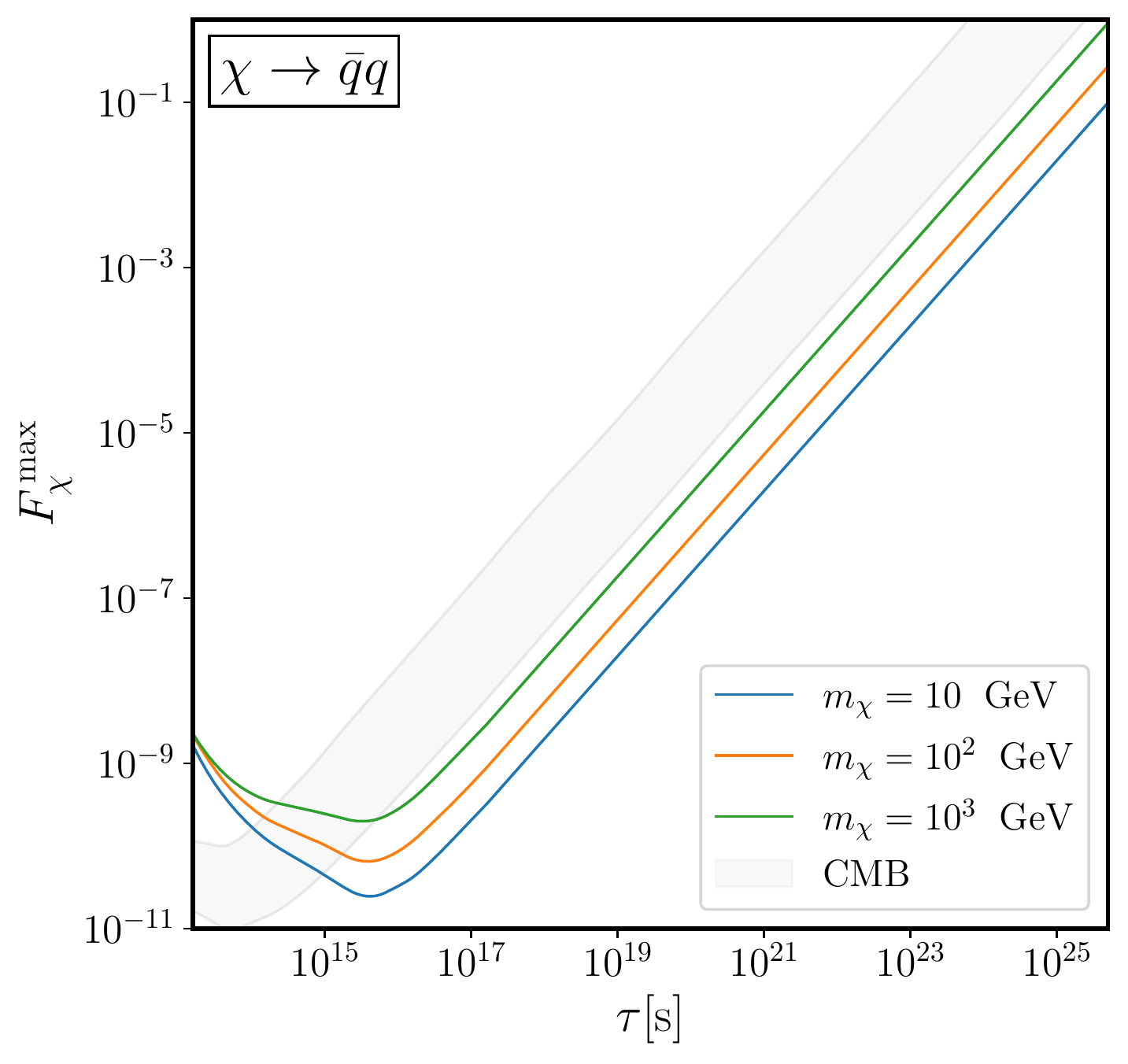}
    \includegraphics[width=.32\linewidth]{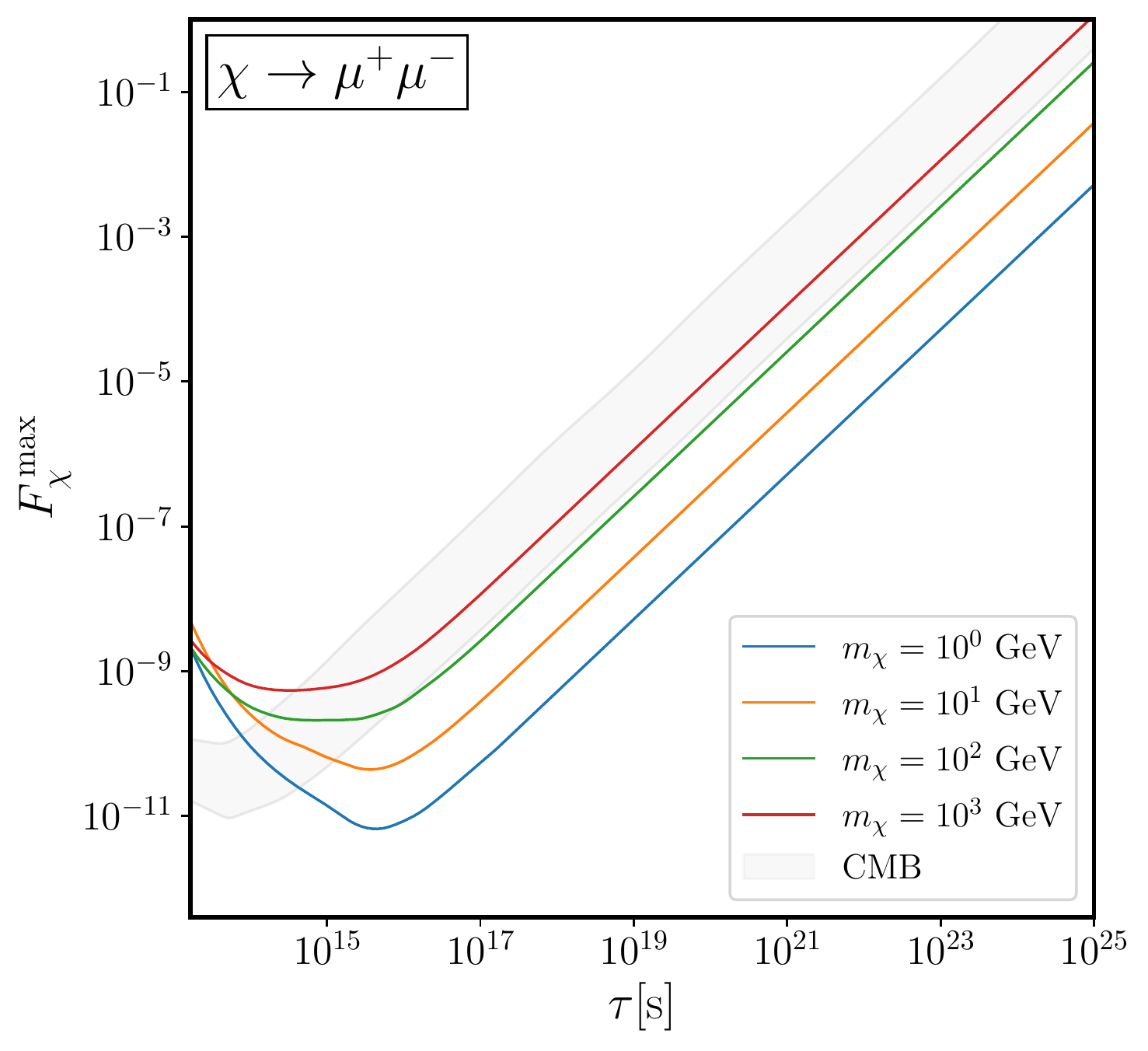} 
    \\
    \includegraphics[width=.32\linewidth]{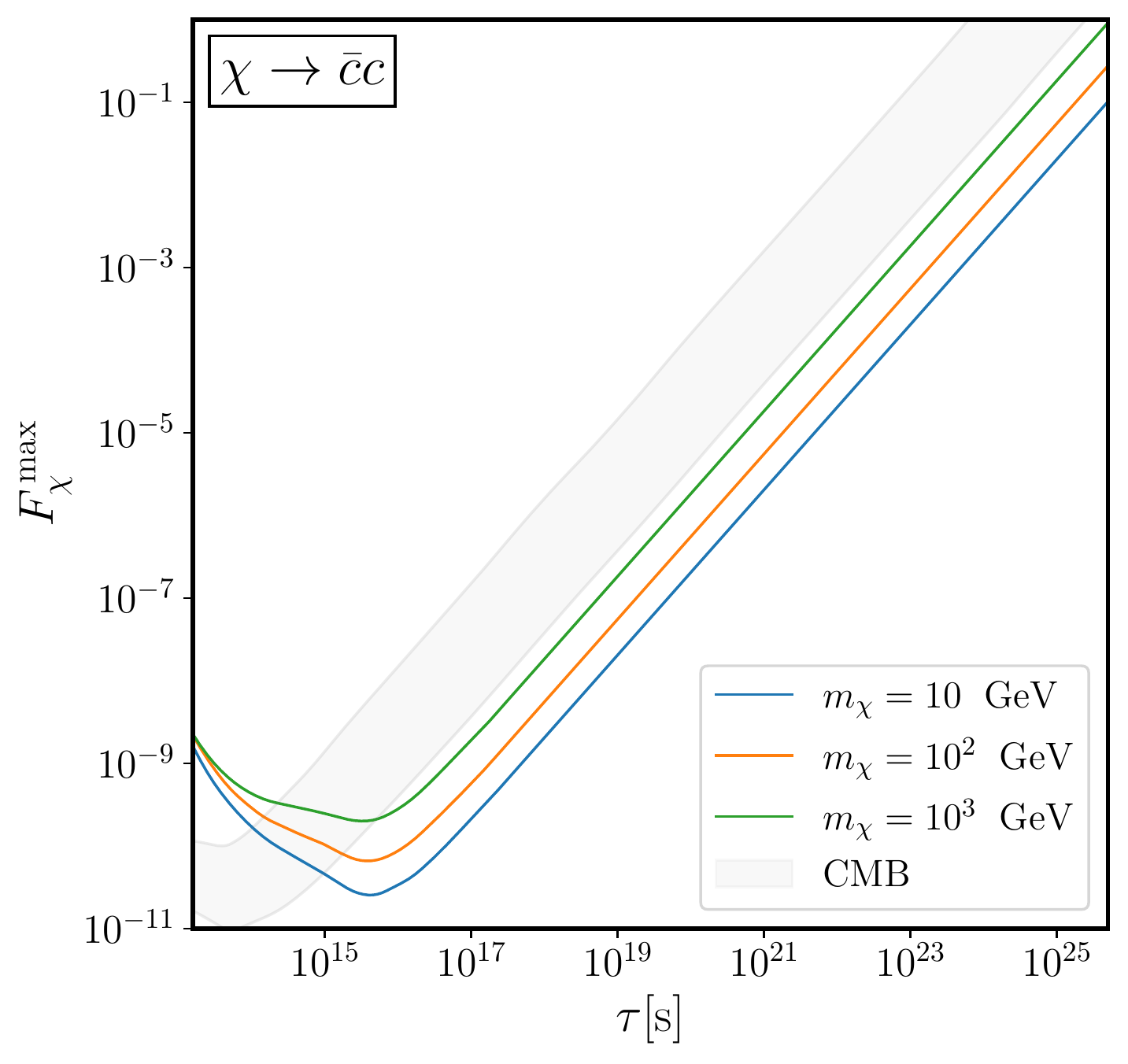}
    \includegraphics[width=.32\linewidth]{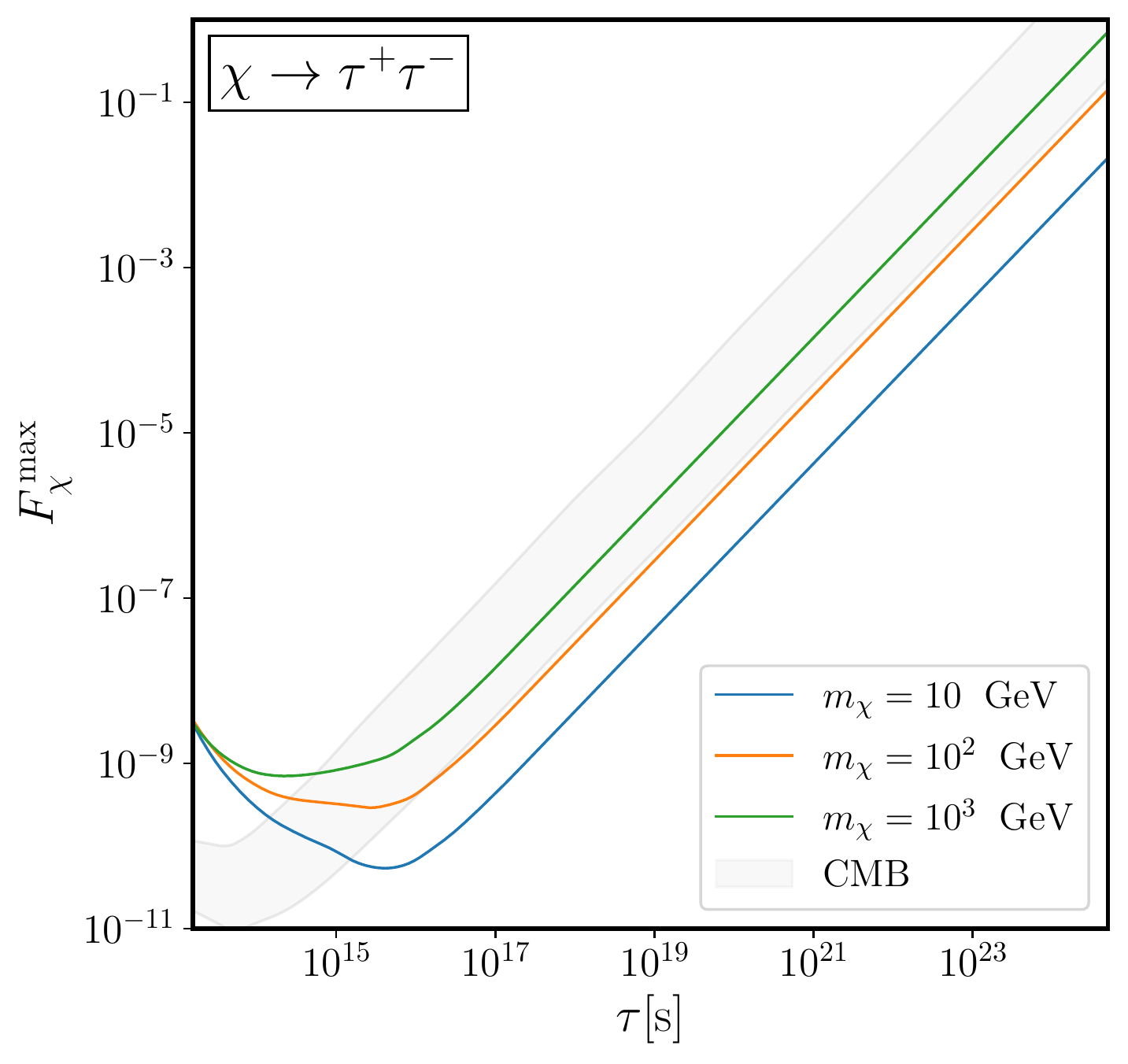}
    \includegraphics[width=.32\linewidth]{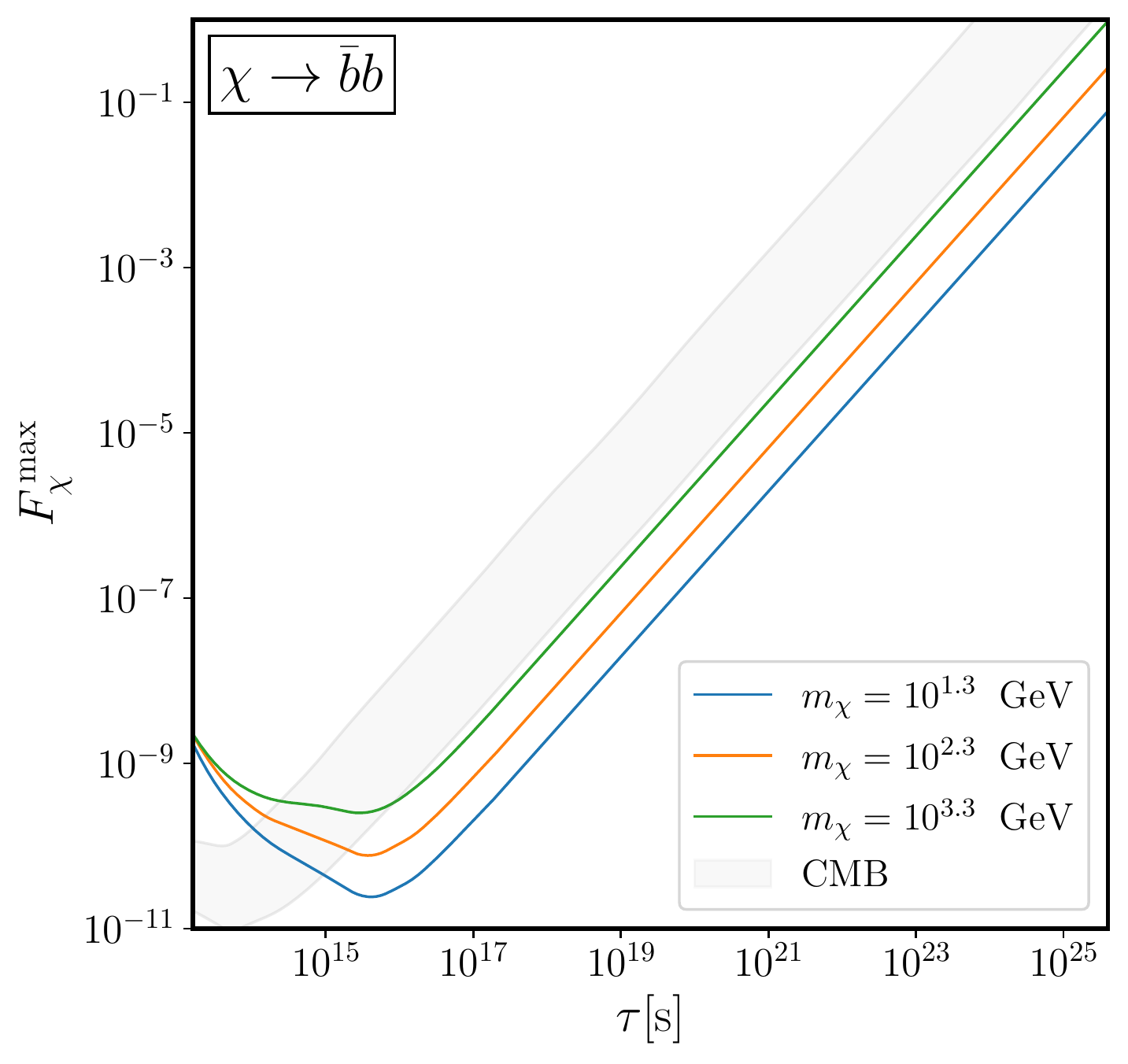}
    \\
    \includegraphics[width=.32\linewidth]{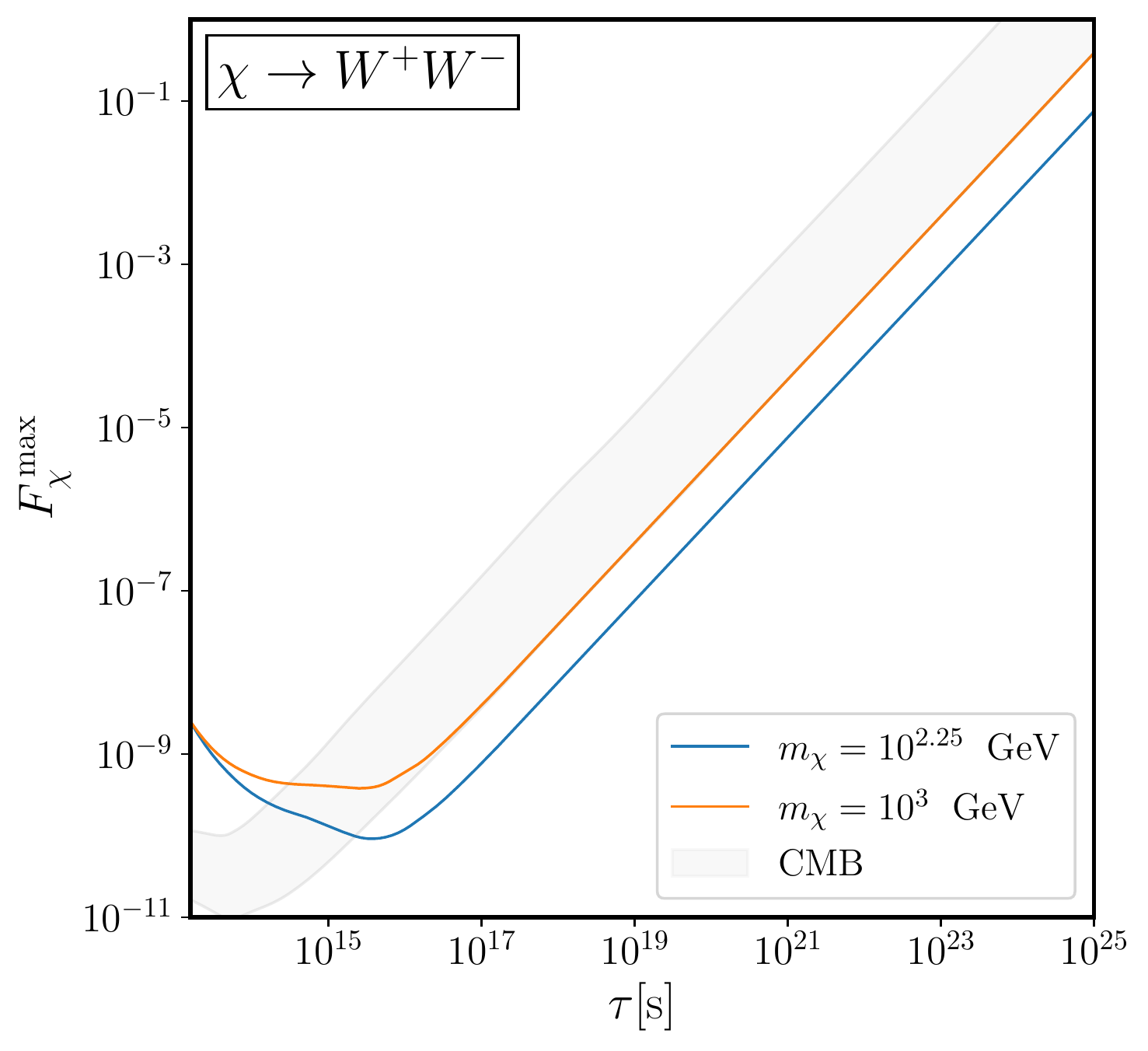}
    \includegraphics[width=.32\linewidth]{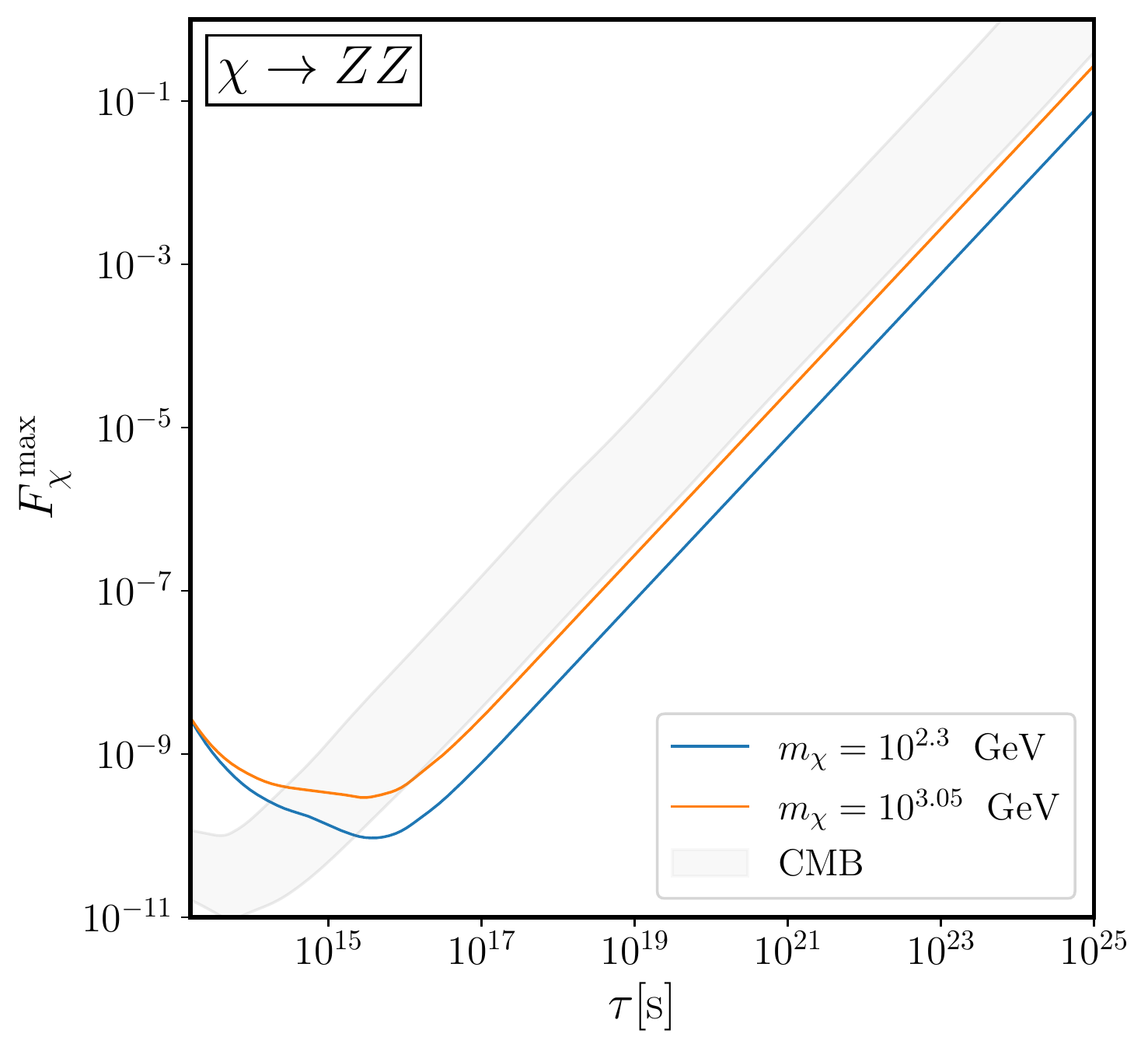}
    \includegraphics[width=.32\linewidth]{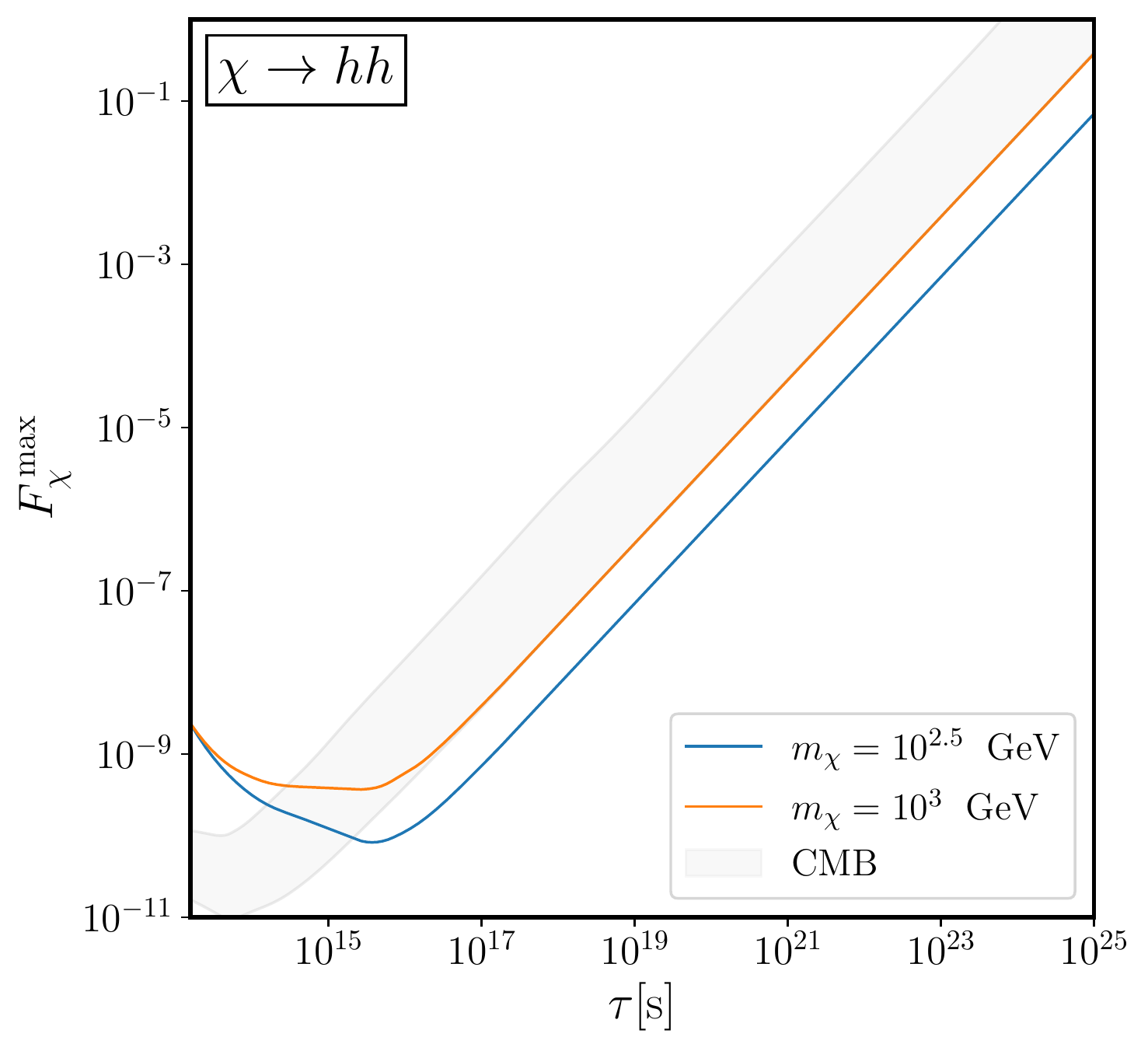}
    \\
    \includegraphics[width=.32\linewidth]{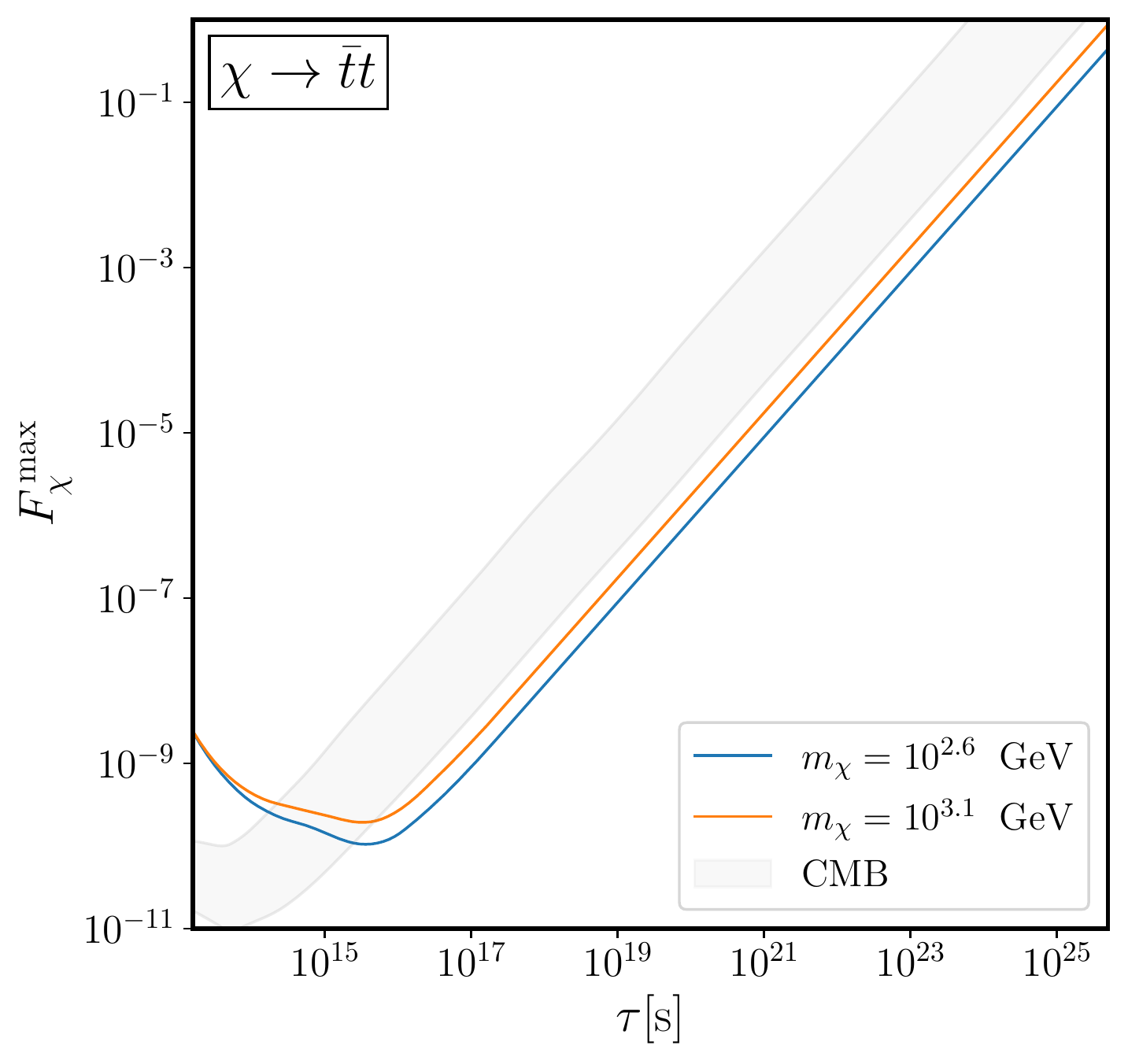}
    \caption{Upper limits on the DM mass fraction $F_\chi^{\rm max}$ as a function of the DM lifetime $\tau$. Each panel corresponds to a specific decay channel, as indicated in the label. All decay channels to SM particles are shown, except for $\chi \to \gamma\gamma$ and $\chi \to e^+ e^-$, which are discussed in the main text. Different colored lines represent different values of the DM mass $m_\chi$. The bounds are computed assuming $\mathcal{T}/\mathcal{A}_{\Lambda\rm CDM}^{\rm max} = 0.50$ and a fixed dip redshift $z_t = 15$. For comparison, the light gray band shows the range of upper bounds on the DM mass fraction derived from CMB anisotropy constraints \cite{sla:gen}, with the band width reflecting variations across injection species and DM masses.}
    \label{fig: sub-component bounds appendix}
\end{figure}

In this appendix, we present additional results not included in the main body of the paper. Specifically, we show bounds for all decay channels into SM particles and examine their dependence on the dip redshift $z_t$ and the threshold amplitude fraction $\mathcal{T}/\mathcal{A}_{\Lambda\rm CDM}^{\rm max}$. The results for the single-component DM scenario are displayed in Fig.~\ref{fig:singlecomponentsboundsAPP}, with the qualitative behavior of the bounds consistent with the discussion in Sec.~\ref{sec:DMbounds}. For the sub-component scenario, Fig.~\ref{fig: sub-component bounds appendix} presents the maximal DM abundance as a function of the DM lifetime for all possible decay channels into SM particles. The qualitative behavior of these bounds also matches that described in Subsec.~\ref{subsec:sub-component}.

\section{Decay Widths and Cross Sections: Useful Results}
\label{app:rates}

We collect useful general expressions to compute decay widths and scattering cross sections. Whether dealing with decays or scatterings, a common feature is the presence of final states with $n$ particles $B_i$ of masses $m_i$. We denote the four-momentum of particle $B_i$ by a capital letter $P_i$, and its associated spatial momentum by the corresponding lowercase letter, $\vec{p}_i$. The full four-momentum thus reads $P_i^\mu = (E_i, \vec{p}_i)$. The energy and magnitude of the spatial momentum (denoted by the same letter without the vector symbol, $|\vec{p}_i| = p_i$) are related via the dispersion relation $E_i^2 = p_i^2 + m_i^2$.

\subsection*{Lorentz invariant phase space}
The Lorentz-invariant $n$-body phase space that counts the final states is defined as
\begin{equation}
d\Phi^{(n)} \equiv (2\pi)^4 \delta^4 (P_{\rm init} - P_{\rm fin} ) \prod_{i=1}^n \frac{d^3 p_i}{2E_i (2\pi)^3} \ .
\label{eq:LIPS}
\end{equation}
Here, $P_{\rm init}$ is the total initial-state four-momentum, and the four-dimensional Dirac delta function enforces four-momentum conservation, with $P_{\rm fin} = \sum_i P_i$. Additional factors (e.g., for two identical particles) must be included manually. We introduce the Lorentz-invariant initial-state squared four-momentum $s \equiv P_{\rm init}^2$, and provide explicit results for two cases.

First, we consider two-body final states
\begin{equation}
    d\Phi^{(2)} = \frac{d\Omega}{32\pi^2} \sqrt{1 - 2\frac{m_1^2 + m_2^2}{s} + \frac{(m_1^2 - m_2^2)^2}{s^2}} \ ,
    \label{eq:2bPS}
\end{equation}
where $d\Omega$ is the infinitesimal solid angle element corresponding to the relative angle between the spatial momenta of the two particles, evaluated in the reference frame where the spatial components of $P_{\rm init}$ vanish. Examples include the rest frame of a decaying particle or the center-of-mass frame of a binary scattering.

The three-body phase space is more involved than the two-body case because the final-state particle momenta are no longer monochromatic in the frame where $P_{\rm init}$ has only a time-like component. Without loss of generality, we perform our analysis in this frame and use the three-dimensional Dirac delta function to eliminate one of the final-state spatial momenta. Choosing to integrate over $d^3 p_3$ leads to the constraint $\vec{p}_3 = - \vec{p}_1 - \vec{p}_2$. Consequently, the energy $E_3$ in the phase space is given by the dispersion relation $E_3 = \sqrt{(\vec{p}_1 + \vec{p}_2)^2 + m_3^2}$. To proceed, we consider how the squared amplitude $|\mathcal{M}_{A \rightarrow B_1 B_2 B_3}|^2$ depends on $\vec{p}_1$ and $\vec{p}_2$. We assume it depends only on the relative angle $\theta$ between these two momenta, such that $\vec{p}_1 \cdot \vec{p}_2 = p_1 p_2 \cos\theta$. This scenario arises, for example, when the decaying particle is a spinless boson. We use polar coordinates to perform the integrations over the spatial momenta. While all angular integrals are straightforward except for the one over $\theta$, we exploit the relation 
$E_3^2 = p_1^2 + p_2^2 + 2 p_1 p_2 \cos\theta + m_3^2$,
to change variables from $d\cos\theta$ to $dE_3$. Including the corresponding Jacobian factor, the three-body phase space element becomes
\begin{equation}
    d\Phi^{(3)} = \frac{dE_1 \, dE_2 \, dE_3}{32\pi^3} \, \delta(\sqrt{s} - E_1 - E_2 - E_3) \ .
    \label{eq:3bPS}
\end{equation}

\subsection*{Decays}

We consider the most general decay of a mother particle $A$ with mass $M$ into an $n$-body final state, $A \rightarrow B_1 \ldots B_n$. For this case, we identify $s = P_{\rm init}^2 = M^2$. The differential decay width is given by
\begin{equation}
    d\Gamma_{A \rightarrow B_1 \ldots B_n} = \frac{|\mathcal{M}_{A \rightarrow B_1 \ldots B_n}|^2}{2 M} \, d \Phi^{(n)} \ .
    \label{eq:GammaGeneral}
\end{equation}
The squared matrix element $|\mathcal{M}_{A \rightarrow B_1 \ldots B_n}|^2$ is understood to be averaged over initial and summed over final states. For two-body decays, the final-state particles are monochromatic, so the squared matrix element cannot depend non-trivially on the final-state momenta. The phase space integration is straightforward, yielding the total decay width
\begin{equation}
    \Gamma_{A \rightarrow B_1 B_2} = \frac{|\mathcal{M}_{A \rightarrow B_1 B_2}|^2}{16\pi M} \sqrt{1 - 2\frac{m_1^2 + m_2^2}{M^2} + \frac{(m_1^2 - m_2^2)^2}{M^4}} \ .
    \label{eq:Gamma2body}
\end{equation}
For three-body decays, we use the Dirac delta function to integrate over $dE_3$, leading to
\begin{equation}
    \frac{d\Gamma_{A \rightarrow B_1 B_2 B_3}}{dE_1 \, dE_2} = \frac{|\mathcal{M}_{A \rightarrow B_1 B_2 B_3}|^2}{64\pi^3 M} \ .
\end{equation}
This expression is non-vanishing only for values of $E_1$ and $E_2$ that satisfy energy-momentum conservation.

\subsection*{Scatterings}

We now consider a generic binary scattering process, $A_1 A_2 \rightarrow B_1 \ldots B_n$, with an arbitrary number of particles in the final state. The differential cross section is given by
\begin{equation}
    d \sigma_{A_1 A_2 \rightarrow B_1 \ldots B_n} = 
    \frac{|\mathcal{M}_{A_1 A_2 \rightarrow B_1 \ldots B_n}|^2}{4\, I(s)} \, d \Phi^{(n)} \ .
\end{equation}
We denote by $K_{1,2}$ and $M_{1,2}$ the four-momenta and masses of the initial-state particles, respectively. The incoming flux factor is defined as $I(s) \equiv \sqrt{(K_1 \cdot K_2)^2 - (M_1 M_2)^2}$, and it depends on the square of the initial state four-momentum, $s = (K_1 + K_2)^2$. As in the case of decays, the squared matrix element $|\mathcal{M}_{A_1 A_2 \rightarrow B_1 \ldots B_n}|^2$ is averaged over initial and summed over final states. The total cross section, obtained by integrating the expression above over all allowed final-state configurations, is a function of $s$ only. In our case of interest, $n = 2$, the corresponding two-body phase space element is given by \Eq{eq:2bPS} with $d\Omega$ denoting the differential element of the solid angle in the center-of-mass frame of the collision. To proceed further, one must determine the angular dependence of the matrix element. Poincaré invariance constrains this dependence to the scattering angle $\theta$. The azimuthal integration is then trivial, and we obtain the general expression for the differential scattering cross section
\begin{equation}
    \frac{d \sigma_{A_1 A_2 \rightarrow B_1 B_2}}{d \cos\theta} =  \frac{|\mathcal{M}_{A_1 A_2 \rightarrow B_1 B_2}|^2}{32 \pi s}  \frac{\sqrt{1 - 2\frac{m_1^2 + m_2^2}{s} + \frac{(m_1^2 - m_2^2)^2}{s^2}}}{\sqrt{1 - 2\frac{M_1^2 + M_2^2}{s} + \frac{(M_1^2 - M_2^2)^2}{s^2}}} \ .
    \label{eq:sigmageneral}
\end{equation}

\section{Boltzmann Equation for Freeze-In Production}
\label{app:BE}

In this Appendix, we summarize the Boltzmann formalism used to compute the abundance of metastable states that subsequently decay and impact the 21\,cm signal. In all cases considered in Sec.~\ref{sec:Lag}, the metastable states are produced via the freeze-in mechanism~\cite{Hall:2009bx}, implying that they never reach thermal equilibrium and that depletion processes can be neglected.

Our goal is to keep track of the number density $n_X$ of a generic metastable state $X$, and for this purpose it is convenient to scale out the effect of the Hubble expansion by introducing the comoving number density $Y_X = n_X / s$. Here, the entropy density of the primordial plasma as a function of temperature $T$ is given by the expression $s(T) = (2\pi^2 / 45) \, g_{*s}(T) \, T^3$, where $g_{*s}(T)$ denotes the effective number of relativistic degrees of freedom contributing to the entropy. The comoving number density evolves according to the Boltzmann equation
\begin{equation}
    \frac{dY_X}{d \ln T} = - \left(1 + \frac{1}{3} \frac{d \ln g_{*s}(T)}{d \ln T}\right) \frac{\sum_i \gamma_i(T)}{s(T) H(T)} \ .
    \label{eq:BE}
\end{equation}
Throughout this work, we assume a standard cosmological background dominated by radiation, in which the Hubble parameter takes the form $H(T) = (\pi \sqrt{g_*(T)} / (3\sqrt{10})) \, T^2 / M_{\rm Pl}$. Here, $g_*(T)$ is the temperature-dependent effective number of relativistic degrees of freedom contributing to the energy density, and $M_{\rm Pl} = (8\pi G)^{-1/2} \simeq 2.4 \times 10^{18} \, \text{GeV}$ is the reduced Planck mass. For the temperature dependence of $g_*(T)$ and $g_{*s}(T)$, we use the results of Ref.~\cite{Laine:2015kra}. Finally, the sum over $i$ includes all microscopic processes responsible for the production of $X$ particles. The quantities $\gamma_i(T)$ represent the temperature-dependent production rates, defined as the number of $X$ particles produced per unit time and per unit volume.

We provide three explicit expressions for the production rates when the underlying microscopic processes are decays, inverse decays, and binary scatterings. 
\begin{itemize}
\item \textbf{Decays.} When freeze-in production occurs via the decay of a bath particle $A$ through the process $A \rightarrow X$ (regardless of the number of additional final-state particles), the production rate is proportional to the partial decay width $\Gamma_{A \rightarrow X}$ and is given by
\begin{equation}
\gamma_{\rm decays}(T) = n_A^{\rm eq}(T) \, \Gamma_{A \rightarrow X} \, \frac{K_1[m_A / T]}{K_2[m_A / T]} \ .
\end{equation}
Here, $n_A^{\rm eq}(T)$ denotes the equilibrium number density of $A$ particles. In deriving its explicit form, we note that the Boltzmann equation for the comoving number density in \Eq{eq:BE} follows from the integro-differential Boltzmann equation for the phase space distribution, under the assumptions of Maxwell-Boltzmann statistics and kinetic equilibrium~\cite{DEramo:2023nzt}. Consistently, we neglect quantum statistical effects and use the Maxwell-Boltzmann expression for the equilibrium number density:
\begin{equation}
n_A^{\rm eq}(T) = \frac{g_A}{2 \pi^2} m_A^2 T \, K_2[m_A / T] \ .
\label{eq:nAeq}
\end{equation}
The thermal motion of the decaying particles leads to a Lorentz dilation of their lifetime, which is accounted for by the ratio of modified Bessel functions appearing in the production rate. In the high-temperature limit $T \gg m_A$, when the particles become ultra-relativistic, this ratio approaches $K_1[m_A / T] / K_2[m_A / T] \simeq m_A / (2 T)$, as expected from the increase of the effective lifetime due to relativistic time dilation.
\item \textbf{Inverse decays.} For freeze-in production via inverse decays, $A_1 \ldots A_n \rightarrow X$, the principle of detailed balance allows us to express the production rate in terms of the equilibrium number density of $X$ particles and the partial decay width of the decay process $X \rightarrow A_1 \ldots A_n$. The resulting expression is
\begin{equation}
\gamma_{\rm inv. decays}(T) = n_X^{\rm eq}(T) \, \Gamma_{X \rightarrow A_1 \ldots A_n} \, \frac{K_1[m_X / T]}{K_2[m_X / T]} \ .
\label{eq:rateinvdecays}
\end{equation}
\item \textbf{Binary scatterings.} For production via binary scatterings through $A_1 A_2 \rightarrow X$, where the final state may include an arbitrary number of bath particles, the production rate can be expressed as
\begin{equation}
\gamma_{\rm scattering}(T) = n^{\rm eq}_{A_1}(T) \, n^{\rm eq}_{A_2}(T) \, \langle \sigma_{A_1 A_2 \rightarrow X} \, v_M \rangle(T) \ .
\label{eq:scatteringrateTEMP}
\end{equation}
The thermally averaged cross section times the Møller velocity is given by
\begin{equation}
\langle \sigma_{A_1 A_2 \rightarrow X} \, v_M \rangle(T) = \frac{\int_{s_{\rm min}}^\infty ds \, \frac{\lambda(s, M_1, M_2)}{s^{1/2}} \, \sigma_{A_1 A_2 \rightarrow X}(s) \, K_1(\sqrt{s}/T)}{8 \, K_2(M_1/T) \, K_2(M_2/T) \, M_1^2 M_2^2 \, T} \ .
\label{eq:sigmaaveraged}
\end{equation}
Here, the Källén function is defined as $\lambda(x,y,z) \equiv [x - (y+z)^2][x - (y-z)^2]$, and $M_{1,2}$ are the masses of the initial-state particles. The integration variable $s$ is the Mandelstam invariant representing the squared center-of-mass energy. The lower integration limit is set by the kinematic threshold, $s_{\rm min} = \max\{(M_1 + M_2)^2, \, M_f^2\}$, where $M_f$ is the sum of the masses of the final-state particles. Combining these expressions and using the Maxwell-Boltzmann form for the equilibrium number densities from \Eq{eq:nAeq}, the final expression for the collision operator reads
\begin{equation}
\gamma_{\rm scattering}(T) = \frac{g_{A_1} g_{A_2}}{32 \pi^4} T \,
\int_{s_{\rm min}}^\infty ds \, \frac{\lambda(s, M_1, M_2)}{s^{1/2}} \, \sigma_{A_1 A_2 \rightarrow X}(s) \, K_1(\sqrt{s}/T) \ .
\label{eq:scatteringrate}
\end{equation}
\end{itemize}

As illustrated by these examples, the right-hand side of \Eq{eq:BE} depends solely on the bath temperature, with the unknown function $Y_X(T)$ not appearing explicitly. This allows us to integrate the Boltzmann equation and determine the comoving abundance. We perform this integration forward in time, starting from the reheat temperature $T_R$, defined as the highest temperature of the primordial bath during the radiation-dominated epoch, down to a generic temperature $T < T_R$. The general solution is
\begin{equation}
Y_X(T) = Y_X(T_R) + \int_{T}^{T_R} \frac{d T^\prime}{T^\prime} \left(1 + \frac{1}{3} \frac{d \ln g_{*s}(T^\prime)}{d \ln T^\prime} \right) \frac{\sum_i \gamma_i(T^\prime)}{s(T^\prime) H(T^\prime)} \ .
\label{eq:YXsol}
\end{equation}
Throughout this work, when setting observational bounds, we assume that inflationary reheating creates a thermal bath with essentially no initial $X$ particles, meaning $Y_X(T_R) = 0$. If reheating instead produces some nonzero initial abundance, $Y_X(T_R) > 0$, then our calculation provides only a minimal estimate of the $X$ particle abundance in the early universe. In that case, the bounds we derive are conservative.

In our analysis, we do not need to track the evolution of $Y_X$ throughout the entire expansion history of the universe. Instead, we focus exclusively on its value at the time when the $X$ particles decay. Particle production is efficient only at very early times, when the age of the universe is much shorter than the $X$ lifetime. Furthermore, the production rates $\gamma_i$ become exponentially suppressed at temperatures below the masses of the particles involved in the relevant processes, which happens well before the decays occur. As a result, extending the integral in \Eq{eq:YXsol} to sufficiently low temperatures makes the comoving abundance insensitive to the lower integration limit. Therefore, the asymptotic comoving density at the decay time can be approximated by  
\begin{equation}
Y_X^\infty \simeq \int_{0}^{T_R} \frac{d T}{T} \left(1 + \frac{1}{3} \frac{d \ln g_{*s}(T)}{d \ln T} \right) \frac{\sum_i \gamma_i(T)}{s(T) H(T)} \ .
\label{eq:YXinfty}
\end{equation}

The temperature dependence of the production rate determines the epoch when particle production is most efficient. In all cases considered in this work, the leading behavior follows a power-law scaling, $\gamma_i(T) \propto T^\alpha$, which holds until the onset of the aforementioned Maxwell-Boltzmann suppression at a characteristic temperature denoted by $T_{\rm FI}$. By inserting this power-law form for $\gamma_i(T)$ into the expression in \Eq{eq:YXinfty}, and neglecting the subleading temperature dependence of $g_{*s}$, we obtain the following scaling for the asymptotic solution
\be
Y_X^\infty \propto \int^{T_R}_{T_{\rm FI}} dT \, T^{(\alpha - 6)} = 
\frac{T_R^{(\alpha - 5)} - T_{\rm FI}^{(\alpha - 5)}}{\alpha - 5}  \qquad \qquad (\alpha \neq 5) \ .
\label{eq:YXscaling}
\ee
From the expression above, we observe a clear separation into two distinct regimes, with the specific case $\alpha = 5$ acting as a dividing line. For $\alpha < 5$, the integral is dominated by values around $T_{\rm FI}$: particle production is most efficient at low temperatures, and we refer to this as IR-dominated freeze-in production~\cite{Hall:2009bx}. In contrast, for $\alpha > 5$, the integral receives its dominant contribution from the region around $T_R$: production is most efficient at high temperatures, and we describe this as UV-dominated freeze-in. The intermediate case $\alpha = 5$ does not arise in our study. Instead, we encounter only three types of freeze-in processes: inverse decays, with $\alpha = 2$ (IR-dominated); binary scatterings mediated by renormalizable interactions, with $\alpha = 4$ (IR-dominated); and binary scatterings mediated by dimension-5 operators, with $\alpha = 6$ (UV-dominated).

\section{ALP Production Rates}
\label{app:ALP}

The ALP production rate via the photon coupling in \Eq{eq:aFFdual} is always proportional to $g_{a\gamma\gamma}^2$. To compute this rate, it is convenient to rewrite the interaction by expanding the photon field strength as
\begin{equation}
    \mathcal{L}_{a\gamma\gamma} = \frac{g_{a\gamma\gamma}}{4} \, a F_{\mu\nu} \tilde{F}^{\mu\nu} =
    \frac{g_{a\gamma\gamma}}{2} \, a \, \epsilon^{\mu\nu\rho\sigma} \, \partial_\mu A_\nu \, \partial_\rho A_\sigma \ .
    \label{eq:aFFdualv2}
\end{equation}
We identify three main production channels: inverse decays, the Primakoff process, and fermion/antifermion annihilations. As discussed in Sec.~\ref{subsec:ALP}, the contribution from $W^{\pm}$ annihilations is Maxwell--Boltzmann suppressed in the reheat temperature range of interest. Similarly, we neglect the contribution from charged pion annihilations in the confined phase, as we evaluate the rate in a temperature regime where this channel is also suppressed.

\subsection*{Inverse Decays}

The production via inverse decays has a rate provided by \Eq{eq:rateinvdecays}. The only missing ingredient is the expression for the decay width of the axion into two photons. First, we evaluate the squared matrix element from the Lagrangian in \Eq{eq:aFFdualv2} and find the result
\begin{equation}
\squared{\mathcal{M}_{a \rightarrow \gamma \gamma}} = 2 \, g_{a\gamma\gamma}^2 \left( P_1 \cdot P_2\right)^2 = \frac{g_{a\gamma\gamma}^2 m_a^4}{2} \ ,
\end{equation}
where $P_{1,2}$ are the four-momenta of the final-state photons. In the second equality we imposed four-momentum conservation and, as anticipated, the squared matrix element does not feature any non-trivial momentum dependence. The partial decay width is obtained from \Eq{eq:Gamma2body} with an additional factor of $1/2$ to account for the two identical photons in the final state
\begin{equation}
\Gamma_{a \rightarrow \gamma \gamma} = \frac{g_{a\gamma\gamma}^2 m_a^3}{64 \pi} \ .
\label{eq:Gammaaggg}
\end{equation}
This equation assumes the final-state photons to be massless. However, as discussed in the main text, thermal effects generate a non-vanishing temperature-dependent mass that kinematically forbids this decay process. Furthermore, ALP production via this process is IR-dominated (see \Eq{eq:YXscaling}) and, for the tiny ALP mass considered here, would be negligible even if we neglect the thermally induced photon mass. For these reasons, we do not include this contribution in our analysis.

\subsection*{Primakoff process}

The rate calculation for the Primakoff production poses severe challenges due to an unpleasant IR behavior of the transition amplitude. In particular, the exchange of a massless photon leads to singularities that need to be regulated by thermal masses. This issue was first addressed by Ref.~\cite{Braaten:1991dd} via a hard thermal loop resummation. We report here the result of Ref.~\cite{bol:the}, valid for a QED plasma with temperature $T$ much larger than the electron mass
\be
\gamma_P^{(e^{\pm})}(T) = \frac{\alpha g_{a\gamma\gamma}^2 \zeta(3) \, T^6}{12\pi^2} \left[\ln \left(\frac{T^2}{m_\gamma^2}\right) + 0.8194\right] \ .
\label{eq:PrateQCD}
\ee
The result above carries two main limitations once one tries to employ it in our framework. First, it includes only the contribution from the electron field, but other charged fermions are present in the primordial bath for the reheat temperature range under investigation. Second, it does not include the electron mass; while this is acceptable for the electron, it is not justified for heavier charged fermions that could contribute to the production process. Thus, we need to extend the expression in \Eq{eq:PrateQCD}, valid for a QED plasma, to a more general one accounting for all charged fermions.

We follow the prescription provided by Ref.~\cite{cad:cos}. The number density for the excitations of a generic fermion field $\psi$, including contributions from both particles and antiparticles, is obtained by integrating its distribution over phase space
\begin{equation}
n_{\psi}(T) + n_{\bar\psi}(T) = 2 \, g_\psi \int\frac{d^3p}{(2\pi)^3} f_\psi(E,T) = 2 \, \frac{g_\psi}{2\pi^2}\int_{m_\psi}^\infty dE \frac{\sqrt{E^2 - m_\psi^2} \, E}{\exp[E/T] + 1} = \mathcal{N}_\psi(T) \frac{\zeta (3)}{\pi^2} T^3 \ ,
\label{eq:NpsiDEF}
\end{equation}
where $g_\psi$ is the number of internal degrees of freedom of the fermion, and the overall factor of $2$ accounts for the identical contribution from the fermion antiparticle. In the second equality, we employ polar coordinates to perform the phase space integral using the dispersion relation $E^2 =  p^2 + m_\psi^2$ and write down the explicit form of the Fermi-Dirac distribution. In the last equality, we define the effective contribution of $\psi$ particles in terms of effective degrees of freedom to the total number density. For a massless Dirac fermion, such as the electron field at temperatures above the MeV scale, we have $\mathcal{N}_{e^\pm}(T \gg m_e) = (3/4) \times (2 + 2) = 3$.

For Primakoff production, the process involves either a fermion or an antifermion in the initial state. The resulting rate is proportional to its number density and takes the form given in \Eq{eq:gammaP}, with the temperature-dependent function $g_P(T)$ defined as
\begin{equation}
g_P(T) = \sum_\psi Q_\psi^2 \, \mathcal{N}_\psi(T) \ ,
\label{eq:gPT}
\end{equation}
where $Q_\psi$ is the electric charge of $\psi$ in units of the absolute value of the electron charge. This procedure accounts only for the Maxwell--Boltzmann suppression of the initial-state fermion, but does not include the effect of its finite mass in the transition amplitude. Evaluating the corresponding mass correction lies beyond the scope of this work, and we therefore restrict ourselves to the expression in \Eq{eq:gPT}.

\begin{figure}
    \centering
    \includegraphics[width=.48\linewidth]{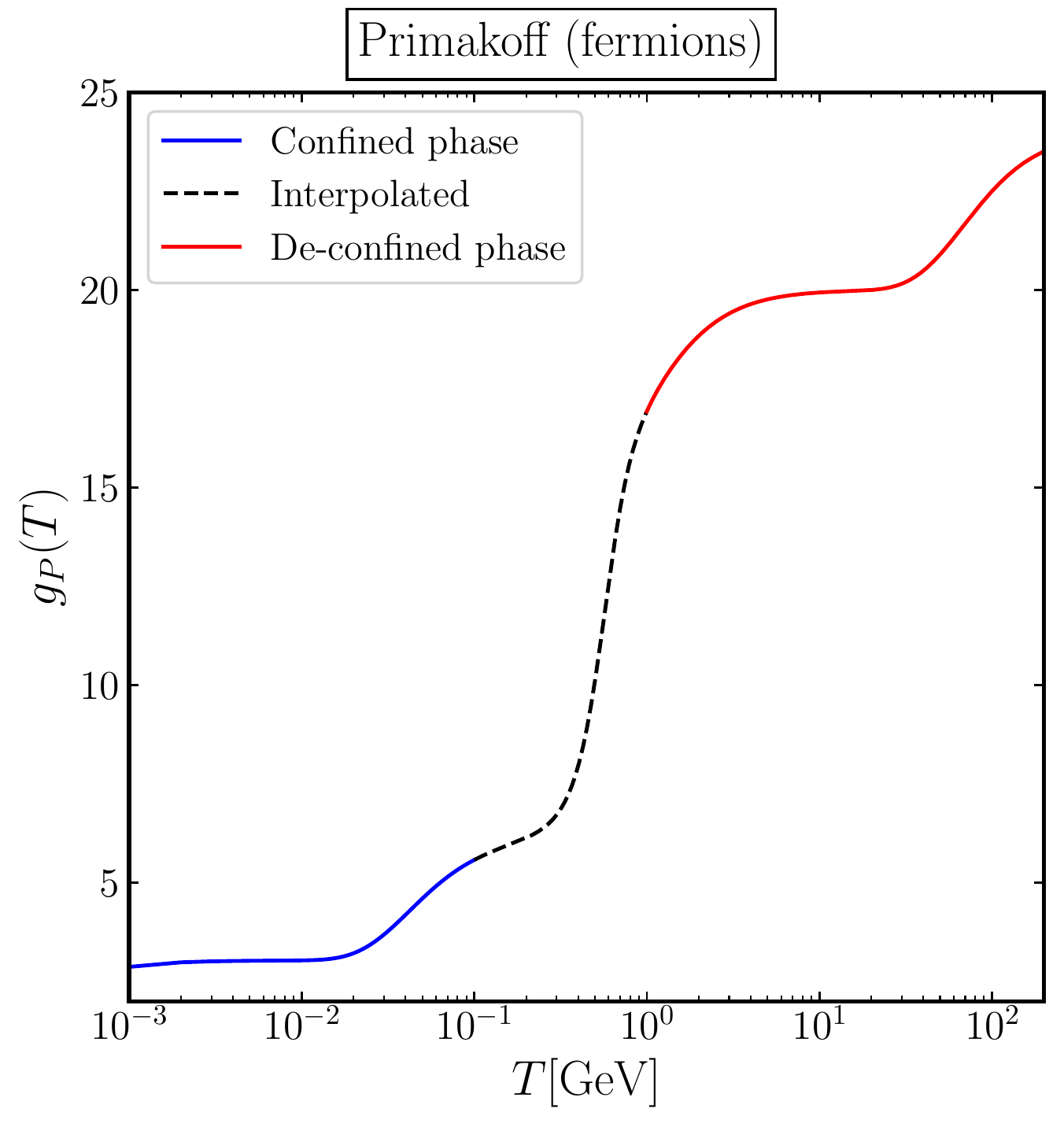}
    \includegraphics[width=.48\linewidth]{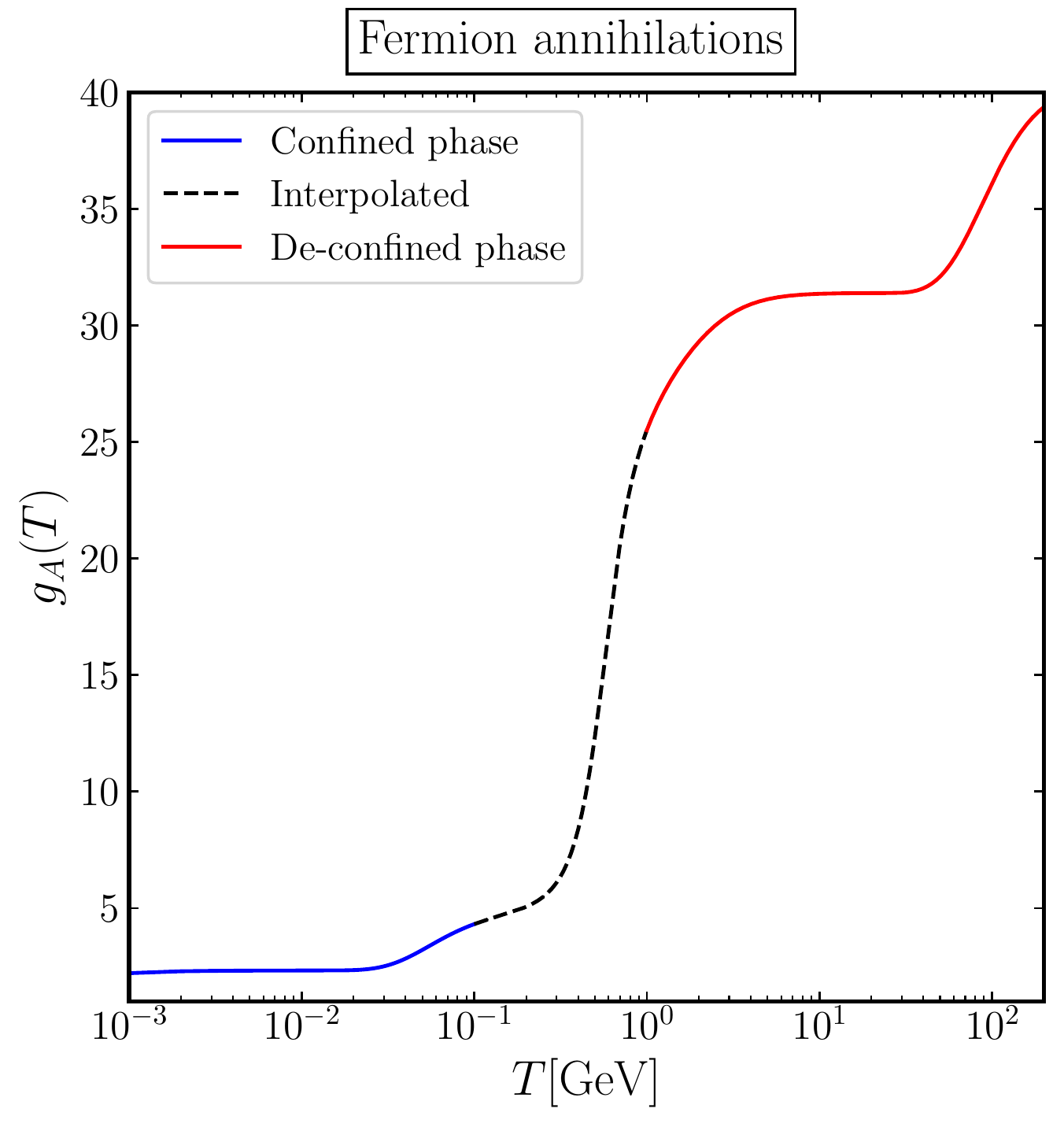}
    \caption{Numerical evaluation for the functions $g_P(T)$ (left panel) and $g_A(T)$ (right panel). For the temperature range $100\text{ MeV}<T<1\text{ GeV}$, the result is obtained via a smooth interpolation between the high and low temperature regimes.}
    \label{fig:gPA}
\end{figure}

It is straightforward to evaluate $g_P(T)$ numerically. First, we compute the temperature-dependent functions $\mathcal{N}_\psi(T)$ for each fermion $\psi$ using the definition provided in \Eq{eq:NpsiDEF}, and then sum over all SM fermions as prescribed by \Eq{eq:gPT}. This procedure is reliable both well above and well below the QCD phase transition, where the strongly interacting SM degrees of freedom are in the deconfined and confined phases, respectively. In contrast, the intermediate temperature range is notoriously difficult to model, as quarks and gluons confine into hadrons during this epoch, leading to a dramatic change in the plasma’s degrees of freedom. A viable approach in such cases is to perform the calculation in the two well-defined regimes and then interpolate smoothly between them to obtain a fiducial description of the transition region. This strategy has been successfully employed in studies of sterile neutrino production~\cite{Venumadhav:2015pla}, axions coupled to gluons~\cite{DEramo:2021psx,DEramo:2021lgb}, and axions coupled to photons~\cite{Caloni:2022uya}. We adopt the same approach here, interpolating our calculations between $100$ MeV and $1$ GeV. The numerical result for $g_P(T)$ is shown in the left panel of Fig.~\ref{fig:gPA}.

\subsection*{Fermion/antifermion annihilations}

The only Feynman diagram contributing to ALP production via fermion--antifermion annihilation involves a photon exchanged in the $s$-channel, and therefore no IR singularity arises. The squared matrix element reads
\begin{equation}
\squared{\mathcal{M}_{\psi \bar\psi \rightarrow \gamma a}} = \pi \alpha Q_\psi^2 g_{a\gamma\gamma}^2 \left[ s + 2t + \frac{2t^2}{s} + 2m_\psi^2 
\frac{m_\psi^2 - 2t}{s} + m_a^2 \frac{m_a^2 (s + 2m_\psi^2) - 2s (s + t + m_\psi^2)}{s^2} \right] \ .
\end{equation}
Here, the Mandelstam variables are defined as usual: $s = (K_1 + K_2)^2$, where $K_1$ and $K_2$ are the four-momenta of the initial-state fermions, and $t = (K_1 - P)^2$, with $P$ the four-momentum of the outgoing photon. The total scattering cross section is obtained by inserting the squared matrix element into the general expression given in \Eq{eq:sigmageneral} and integrating over the scattering angle. We find
\begin{equation}
\sigma_{\psi \bar\psi \rightarrow \gamma a}(s) = 
\frac{\alpha Q_\psi^2 g_{a\gamma\gamma}^2}{24} \frac{\left(1 + \frac{2 m_\psi^2}{s}\right) \left(1 - \frac{m_a^2}{s}\right)^3}{\sqrt{1 - \frac{4 m_\psi^2}{s}}} \ .
\end{equation}

The resulting interaction rate for this specific channel can be obtained from the combination of the results in \Eqs{eq:scatteringrate}{eq:sigmaaveraged}. We use Maxwell-Boltzmann statistics to evaluate the equilibrium number density of the initial-state fermions, and we find the expression
\begin{equation}
\gamma_A^{(\psi/\bar\psi)}(T) = g_\psi g_{\bar\psi}
\frac{T}{32\pi^4} \int_{4 m_\psi^2}^\infty ds \, \sqrt{s} (s - 4 m_\psi^2) \, \sigma_{\psi \bar\psi \rightarrow \gamma a}(s) \, K_1(\sqrt{s}/T) \ .
\end{equation}
The number of internal degrees of freedom for particles and antiparticles is the same, \(g_\psi = g_{\bar\psi}\), but it is convenient to keep them separate in order to count correctly the multiplicity factors in our final expression. It is useful to perform the integration analytically in the limit where all particles are massless. This approximation is certainly justified for the ALP mass, and it is also valid for the fermion mass when the temperature satisfies \(T \gtrsim m_\psi\). Since the production is UV dominated, finite fermion mass corrections can be relevant only in the regime where \(T_R \simeq m_\psi\). This limit is analogous to what we have done for the previous case with Primakoff production. We find the result
\begin{equation}
\gamma_A^{(\psi/\bar\psi)}(T) \simeq g_\psi g_{\bar\psi}
\frac{\alpha Q_\psi^2 g_{a\gamma\gamma}^2 T}{768 \pi^4} \int_{0}^\infty ds \, s^{3/2} \, K_1(\sqrt{s}/T) = g_\psi g_{\bar\psi} \frac{\alpha Q_\psi^2 g_{a\gamma\gamma}^2 T^6}{24 \pi^4} \ .
\end{equation}
The total rate is given by the sum of the above expression over all SM fermions. Since the initial state includes a fermion and an antifermion, the effective number of degrees of freedom must be computed accordingly. The definition of \(\mathcal{N}_\psi(T)\) in \Eq{eq:NpsiDEF} is such that this factor includes both particles and antiparticles, and for this reason the total interaction rate is given by the expression
\begin{equation}
\gamma_A(T) \simeq g_A(T) \frac{2 \, \alpha g_{a\gamma\gamma}^2 T^6}{27 \pi^4} \ ,
\end{equation}
where the effective coupling \(g_A(T)\) is defined as
\begin{equation}
g_A(T) = \sum_\psi Q_\psi^2 \left(\frac{\mathcal{N}_\psi(T)}{2}\right)^2 \ .
\label{eq:gAT}
\end{equation}
Once the overall normalization is adjusted, the total rate for this channel is given by the expression in \Eq{eq:gammaA}.

\bibliographystyle{JHEP}
\bibliography{biblio.bib}

\end{document}